\documentclass[%
reprint,
superscriptaddress,
 amsmath,amssymb,
 aps,
]{revtex4-2}
\usepackage[english]{babel}
\usepackage{color}
\usepackage{gensymb}
\usepackage{graphicx}% Include figure files
\usepackage{dcolumn}% Align table columns on decimal point
\usepackage{bm}% bold math
\usepackage{hyperref}% add hypertext capabilities
\usepackage{epstopdf}
\usepackage{float}
\usepackage{listings}
\usepackage{times,mathptm,bm}
\usepackage{xcolor}
\usepackage[export]{adjustbox}
\usepackage{mhchem}[version=4]

\begin{document}

%\preprint{APS/123-QED}
\title{Collecting Particles in Confined Spaces by Active Filamentous Matter}

\author{R.~Sinaasappel}
\thanks{These two authors contributed equally}
\affiliation{Van der Waals-Zeeman Institute, Institute of Physics, University of Amsterdam, 1098XH Amsterdam, The Netherlands.}
\author{K.~R.~Prathyusha}
\thanks{These two authors contributed equally}
\affiliation{School of Chemical and Biomolecular Engineering, Georgia Institute of Technology, Atlanta, GA 30332, USA.}
\author{Harry Tuazon}
\affiliation{School of Chemical and Biomolecular Engineering, Georgia Institute of Technology, Atlanta, GA 30332, USA.}
\author{E.~Mirzahossein}
\affiliation{Van der Waals-Zeeman Institute, Institute of Physics, University of Amsterdam, 1098XH Amsterdam, The Netherlands.}
\author{P.~Illien}
\affiliation{Sorbonne Universit\'e, CNRS, Physicochimie des Electrolytes et Nanosyst\`emes Interfaciaux (PHENIX), Paris, France.}
\author{Saad Bhamla}
\email{saadb@chbe.gatech.edu}
\affiliation{School of Chemical and Biomolecular Engineering, Georgia Institute of Technology, Atlanta, GA 30332, USA.}
\author{A.~Deblais}
\email{a.deblais@uva.nl}
\affiliation{Van der Waals-Zeeman Institute, Institute of Physics, University of Amsterdam, 1098XH Amsterdam, The Netherlands.}

\date{\today}
\begin{abstract}
Biological and robotic systems often operate in confined environments where material must be gathered without centralized control. Inspired by the effective collection strategies of aquatic worms (\textit{Lumbriculus variegatus} and \textit{Tubifex tubifex}), we investigate how active filaments autonomously aggregate dispersed particles. We study this process across four platforms: living worms, a robotic chain, Brownian dynamics simulations of active polymers, and a coarse-grained toy model. We show that aggregation emerges from repeated contact and body deformation, and demonstrate that clustering dynamics are governed by filament length and bending stiffness. Across systems, particle gathering follows a shared aggregation–fragmentation process, with the steady-state cluster size scaling as $\langle s\rangle_L\sim (W/D)^2$, where $W$ is the effective width of the path cleared by the filament and $D$ the domain size. We find that filament flexibility modulates $W$, enabling more flexible filaments to sweep larger areas and collect more particles. These results establish a unifying framework for understanding how shape and flexibility influence transport and organization in active filament systems and filamentous robots.
\end{abstract}

\pacs{Valid PACS appear here}% PACS
\maketitle

%\section*{Introduction}
Active, flexible filamentous materials are essential to a variety of biological and synthetic systems, where their capacity for self-organization, mechanical force generation, and adaptability drives critical functions such as locomotion, resource collection, and environmental restructuring~\cite{ReviewWorm2023,winkler2020}. Examples span from microscopic systems—motor-driven cytoskeletal filaments~\cite{ganguly2012cytoplasmic,Nedelec1997,sumino-nature-12,LeGoff2002,dogic-activenematic-nature-12,kirchenbuechler2014direct} and cilia~\cite{biology_of_cilia_flagella,shah2009motile}—to macroscopic organisms, including flagella~\cite{bardy2003prokaryotic,mayfield1977rapid,friedrich2010sperm}, worms~\cite{ReviewWorm2023,patiltuazon2023}, and snakes~\cite{gray1946mechanism}. Synthetic analogs such as self-propelled robots extend these concepts, utilizing flexibility and activity to adaptively explore their surroundings~\cite{lopezactivesolids2024,zhengrobot-anchor2023,brun2024emergent}.

One striking manifestation of active material functionality is particle aggregation, a process observed in both natural and synthetic systems. For example, in freshwater ecosystems, oligochaetes like \textit{Tubifex~tubifex} and \textit{Lumbriculus variegatus} (California blackworms) use coordinated motion and mucus secretions to aggregate dispersed particles~\cite{tuazon2023collecting}, enabling nutrient cycling and habitat stabilization. Similarly, robotic systems exploit activity and flexibility to manipulate inert (i.e., passive) materials, mimicking these natural behaviors to perform tasks like debris management and targeted delivery~\cite{brun2024emergent}. These aggregation dynamics are governed by the interplay between filament flexibility, activity, and the environmental context; yet, the precise mechanisms remain elusive.

Beyond biological systems, active-passive mixtures provide a simplified yet insightful framework to explore aggregation phenomena from an active matter perspective. Active agents, such as bacteria, colloidal Janus particles, or robotic systems, impose stresses and collisions on passive components, giving rise to emergent behaviors like clustering~\cite{gokhale2022dynamic}, phase separation~\cite{pritha_phase_2018,Cates-mips-AP-mixture-PRL-15}, and laning~\cite{aparna2012spontaneous}. These processes may influence nutrient flows, structural formation, and ecosystem dynamics in natural habitats~\cite{visser2007biomixing} while inspiring novel applications in material assembly and robotics~\cite{brun2024emergent}. Notably, the role of filament flexibility in such mixtures remains underexplored~\cite{prathyushatransvercargo,zhangpolymerloop-21}, despite its relevance to both biological polymers~\cite{smrek_small_2017} and synthetic systems~\cite{deblais2018boundaries,brun2024emergent}.

In this study, we bridge the gap between biological and synthetic active systems by investigating the particle collection dynamics of slender and flexible \textit{Tubifex~tubifex} and \textit{Lumbriculus~variegatus} worms in a controlled environment. We quantify the impact of their conformation and dynamics on particle aggregation. Extending these insights, we construct larger-scale robotic filaments and perform simulations to examine how variations in filament flexibility and length influence performance across biological, robotic, and simulated systems. Our findings suggest that flexibility and activity are central to enhancing particle aggregation, offering a general framework for designing adaptable, soft robotic systems that can autonomously manipulate materials in constrained or complex environments.

\section*{Collecting Experiments}
The particle collection ability of the worms in confined environments presents both a biological puzzle and an exciting potential application for tasks such as microplastic collection, sorting, and cleaning~\cite{tuazon2023collecting}. To explore how this natural behavior can be generalized to active filaments, we investigate the collecting dynamics in three distinct active filament systems: two living, centimeter-scale biological worms, \textit{T.~tubifex} and \textit{L.~variegatus}, which differ in aspect ratio; a meter-scale robotic filament, designed to isolate the effects of filament parameters such as elasticity and length; and a Brownian dynamics simulation of an active filament with a tangentially propelling force~\cite{prathyusha2018PRE,riseleholder-15,fazelzadeh2023} to validate and extend our experimental observations. Despite the fact that these two worms have similar appearances and quantitatively comparable collective behaviors \cite{ReviewWorm2023,Ozkan2021}, they exhibit distinct locomotion-diffusion-dominated random walks in \textit{T.~tubifex} and ballistic motion in \textit{L.~variegatus}, providing the opportunity to study and compare their mechanisms of dispersed passive particle collection \cite{ReviewWorm2023}.

\begin{figure*}
    \centering
    \includegraphics[width=\linewidth]{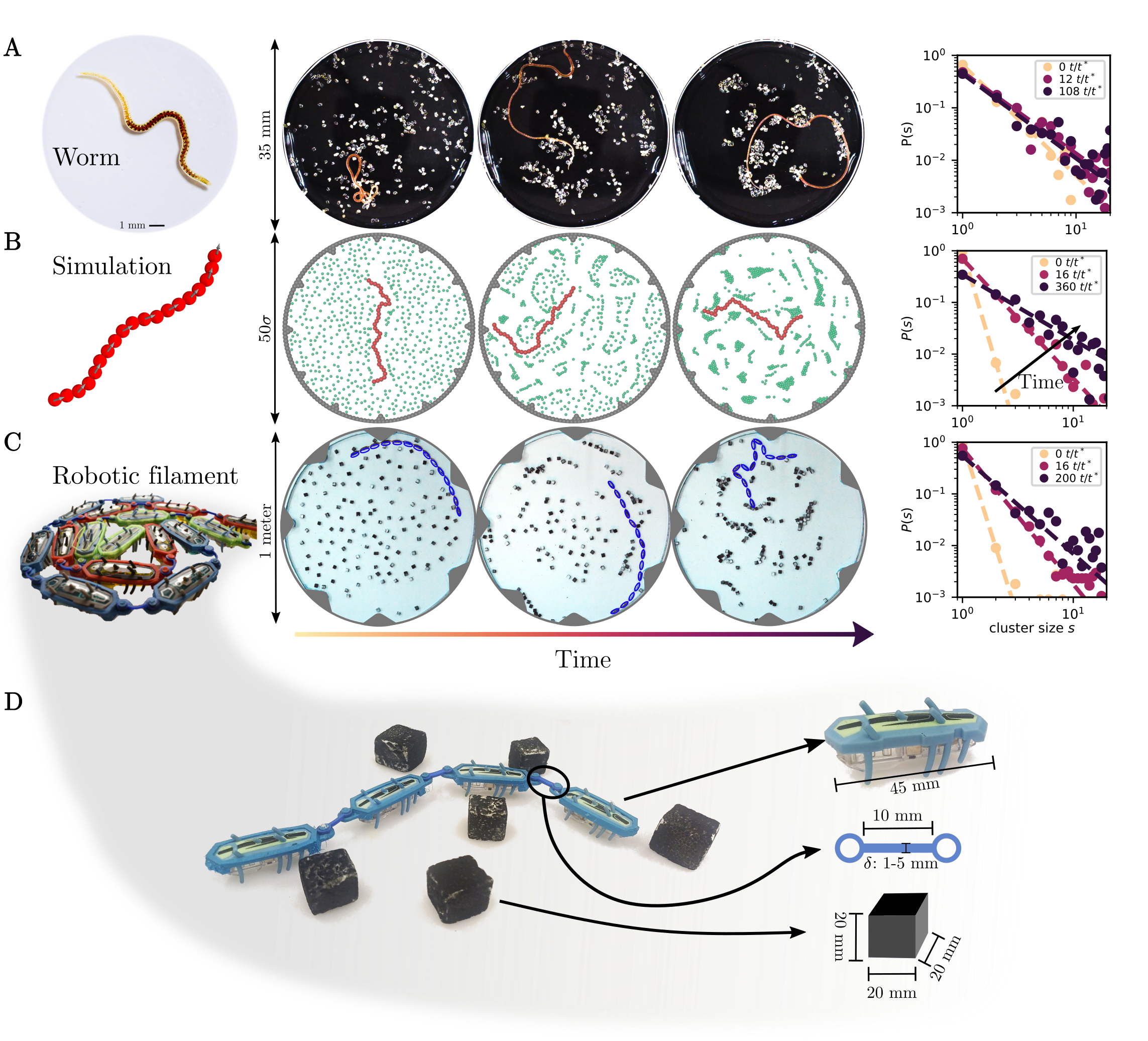}
\caption{\textbf{Collection of passive particles by living, in silico, and robotic active filaments.}  
(\textbf{A})~A California blackworm (\textit{Lumbriculus variegatus}) (left panel) and a \textit{T.~tubifex} worm (right panels) in a Petri dish with sand particles. Over time, the worm gathers the sand into larger clusters, eventually reaching a steady state, as shown by the evolution of the cluster size distribution over time in the right panel. Time is non-dimensionalized by the characteristic time of one full passage of the filament's center-of-mass across the arena, \( t^*\approx 120\) s, allowing direct comparison across systems.  
(\textbf{B})~An active, tangentially-driven filament interacting with passive particles exhibits similar clustering behavior, where the cluster size distribution grows over time until reaching a steady state. The active force is applied to all monomers; for visual clarity, its direction is indicated by arrows on alternate monomers in the left panel.  \(t^*\approx70~\tilde{\tau}\)
(\textbf{C})~A robotic filament composed of connected Hexbug robots moves within a circular arena, interacting with passive styrofoam cubes. The robotic filament collects particles into clusters, ultimately reaching a steady state. \( t^*\approx 5\) s 
(\textbf{D})~Design details of the robotic filament and a close-up view of the styrofoam particles. The robotic units are connected via elastic rubber bands of tunable width \( \delta \), allowing control over the persistence length (\(\ell_p\)) and contour length (\(\ell_c\)) of the filament.}  
\label{fig:WormRobotSimulations_Cleaning}
\end{figure*}
\clearpage

\textbf{Living biological filamentous worms.} In our first quasi-2D experimental setup, we place a single \textit{Tubifex tubifex} (with typical lengths \( L_c = 25-45 \) cm) or \textit{Lumbriculus variegatus} (\( L_c = 15-35 \) cm) into a 35 mm diameter ($D$) Petri dish filled with thermostated water (\( T = 21^\circ \)C) and \( N_{\text{t}} \) randomly dispersed sand particles (120 particles, \( d_{p,sand}\) = 0.7 mm, total weight 50 mg, \textit{SI Appendix}, Fig.~S1). The particles are light enough so that inertia does not influence their dynamics.  

Once in the Petri dish, the worms, denser than water, crawl along the bottom via wriggling motion. Unlike algae or microplastics, which adhere to the worm’s mucus~\cite{tuazon2023collecting}, sand particles do not stick to the worm's body. Instead, the worm's self-propulsion displaces sand particles upon contact; otherwise, the particles remain stationary. This displacement occurs only through direct physical interactions, as the motion of the worm does not generate significant fluid flow in the medium. As the worm moves through the dish and reorients its direction at the boundaries, it actively gathers particles into progressively larger clusters over time, as shown in Fig.~\ref{fig:WormRobotSimulations_Cleaning}\textbf{A} (See Movie S1~\cite{Movies}, \textit{SI Appendix}, Fig.~S2). The continuous energy input from the worm also fragments these clusters, preventing unrestricted growth and maintaining a quasi-steady state in which clusters continuously form and break. This clustering behavior appears to emerge from the interplay between the worm's flexibility and its dynamic adaptability, suggesting a mechanism distinct from the phase separation typically observed in dense active–passive mixtures~\cite{pritha_phase_2018,gokhale2022dynamic}.

To further analyze the clustering behavior and the role of filament flexibility, we use image analysis to track worm conformations and particle positions, allowing us to characterize the distribution of cluster sizes over time (see \textit{SI Appendix} section~1, Fig.~S3). We define a group of $n$ number particles as a cluster of size $s_n$ (for simplicity, we refer to the cluster size as $s$ unless otherwise specified), where each particle is within a distance of one particle size $d_{p,sand}$ from at least one other particle in the group. 

A typical result of the cluster size distribution at different times, $P(s)$ is shown in Fig.~\ref{fig:WormRobotSimulations_Cleaning}\textbf{A}, right. We find, as in previous studies looking at clustering in active systems \cite{ginot2018,Redner2016,gokhale2022dynamic}, that a power law with an exponential cut-off, $P(s) = s^{-\gamma} \exp{(-s/s^{*})}$, describes our data well. The fitted exponent $\gamma$ increases over time (Sup.~Table.~S2), indicating reduced coarsening of the clusters.

Within a characteristic timescale of approximately 30 minutes, the cluster size distribution reaches a steady state. The average cluster size, $\langle s\rangle=\sum s P(s)$, at long times is typically $\langle s \rangle_L \approx$ 10 particles for both species of worms. Fig.~\ref{fig:WormRobotSimulations_Cleaning}\textbf{A} shows results for one \textit{T.~tubifex} worm, but our results are comparable between worms. See Movie~S1~\cite{Movies} for an example of the experiments and a comparison between the two studied worm species.

\textbf{Active polymer simulation.} We perform Brownian dynamics simulations using an active polymer model~\cite{prathyusha2018PRE,riseleholder-15,locatelli-24} (see Materials and Methods Section and \textit{SI Appendix} section~2 for details)%on the numerical methods and model)
, which has been shown to effectively capture the behavior of these worms~\cite{sinaasappel2025}. Each monomer of size $\sigma$ in the polymer follows overdamped Langevin dynamics, driven by an active force of amplitude $f^a$ (magnitude is 1) applied tangentially along the filament backbone. The flexibility of the active polymer is controlled by the bending stiffness $\kappa$ between neighboring bonds.
(Fig.~\ref{fig:WormRobotSimulations_Cleaning}\textbf{B}). In addition to the polymer length, the key parameters in our model are therefore reduced to $f^a$ and $\kappa$. Passive particles, with diameter $d_{p,\mathrm{sim}} = 0.5\,\sigma$, are initially placed at a uniform concentration throughout the simulation domain. As observed in the worm experiments, there is no adhesion between the particles and the worms, nor any fluid flow generated by the worm motion. This suggests that interactions are limited to short-range steric repulsion. In the simulation, passive particles remain immobile until pushed by an active polymer segment and interact repulsively with other particles within a cutoff distance. The system thus represents a dry active environment, where the medium contributes only through single-particle friction ($\gamma$), and hydrodynamic interactions are neglected. The time unit of the simulations is set by \(\tilde{\tau} = \sigma^2\gamma/(k_BT)\). We systematically vary the number of monomers \( N = 9,\,18,\,28,\,37 \), the number of passive particles \( N_t = 50,\,100,\,250,\,500,\,600 \), and the magnitude of bending stiffness \( \kappa \) in the range 0.01–10 to explore different filament flexibilities. These variations allow us to explore a broad range of contour lengths, particle densities, and filament flexibilities.

To ensure realistic boundary conditions, we place stationary particles along the circumference of a circular boundary of diameter, $D$ = $50 \sigma$. These boundary particles impart a steric repulsion to the active polymer and passive particles. Additionally, three-bead triangular ``re-injectors'' are positioned along the circular boundary at regular angular intervals $\theta= 30^{\circ}$, with their apexes pointing toward the center of the confining circle, (See Fig.~S4). This approach, commonly used in persistent active systems, prevents the filament from getting trapped along the walls \cite{Deseigne2010a,deseigne2012}. When gliding near the wall, the tangentially driven filament is reoriented into the bulk of the arena by these triangular re-injectors, closely mimicking the behavior observed in biological worms (\textit{SI Appendix}, Fig.~S2); different spacing of the re-injectors has no significant effect on the cluster formation (\textit{SI Appendix}, Fig.~S4).

We find that the dynamics of the simulated active filament closely resemble those of the living worms, and the particles aggregate into clusters over time in a similar fashion (see the sequence of pictures in Fig.~\ref{fig:WormRobotSimulations_Cleaning}\textbf{B} \& Movie S2~\cite{Movies}). This is confirmed by the cluster size distribution, which exhibits the same power-law behavior as in the worm experiments, with the distribution shifting toward larger values as time progresses (Fig.~\ref{fig:WormRobotSimulations_Cleaning}\textbf{B}, right). 

\begin{figure*}
    \centering
    \includegraphics[width=0.9\textwidth]{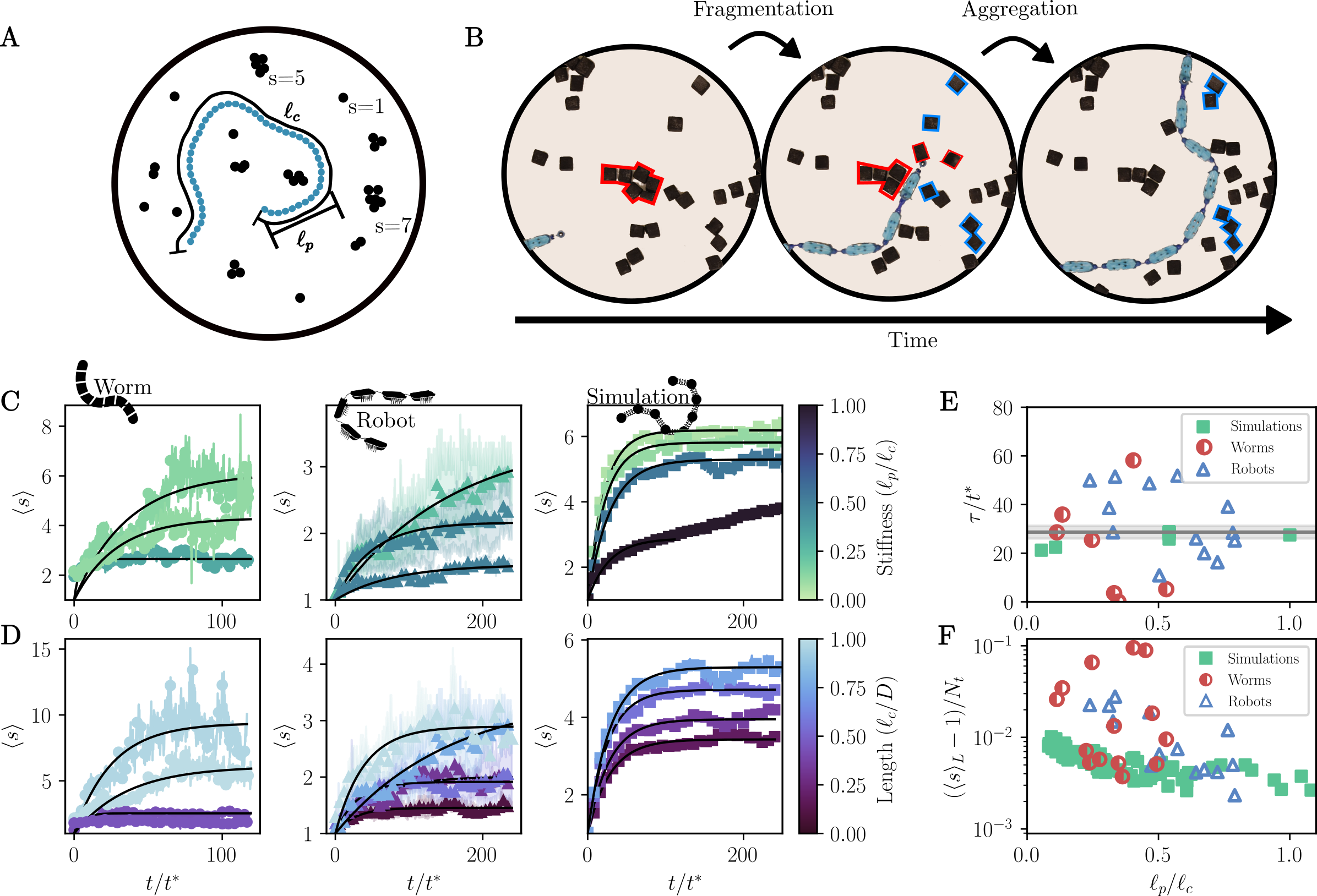}
\caption{\textbf{Effect of filament length and flexibility on collecting dynamics and long-time average cluster size.}
    (\textbf{A})~Active filamentous systems are characterized by their flexibility (\(\ell_p\)) and contour length (\(\ell_c\)). The cluster size $s$ is defined by the number of touching particles.  
    (\textbf{B})~Larger cluster formation results from successive \textit{fragmentation} (shown in red), and \textit{aggregation} (shown in blue) processes, driven by interactions between the particles and the active filaments' conformations as they move through the arena.  
    (\textbf{C})~Effect of the active filament’s normalized stiffness (\(\ell_p/\ell_c\)) on the average cluster size. For all systems, the average cluster size grows over time before reaching a steady-state value \(\langle s \rangle_L\) at long timescales, following Eq.~\ref{eq:growth}. Lower filament stiffness leads to larger final cluster sizes.  
    (\textbf{D})~For a fixed flexibility, the steady-state average cluster size increases with the normalized filament length (\(\ell_c/D\)). The shaded area in both (\textbf{E}, \textbf{F})~Long-time steady-state cluster size \( \langle s \rangle_L \) and characteristic aggregation timescale \( \tau = k_{\text{eff}}^{-1} \), plotted as a function of filament stiffness (\( \ell_p/\ell_c \)). Values are obtained by fitting Eq.~\ref{eq:growth} to the data shown in Panels~(\textbf{C}) and (\textbf{D}). In (\textbf{F}), \( \langle s \rangle_L \) is normalized by the total number of particles \( N_t \) to enable direct comparison across systems. The gray line in (\textbf{E}) shows the mean value across all data points (\( n = 25 \)); shaded regions indicate standard deviation.}  
    \label{fig:Effect_Length_and_Flexibility}
\end{figure*}

\begin{figure*}
    \centering
    \includegraphics[width=0.8\textwidth]{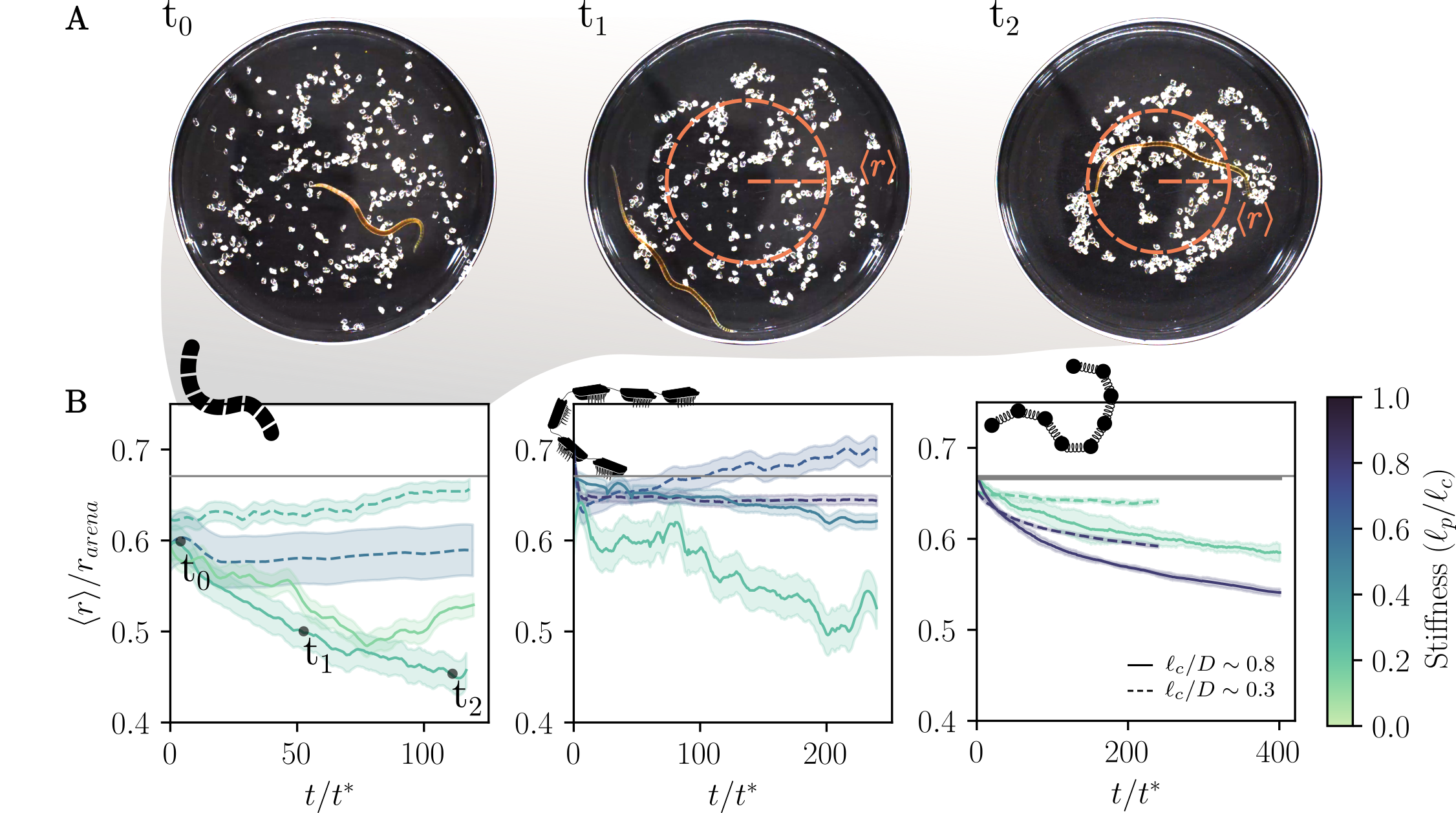}
    \caption{\textbf{Evolution of the cluster’s average location over time.}
    (\textbf{A})~Sequence of images (\(t_i = t/t^{*}\) = 0; 52; 120) showing the spatial evolution of particle positions over time.  
    (\textbf{B})~Average radial position of the cluster as a function of time for the three different active filamentous systems. Long filaments ($\ell_c/D \sim 1$) tend to accumulate at the center of the arena compared to short ones ($\ell_c/D \sim 0.3$), stiffer filaments seem to be slightly more efficient at gathering particles to the center. Data shown for worms and robots are for representative single experiments, smoothed with a Savitzky–Golay filter. The shaded area indicates the magnitude of the noise. The horizontal line indicates the theoretical value for an uniform distribution in the arena, and the shaded area on the simulation data indicates the standard error.}  
    \label{fig:Spatial_Distribution}
\end{figure*}

\textbf{Robotic filament.} In our second experimental setup, we scaled up the worm experimental platform to the meter scale and turned the worm into a robotic filament enclosed within a fixed arena of meter diameter (Fig.~\ref{fig:WormRobotSimulations_Cleaning}\textbf{C}). Our active filament is composed of $N$ commercially available self-propelled microbots (Hexbug Nano v2 with a characteristic size, $\sigma_{unit}$ = 45 mm) \cite{patterson2017,deblais2018boundaries,dauchot2019,zhengrobot-anchor2023,brun2024emergent}, encased in a 3D-printed frame around each individual bot and elastically coupled by laser-cut silicone rubber connectors (Fig.~\ref{fig:WormRobotSimulations_Cleaning}\textbf{C}; see also \textit{SI Appendix} section 3 and Figs.~S6 \& S7). 
Adjusting the width of the connections ($\delta$ in the close-up of the Fig.~\ref{fig:WormRobotSimulations_Cleaning}\textbf{C}), we can fine-tune the bending stiffness of the filament, $\kappa$. 

We enclosed the robotic filament using a rigid circular metal boundary with a 120~cm diameter. To ensure consistency with our simulations, we incorporated re-injectors at the boundary to redirect the robotic filament into the bulk of the arena. The initially dispersed $N_{t}=120$ passive cubic particles of size $d_{p,cube}$ = 20 mm, made of lightweight styrofoam, ensure that their weight does not play a role in the dynamics of the active robotic filament.

Interestingly, we observe that the robotic filament interacts with the particles in a manner similar to the biological and simulated systems, as confirmed by the probability distribution of cluster sizes over time (see Movie S3~\cite{Movies} and Fig. \ref{fig:WormRobotSimulations_Cleaning}\textbf{C}, right). This raises the question of why and how these active filaments, seemingly displaying universal behavior, manage to collect particles when confined in a circular arena. To address this, we next investigate the underlying factors that govern this particle collection process. 

\section*{Dynamics of aggregation \& spatial distribution}
\textbf{Aggregation-fragmentation dynamics and long-time average cluster size.} To investigate the mechanisms driving particle clustering across our different active filament systems, we tracked the aggregation dynamics by measuring the average cluster size, $\langle s \rangle$, over time. We also examined the influence of two key filament parameters: contour length ($\ell_c$) and bending stiffness ($\kappa$) (Fig.~\ref{fig:Effect_Length_and_Flexibility}\textbf{A}). In our robotic system, the bending stiffness $\kappa$ is controlled by adjusting the width of the elastic bonds ($\delta$), with larger $\delta$ corresponding to increased stiffness. In simulations, in contrast, stiffness is controlled directly by tuning the bending potential parameter $\kappa$. In all systems, we measured the effective persistence length ($\ell_p$) of the filaments by analyzing tangent–tangent correlations along the contour of the filament. In experiments, $\ell_p$ was extracted from image sequences, while in simulations, it was computed directly from the filament configuration data. This quantity correlates non-trivially with the initial bending stiffness $\kappa$~\cite{sinaasappel2025}, but also captures changes in filament conformation due to activity and boundary interactions, where filaments tend to curl while following the arena edge. By normalizing the persistence length by the contour length of the filament ($\ell_p/\ell_c$), we obtain a common stiffness quantity that allows direct comparison between different systems, particularly useful for worms, whose bending stiffness cannot be controlled externally.

\begin{figure*}
    \centering
    \includegraphics[width=0.9\textwidth]{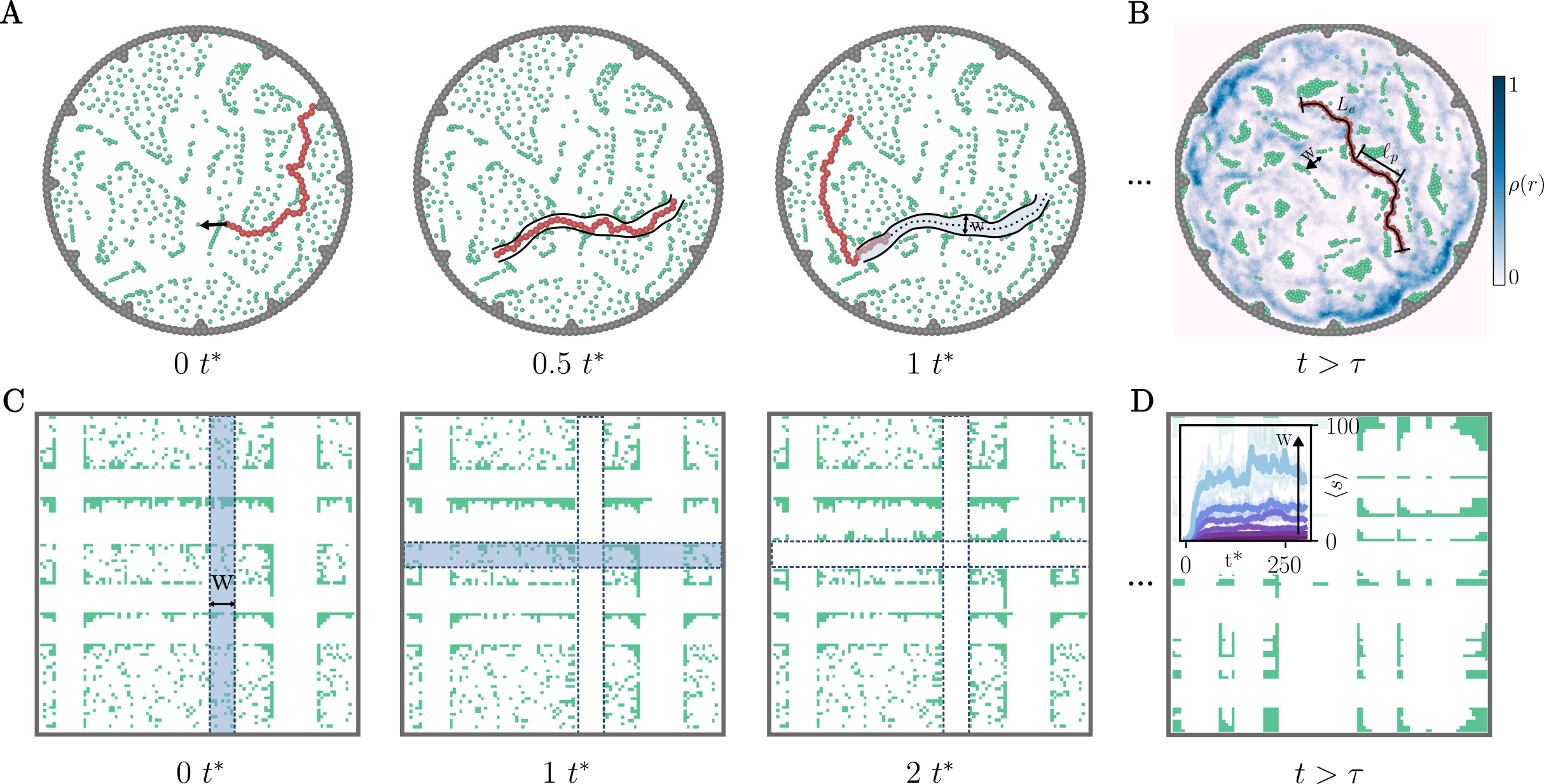}
\caption{\textbf{A simplified model for particle aggregation dynamics.}
(\textbf{A})~Brownian dynamics simulation of an active filament sweeping through a field of particles, progressively clustering them into larger aggregates over time.  
(\textbf{B})~Heatmap of the filament’s trajectory, showing the regions it has traversed (color bar indicate the probability density of the center-of-mass of the filament, \(\rho(r)\)). As the filament moves, it clears particles along its path of width \( W \), leaving unvisited areas where particles accumulate.
(\textbf{C})~To model this process, we consider particle displacement using a simplified system where bands of width \( W \) (representing the filament) sweep particles at each timestep within a confined arena. The sweeping bands are randomly placed and oriented either horizontally or vertically (see methods section).  
(\textbf{D})~Steady-state cluster formation. The inset shows the evolution of the average cluster size over time, showing that this simplified model captures the aggregation dynamics observed in experiments and simulations [Fig.~\ref{fig:Effect_Length_and_Flexibility}(\textbf{C},\textbf{D})] (see also \textit{SI Appendix}, Fig.~S12).} 
    \label{fig:Toy_Model}
\end{figure*}

As shown in Fig.~\ref{fig:WormRobotSimulations_Cleaning}, all three systems exhibit strikingly similar behavior. When an active filament is introduced into an enclosed arena with randomly distributed particles, it sweeps particles along its path, driving both aggregation and fragmentation events (Fig.~\ref{fig:Effect_Length_and_Flexibility}\textbf{B}). This dynamic leads to the progressive formation of larger clusters, reminiscent of an aggregation–fragmentation process similar to polymerization reactions~\cite{Krapivsky2010,ziff1980}. After a transient period, the system reaches a steady state where the average cluster size stabilizes, indicating that the aggregation and fragmentation rates have been balanced. In this analogy, the active filament acts as a dynamic reactor, driving both the aggregation and fragmentation of particle clusters. 

We find that the evolution of the average cluster size, $\langle s \rangle$, across all systems follows a saturating exponential behavior (Figs.~\ref{fig:Effect_Length_and_Flexibility}\textbf{C}, \textbf{D}), which is well described by the expression:  
\begin{equation}
\langle s(t) \rangle = 1 + \langle s \rangle_L\left(1 - e^{-k_{\text{eff}}t}\right).
\label{eq:growth}
\end{equation}
This functional form suggests an underlying aggregation-fragmentation process, where clusters grow and break apart at effective rates \( k_a \) and \( k_f \), leading to a characteristic timescale \( \tau = 1/k_{\text{eff}} \). To model this behavior, we treat the system as a binary process, where aggregates of any size \( s_n \) undergo aggregation (\( k_a \)) and fragmentation (\( k_f \)) at the same rates, independent of size:  
\begin{equation}
[ s_1 ] + [ s_2 ] \underset{k_f}{\stackrel{k_a}{\rightleftharpoons}} [ s_1 + s_2].
\end{equation}
By fitting our data to Eq.~\ref{eq:growth}, we extract both the long-term average cluster size, \( \langle s \rangle_L \), and the effective growth rate, \( k_{\text{eff}} = k_a + k_f \), which determines the timescale for the system to reach the steady state. In particular, we find that this characteristic timescale \( \tau \) is nearly independent of filament length and stiffness across the different active filament systems studied here (Fig.~\ref{fig:Effect_Length_and_Flexibility}\textbf{E}). The system reaches steady state after approximately \( \tau/t^* \approx 35 \) sweeps, where \( t^* \) is the typical time for a filament to cross the circular confinement (Worms: \( t^* \approx 120\) s, Robots: \( t^* \approx 5\) s, Simulations: \(t^*\approx 70~\tilde{\tau}\)).  

Interestingly, both the flexibility and the contour length of the active filaments influence the final cluster size \( \langle s \rangle_L \). As shown in Figs.~\ref{fig:Effect_Length_and_Flexibility}\textbf{C, D}, the filament stiffness (\( \ell_p/\ell_c \)) and the ratio of filament contour length to system size (\( \ell_c/D \)) both affect the long-term cluster size: longer and more flexible filaments tend to generate larger clusters. 

While particle density and size also impact steady-state cluster sizes, we normalize these values by removing  single-particle clusters and scaling by the total number of particles, defining ($\langle \tilde{s}\rangle_L=\frac{\langle s\rangle-1}{N_t}$), to facilitate comparison across systems in the dilute limit (\textit{SI Appendix}, Fig.~S5). Plotting this normalized long-time average cluster size allows us to compare biological, robotic, and simulated systems on equal footing. Fig.~\ref{fig:Effect_Length_and_Flexibility}\textbf{F} shows that the ratio \( \ell_p / \ell_c \) captures a general trend: more flexible active filaments (i.e., lower \( \ell_p / \ell_c \)) tend to form larger steady-state clusters. However, this trend does not yield a single collapsed curve across all systems, suggesting that an additional parameter governs clustering dynamics, which we explore in the next sections below. Beyond characterizing the average cluster size, we also investigated how particles redistribute spatially within the arena as clustering evolves.

\textbf{Spatial distribution.} Fig.~\ref{fig:Spatial_Distribution}\textbf{A} shows a typical sequence illustrating how clusters form preferentially near the arena center. To quantify this spatial distribution, we measured the mean radial position $\langle r \rangle$ of particles over time and compared it to the expected values for two limiting cases. In the case of full centralization, the mean radial position of particles is \( \langle r \rangle \approx \frac{1}{3}d_p\sqrt{N_t} \) (with $d_{p}$ the particle diameter), compared to \( \langle r \rangle = \frac{2}{3}r_{\text{arena}} \) for an initially uniform distribution, which is highlighted by the gray line.

As shown in Fig.~\ref{fig:Spatial_Distribution}\textbf{B}, long active filaments (\( \ell_c/D \gtrsim 0.8 \)) tend to aggregate particles toward the center of the arena across all active filamentous systems studied. In contrast, shorter filaments (\( \ell_c/D \lesssim 0.3 \)) tend to accumulate particles away from the center. In the limit of short filament length, this behavior resembles that of active point-like particles in a bath of passive particles, which typically accumulate near boundaries or form a dispersed, gas-like state \cite{deblais2018boundaries,bouvard2023}. Stiffer filaments (\( \ell_p/\ell_c \gtrsim 0.5 \)) appear more effective at centralizing clusters than flexible filaments of the same length, likely because more flexible filaments (\( \ell_p/\ell_c \lesssim 0.3\)) spend more time in the center of the arena, thereby pushing clusters outward (\textit{SI Appendix}, Fig.~S8 for more details on typical trajectories of the filaments and Fig.~S9 for snapshots of all experiments depicted in Fig.~\ref{fig:Spatial_Distribution}).

\section*{Active Sweeping Collecting Mechanism}
The striking similarities between the three filamentous systems suggest a general underlying physical process, which we aim to elucidate in this section.
A detailed analysis of the conformation of active filaments during exploration reveals that they actively sweep particles away from their path as they move through the arena. As shown in the sequence of snapshots obtained from the simulation in Fig.~\ref{fig:Toy_Model}\textbf{A}, the trajectory of an active filament over time leaves behind a cleared path of effective width \(W\). A probability density map of the filament's center-of-mass position over the entire course of its motion further reveals that the filaments spend a significant portion of their time near the boundary, effectively pushing particles away from it (Fig.~\ref{fig:Toy_Model}\textbf{B}).

As mentioned before, despite the influence of filament length and flexibility on clustering (Fig.~\ref{fig:Effect_Length_and_Flexibility}\textbf{F}), this relationship does not lead to a system-independent collapse across our different systems we studied here. This suggests that an additional parameter more directly controls the clustering dynamics.

We propose that the clustering process is better characterized by the effective width of the footprint, \(W\), defined as the (average) transverse extent of the pathway cleared by the trajectory of the filament. As shown in Fig.~\ref{fig:Toy_Model}\textbf{A}, \(W\) depends on the filament's flexibility and length, with longer and more flexible filaments generally generating wider pathways. To quantify \(W\), we first superimpose all consecutive contours of a filament in the period of time that the filament takes to cross the arena, to find the footprint of the filament. Next, we fit the largest circle that can be inscribed inside this footprint, and take the diameter of this circle to be the footprint width (\textit{SI Appendix}, Fig.~S10). Systematic measurements (\textit{SI Appendix}, Fig.~S11) confirm that more flexible filaments tend to exhibit larger transverse fluctuations relative to their tangential motion, leading to wider footprints. These fluctuations result from the complex conformation of the active polymer due to the interplay between filament activity, flexibility, and interactions with the arena boundaries.

\begin{figure}
    \centering
    \includegraphics[width=0.8\linewidth]{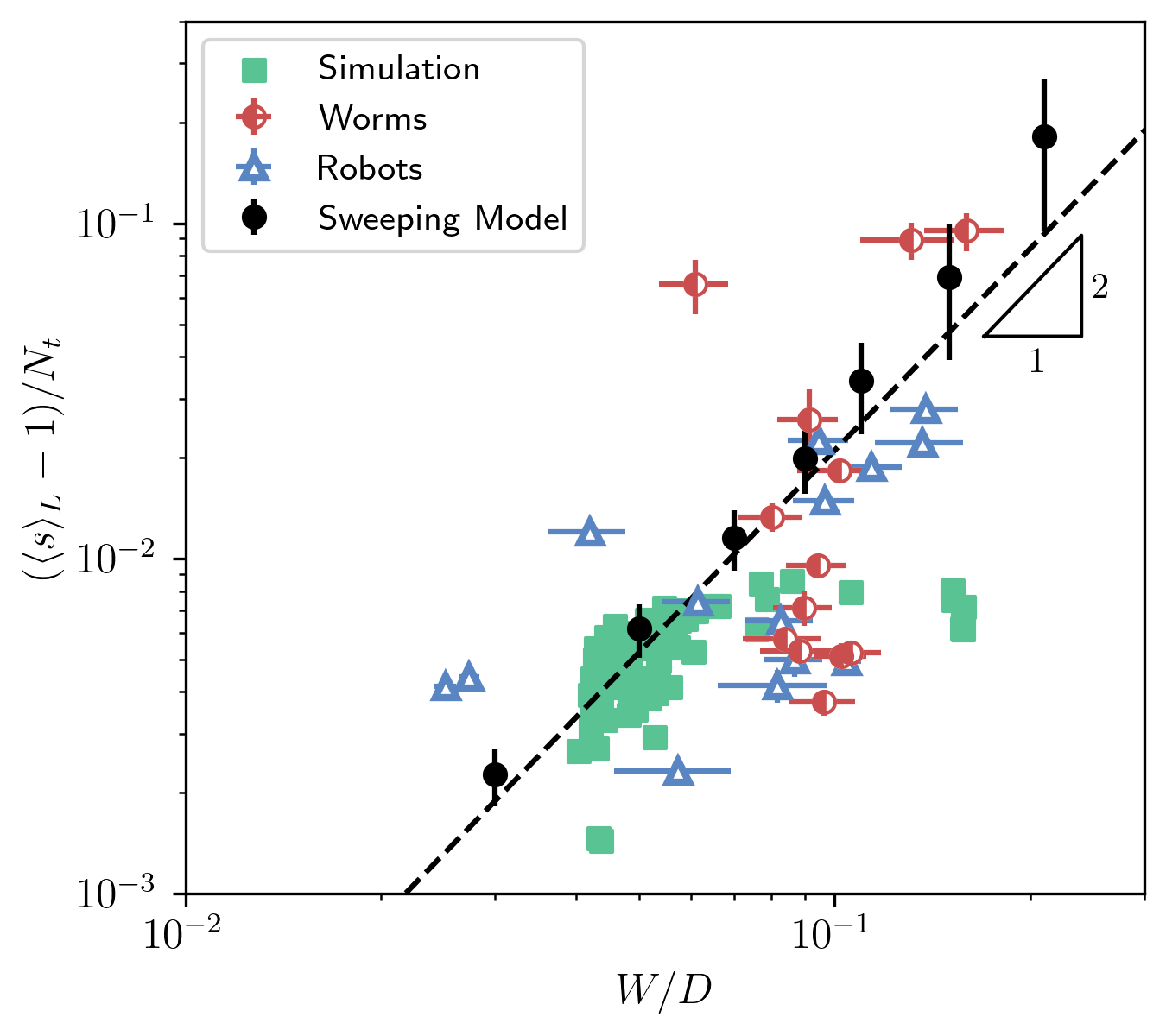}
\caption{\textbf{Steady-state normalized cluster size as a function of filament footprint width.} The long-time average cluster sizes from all experiments and simulations collapse onto a single master curve when plotted against the average active filament sweep width \( W \). The dashed line shows the predicted scaling \( \langle s \rangle_L \sim W^2 \).}  
    \label{fig:MasterCurve}
\end{figure}

When the final average cluster size \( \langle s \rangle_L \) is plotted as a function of \(W\), normalized by the system dimension \(D\), all data from Fig.~\ref{fig:Effect_Length_and_Flexibility}\textbf{F} collapse  onto a shared scaling curve, Fig.~\ref{fig:MasterCurve}. This establishes \(W\) as the key parameter governing the clustering process: the larger the effective footprint width, the larger the final average cluster size. Since \(W\) is set by filament flexibility and length, these properties indirectly dictate the clustering dynamics.

\begin{figure*}
    \centering
    \includegraphics[width=0.9\linewidth]{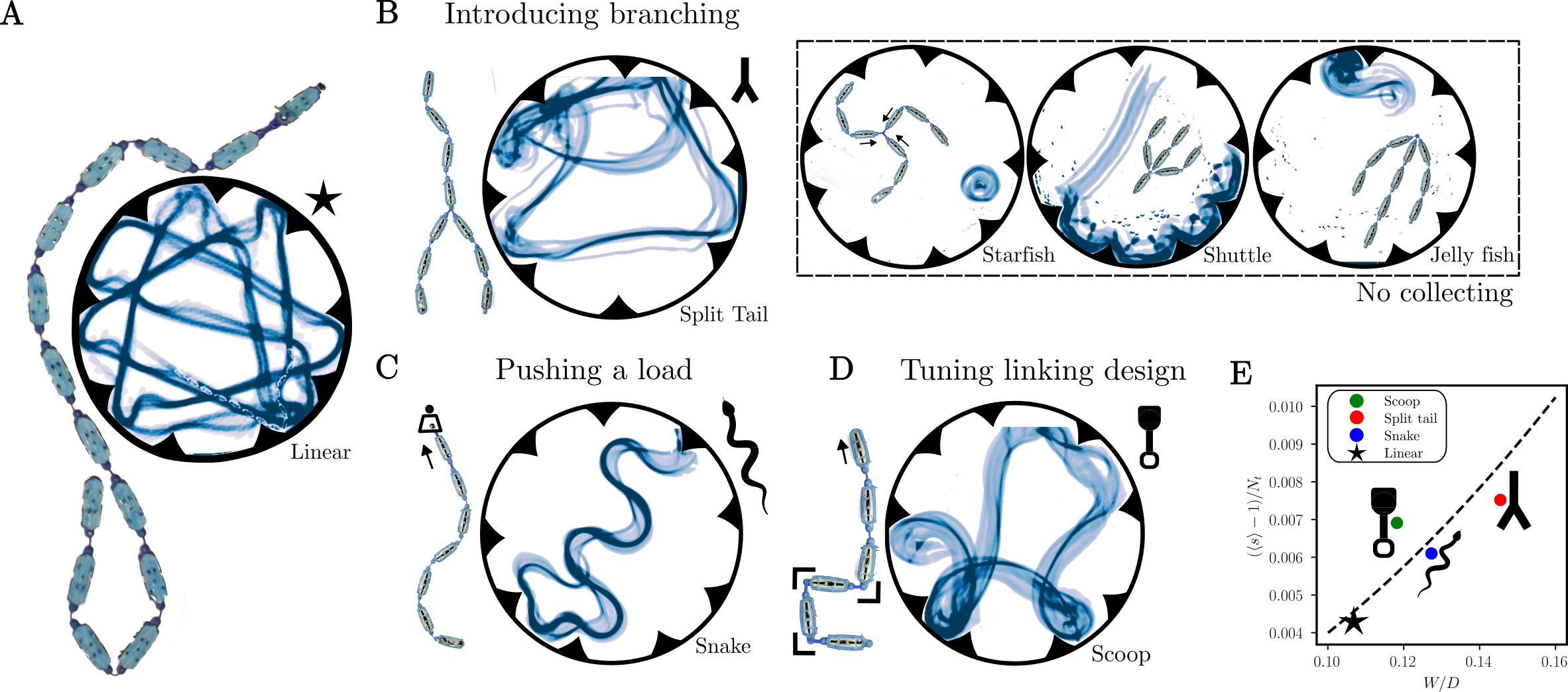}
\caption{\textbf{Active filament Atlas: A design space for optimizing particle collection through active filament chain properties.}  
(\textbf{A})~Baseline linear polymer configuration. The dark blue trails represent the trajectory of robot over time, indicating the area explored. 
(\textbf{B})~A first strategy modifies chain topology by introducing branches, altering sweeping dynamics. Trajectories of each robotic unit within the chain illustrate distinct collection strategies. The split tail design, increases the effective footprint via a rear branch. The inset shows other branched configurations (starfish, shuttle, jellyfish) that exhibit disorganized motion or rotational motion and fail to collect particles effectively.
(\textbf{C})~A second strategy adjusts the speed of the leading bot by increasing its weight, inducing self-oscillatory motion along the path. This snake configuration dynamically broadens the swept area, as reflected in the curved trajectories. 
(\textbf{D})~Tuning the elasticity of individual bonds enables the imposition of complex, pre-programmed curvatures during motion. This scoop configuration uses its intrinsic shape to steer particles along the inner arc as it moves through the arena.
(\textbf{E})~The resulting collection efficacy, quantified by the long time average cluster size (\(\langle s \rangle_L\)), is compared to those of the linear chain (star symbol).  
This active robotic filament at last establishes a taxonomy for emergent, multi-filament, heterogeneous flexible structures, paving new directions for engineering artificial active materials with programmable shape adaptation and collective functionality.  
}  
\label{fig:TamingRoboticChain}
\end{figure*}

To understand the scaling behavior of \(\langle s \rangle_L\) with \(W\) observed in Fig.~\ref{fig:MasterCurve}, we propose a coarse-grained Sweeping Model (see Materials and Methods section). In this model, we start with a box of size \(L \times L\) with $N_t$ passive particles. Clustering occurs as particles are swept away from a region of size \(L \times W\), representing the effective path cleared by the active filaments (see Fig.~\ref{fig:Toy_Model}\textbf{C}). Initially, the particles are uniformly distributed throughout the arena. Over time, repeated filament sweeps along both axes push particles away from these regions, leading to two competing effects: \textit{Aggregation}, where particles accumulate into larger clusters as they are displaced from swept regions, and \textit{Fragmentation}, where clusters can be fragmented when the filament path crosses an aggregate (see Movie~S4~\cite{Movies}). 

This minimal particle sweeping model quantitatively reproduces the experimentally observed clustering dynamics, as confirmed in the inset of Fig.~\ref{fig:Toy_Model}\textbf{D}, where the predicted cluster growth dynamics aligns with the experimental and simulation data of Figs.~\ref{fig:Effect_Length_and_Flexibility}\textbf{C} and {D}. Additionally, the time scale to reach a steady state also aligns very well with our experimental findings: after $\sim$ 35 iterative sweeps, we converge to the steady state (see \textit{SI Appendix} section~5 and Fig.~S12 for more details on the results of this model).

Aggregation and fragmentation of particle clusters can generally be described by the Smoluchowski aggregation equation \cite{Smoluchowski1906,Krapivsky2010}, which applies to irreversible aggregation and thus has inherent limitations in the context considered here (\textit{SI Appendix} section~6, Fig.~S13). This theory relies on a mean-field approximation, assuming that each cluster interacts equally with all others, regardless of their relative separation. However, this assumption breaks down in our system, where the aggregation and fragmentation processes are governed by the motion of the active filament, which introduces spatial correlations and disrupts the equivalence between clusters. To account for these spatial effects, we propose a scaling relation using typical cluster separation and mass conservation. 

Let $a$ be the typical distance between two clusters and $R$ the typical cluster size. If the cluster size distribution is sufficiently narrow, that is, the clusters are approximately the same size, the typical mass of a cluster M scales as $R^2$. Denoting by $n_c$ the number density of clusters, mass conservation implies that $n_c M$ is constant and that $n_c \sim 1/R^2$. The typical distance between two clusters is typically $1/\sqrt{n_c}$, which yields $a\sim R$. In steady state, the clusters remain unaffected by the sweep process when separated by a distance on the order of $W$, giving $a \sim W$. Combining these two conditions, we obtain $R \sim W$. Since the number of particles per cluster scales as $R^2$, this immediately leads to the scaling law $\langle s \rangle_L \sim W^2$, which rationalizes the experimental and simulation data collapse shown in Fig.~\ref{fig:MasterCurve}. 

As shown in Fig.~\ref{fig:MasterCurve}, our sweeping model (black circles) and scaling argument (dashed line) predict that the long-time average cluster size scales as $W^2$, closely matching experimental results. This good agreement between theory, simulations, and experiments confirms that the effective sweeping width $W$ is the key parameter governing the clustering process. We also note that our simulation results deviate from the predicted scaling at larger values of $W$, likely because our argument holds only when the sweep width remains small compared to the domain size, i.e., for $W/D \lesssim 1$. Additionally, our sweeping model represents an idealized situation in which sweeping bands are randomly placed and oriented orthogonally across the domain, sweeping particles in a simplified manner. In contrast, the worms, robotic filament, and tangentially driven filament sweep their surroundings in a more complex way, involving finite-size effects associated with filament conformation, cluster size, and interactions with boundaries. These aspects warrant further detailed investigation.

\section*{Discussion and Concluding Remarks}

\textbf{Summary.} Our work integrates biological experiments, robotic realizations, and computational modeling to uncover how flexible active filaments collect passive particles in confined environments. Through our investigation of \textit{Tubifex~tubifex} and \textit{Lumbriculus~variegatus} worms, as well as robotic analogs and simulations, we demonstrate that active filaments can autonomously cluster dispersed particles by dynamically sweeping across their surroundings. Our results highlight how filament flexibility, activity, and the dynamic footprint width \( W \) collectively govern the efficacy of particle aggregation, revealing a consistent scaling behavior. These findings not only highlight fundamental mechanisms underlying biological organization but also suggest design strategies for adaptable robotic systems and programmable active materials. It is therefore worthwhile to explore the broader implications of our study, particularly in the contexts of living systems, active–passive polymer mixtures, flexible active filament systems, and robotics, as discussed below.

\textbf{Implications for living systems.} Ecosystems continuously evolve through dynamic interactions between living organisms and their surrounding physical structures, as first illustrated by Darwin’s seminal observations of earthworms gradually reshaping landscapes through sustained biological activity~\cite{darwin1898formation}. 

Across diverse ecosystems, the activity of animals—including trampling~\cite{seleb2025moving}, grazing~\cite{menaut2001banded}, and burrowing~\cite{louw2017mammal}—can generate large-scale patterns, illustrating how biological interactions with the environment can give rise to ordered structures without centralized control~\cite{pringle2017spatial}. Among aquatic ecosystems, similar spatial patterns emerge from the activity of benthic organisms that modify sediments through burrowing and feeding. For instance, aquatic worms such as \textit{Tubifex tubifex} and \textit{Lumbriculus variegatus} naturally engage in particle aggregation, influencing sediment structure, nutrient cycling, and oxygenation processes~\cite{cummins1979feeding,kudrolli2019burrowing}. 

Previous studies have largely attributed this behavior in benthic macroinvertebrates to biological adhesion mechanisms, such as mucous secretions that facilitate the collection of organic and inorganic matter, including microplastics~\cite{tuazon2023collecting}. Our results suggest that the flexible body and active motion of worms enable effective environmental reorganization without requiring adhesion or long-range attractive interactions with sand particles. Similar dynamics are observed in other systems, where organisms such as marine polychaete worms~\cite{grill2015burrowing}, snakes~\cite{marvi2014sidewinding}, and nematodes (\textit{C.~elegans})~\cite{juarez-celegance-10} interact with granular or soft media, actively restructuring their surroundings through their body undulations and locomotion. The interplay between flexibility, activity, and physical interactions with the environment may thus represent a general mechanism for environmental reorganization.

\textbf{Implications for active--passive mixtures.}  
From an active matter perspective, the results presented here offer new insights into the behavior of active–passive mixtures under confinement, where self-propelled agents navigate among passive, non-motile components. In classical active matter systems, such mixtures often display clustering~\cite{gokhale2022dynamic}, phase separation~\cite{pritha_phase_2018,Cates-mips-AP-mixture-PRL-15}, and pattern formation~\cite{aparna2012spontaneous}. Here, we examined a distinct regime in which activity is embedded within flexible chains—active filaments—that interact mechanically with passive particles. This filamentous architecture introduces internal degrees of freedom such as bending and lateral fluctuations, which are absent in systems of point-like or rigid active particles.

These transverse fluctuations play a central role in the reorganization of passive particles. The dynamic undulations of the filament laterally extend the effective swept area over time, continuously displacing and aggregating passive particles across the confined space. In the absence of adhesion, attractive interactions, or external control, this motion alone drives the emergence of structures such as voids, clusters, and heterogeneity, in an otherwise dilute and disordered environment.

This sweeping mechanism extends previous strategies that used rigid active colloids~\cite{amin-knead-prl-24} or chiral bacterial baths~\cite{aggregation-bacteria-23} to control passive assembly. Our results demonstrate that flexibility enhances the capacity of active agents to manipulate their surroundings. This has implications for biological systems where active filaments must navigate and reorganize dense passive environments. In cells, for instance, cytoskeletal filaments such as actin and microtubules operate within crowded spaces filled with organelles and macromolecular assemblies. While motor-driven transport along filaments is well-studied, the role of intrinsic filament motion in spatial reorganization remains poorly understood~\cite{peterson2021vesicle}. In addition, our work adds to the existing literature on active polymer research that has shown how the interplay between activity, flexibility, and propulsion mechanisms~\cite{prathyusha2022transverse} can lead to conformational changes and facilitate exploration~\cite{ligesh-rajarshi-2022}, contributing to dynamic organization in crowded ~\cite{wen2022collective} and complex environments~\cite{mokhtari2019dynamics,kurzthaler2021geometric,ligesh-rajarshi2023active}, by revealing how these effects emerge under confinement and through interactions with passive particles.

By showing that activity and flexibility alone can drive large-scale restructuring, our findings suggest a mechanical route to organization in both biological and synthetic systems. More broadly, they highlight how flexible active filaments reshape active–passive interactions, offering principles for designing self-organizing materials that adapt to and reconfigure their environment.

\textbf{Implications for soft robotics.}  
\textit{Tubifex~tubifex} and \textit{Lumbriculus~variegatus} worms occupy a region of relatively large final average cluster size (Fig.~\ref{fig:MasterCurve}), indicating that their particle aggregation strategies are mechanically well optimized. Our sweeping model predicts that active filaments with a larger effective footprint width \( W \) can form even larger clusters. Guided by this principle, we extended the approach to our robotic filaments. By tuning their flexibility, we effectively control their footprint width \( W \), offering a direct strategy to influence the final cluster size. This mechanical adaptability highlights a simple design rule for improving the collecting performance of soft robotic filaments.

Additional strategies can be employed to enhance the collection efficacy of an active filament, as illustrated in Fig.~\ref{fig:TamingRoboticChain} and Movie~S5~\cite{Movies}. One approach involves modifying the filament topology by introducing branching structures, which increases the overall footprint of the active filament (Fig.~\ref{fig:TamingRoboticChain}\textbf{B}). Another approach consists of tuning the self-propulsion velocity along the filament. This can be achieved by locally varying the activity of the bots or, in our case, by changing the weight distribution of the leading bots. This induces dynamic oscillations of the filament, resembling flagellar beating or snake-like motion, which effectively increases its footprint width as it moves through the arena (Fig.~\ref{fig:TamingRoboticChain}\textbf{C}). Finally, tuning the bending stiffness of the filament to impose a local curvature allows particles to be scooped along the path of the active filament (Fig.~\ref{fig:TamingRoboticChain}\textbf{D}).

As a result, these different strategies demonstrate that the robotic filament can be tuned to collect more particles, with the long-time average cluster size \(\langle s \rangle_L\) increasing compared to that of a simple linear chain, as shown in Fig.~\ref{fig:TamingRoboticChain}\textbf{E}.

We also note that there are limits to this design space. When the filament becomes too wide, for instance by adding branches at its front or by increasing the size of its monomers, it can become irreversibly stuck at the boundaries (see inset of (Fig.~\ref{fig:TamingRoboticChain}\textbf{B}). This limitation points to another way in which worms appear to be well adapted to their environment. The footprint width of a worm is not fixed but adjusts dynamically through transverse fluctuations. This polymer-like flexibility allows the worm to remain narrow in confined regions and broaden when space permits, enabling effective exploration and collection.

The strategies that we propose here open several directions for future work, and extend recent studies showing how geometry and internal activity can enable soft robotic materials to perform tasks without the need for external control or feedback \cite{boudet2021,brun2024emergent,veenstra2025}. While our results show that active filaments hold promise for simple environmental manipulations, achieving reliable and targeted control remains a key challenge. For instance, guiding the transport of passive particles requires control over the filament’s trajectory. Strategies such as creating predefined pathways for filament motion or using controlled stimuli, such as light, to locally actuate the filament represent promising, yet largely unexplored, solutions.

\small{\section*{Materials and Methods}
\subsection*{Living Worms}
Two different species of annelid worms where used. These species are widely available from commercial suppliers.
\textit{California blackworms.}
We purchased California blackworms (\textit{Lumbriculus variegatus}) from Ward's Science and were reared similarly as described in past publications (Tuazon \textit{et al.}\cite{tuazon2022oxygenation, tuazon2023collecting, 2024Tuazon}). 
\textit{Tubifex Tubifex} worms.
All batches of \textit{T.~tubifex} worms analyzed in this work were purchased from the provider Aquadip (\url{https://www.aquadip.nl/}) and ordered in a prepacked configuration, where the worms were at adult size. 
The worms were maintained in an aquarium at room temperature, constantly under filtered flow, with water consisting of demineralized water mixed with salt solutions optimized for their needs. 
The salt solution consist of a mixture of: 50~g/L of \ce{NaHCO3}; 10~g/L of \ce{KHCO3}; 100~g/L of \ce{CaCl2}; 90~g/L of \ce{MgSO4}.  
Worms were fed weekly with standard goldfish food, and the water was refreshed once per week or more frequently if needed.

Experiments were recorded with an ImageSource DFK 33UX264 camera (Charlotte, NC) on an optical table using 80/20 parts. The experimental protocol is described similarly in Tuazon, et al.~\cite{tuazon2023collecting}. We recorded each experiment at a frame rate of 20~FPS for four hours with fixed lighting, where the first three hours were used for data analysis. 50$\pm$0.01 mg of 20\# palmetto pool filtered sand (Woodruff, SC) were used for the test materials where $\sim$0.7 mm grains were isolated using nylon mesh. The sand grains were transferred in a 35 mm Petri Dish and submerged in 2 mm of filtered water. Before worm transfer, the sand grains were manually dispersed using a pipet (to ensure that the grains were evenly distributed). After undergoing a one hour habituation period, worms were transferred in the Petri Dish with the sand grains. We repeated this experiment 11 times for blackworms and 3 times for \textit{T.tubifex}. 

\subsection*{Robotic filaments}
Robotic filaments were assembled from commercially available bristle bots (Hexbug Nano Nitro, \url{https://www.hexbug.com/}). Each bot was enclosed in a custom 3D-printed housing featuring front and rear protrusions for attaching flexible silicone rubber connectors. The bending stiffness of these connectors was tuned by varying their width, and filament length was adjusted by connecting different numbers of bots. This modular design enabled exploration of a wide range of persistence lengths, summarized in Sup.~Table~S1 and Sup.~Fig.~S7. To prevent self-induced spiraling in long, flexible filaments, the front connector was made stiffer, reducing excessive bending at the leading end. Despite careful preselection, individual bots may exhibit a slight turning bias due to internal asymmetries. While visible in some trajectories, this bias has negligible effect on cluster formation.
Experiments were performed in a circular arena enclosed by a metal barrier, with reinjection defects along the perimeter. The arena contained 120 black-painted styrofoam cubes (20 mm per side) as passive tracer particles. A top-down camera (Sup.~Fig.~S6) recorded the experiments. Before each run, tracer particles were evenly distributed. The robotic filament was then switched on and gently introduced into the arena to minimize disturbance. Recording began immediately upon introduction.

\subsection*{Image analysis}
Image analysis was mainly done using a custom python code using the Python OpenCV library (\url{https://github.com/opencv/opencv-python}) and scikit-image library (\url{https://github.com/scikit-image/scikit-image}). However, in certain worm experiments, segmenting the worm proved challenging. In these cases, we employed machine-learning-based segmentation using the YOLO package developed by Ultralytics (\url{https://github.com/ultralytics/ultralytics}), which enabled high-precision filament contour detection (model available upon request).

To determine the sweep width $W$ as defined in the main text, we superimposed consecutive images until the filament had traversed its full contour length. In the resulting composite image, we identified the largest inscribed circle fully contained within the contour, taking its diameter as the relevant measure of the sweep width. Repeating this procedure over the full duration of an experiment yielded a distribution of $W$ values, from which we report the median as the characteristic sweep width in the main text.

\subsection*{Brownian dynamics simulation of Active Polymer and tracers}
We employed Brownian dynamics simulations to study the particle aggregation. We model the worm as a self-propelling active filament composed of $
N$ monomers, and the sand particles as non-motile, athermal disks. The system is confined within a circular boundary, implemented using stationary particles arranged along the perimeter, which interact with both the filament and the sand particles to impose confinement. All particles interact sterically with each other using  Weeks-Chandler-Andersen potential, \cite{weeks-jcp-1971}, 
$
U^{WCA}(r_{ij})=4\varepsilon\left[\left(\frac{\sigma}{r_{ij}}\right)^{12}-\left(\frac{\sigma}{r_{ij}}\right)^{6}+\frac{1}{4}\right]
\label{eqn:wca}$, which vanishes beyond any distance greater than  $2^{1/6}\sigma$.
Here $\varepsilon$ measures the strength of the repulsive interaction, and is set to $1.0$.  $\sigma$ is the characteristic size of the interacting particles and we set the monomer size to $\sigma=1$ as the basic length scale in the simulation. For interactions between particles of different sizes, the effective interaction length is computed using the arithmetic mean $\sigma_{ij}=\frac{\sigma_i+\sigma_j}{2.0}$.  $r$=$r_{ij}\equiv\left|\mathbf{r}_{ij}\right|=\left|\mathbf{r}_i-\mathbf{r}_j\right|$ is the distance between two beads.

The equation of motion of the self-propelling active filament is given by, \\
\begin{equation}
\gamma{{\dot {\bf r}}_i}=-\nabla_i{{ U}^{WCA}} +( -\nabla_i{U}^{stretch}{-\nabla_i U}^{bend}+{\bf F}_i^{noise}+ {\bf F}_i^{active})  
 \label{eq:eqofmotion}
\end{equation}
With $\gamma$ the friction coefficient of the monomers with the substrate.

Now, we discuss the last term in Eqn.~\ref{eq:eqofmotion} containing contributions from four interactions 
and is experienced by the polymer beads solely as $\mathcal A$ is set to 1 for the polymers and $\mathcal A = 0$ for tracer particles. 
Since these interactions are absent for passive particles, we set ${\mathcal A}$ to zero for such particles. Each connected pair of the polymer interacts via a harmonic potential \cite{kkremer-90}, 
\begin{equation}
U_{stretch}(r)=k_b \big(r- R_0\big)^2 \label{eq:bond}.
\end{equation}
Here $R_0=1.0\sigma$ is the maximum bond length, and $k_b=4000~k_BT/\sigma^2$
is the bond stiffness. 
Bending interaction is experienced by any connected triplet in the polymer, and it is given by the potential,
\begin{equation}
U_{bend}= \frac{\kappa}{2}\big(\theta -\theta_0 \big)^2
\end{equation}
Here $\kappa$ is the bending stiffness, $\theta$ is the angle between any consecutive bond vectors and $\theta_0$ is set to $\pi$. These parameters make the chain effectively inextensible with bond length $b\sim 1 \sigma$ and polymer length $ L =(N-1)b \sigma$.
 
The stochastic term ${\bf F}^{noise}_i$  represents the thermal fluctuations and is modelled as white noise with zero mean and variance proportional to $ \sqrt{k_BT\gamma/\Delta t }$.  

Finally, the last term represents the self-propulsion force for the polymer, 
\begin{equation}
{\bf F}_i^{active}=\frac{f_p}{2}({\bf \hat  t}_{i-1,i}+{\bf \hat  t}_{i,i+1})
\end{equation}
Here, $f_p$ is the strength of the force and ${\bf \hat  t}_{i,i +1} = {\bf r}_{i,i +1} / {\bf r}_{i,i +1}$ is the unit tangent vector along the bond connecting beads i and i + 1. This is experienced by  the beads ($i=2,3,4... N-1$) of the polymer. 
The active forces for the end beads ($i=1$ and $i=N$), as they have only one nearest neighbor, are 
\begin{equation}
{\bf F}_{1}^{active}=   \ \hat{\bf t}_{1}   f_{p} \; \& \; \;
 \\
 {\bf F}_{N}^{active}=   \hat{\bf t}_{N-1} f_{p}
\label{activeforce-end}
\end{equation}
In experiments, sand particles are stationary until moved by the worm. Thus, the equation of motion of such particles, modeled as an athermal passive particle of size $0.5 \sigma$, is given by,
\begin{equation}
\gamma \frac{d\mathbf{r}_i}{dt} = 
\sum_j 
\begin{cases}
-\nabla_{\mathbf{r}_i} U^{\mathrm{WCA}}(r_{ij}),& r_{ij} < 2^{1/6} \sigma_{ij} \\
0, & r_{ij} \ge 2^{1/6} \sigma_{ij}
\end{cases}
\end{equation}
In our simulations, the circular confinement is modeled using stationary boundary particles that interact with active filament and passive particles via $U^{WCA}$ potential. These boundary particles are fixed in space and do not respond to collisions, effectively acting as a rigid wall. It captures the essential physics of the experimental setup involving worms in a Petri dish, where the walls exert forces without recoiling and the reaction forces are absorbed by the surrounding medium.

In our simulation, we used $\Delta t=0.0001 \tilde{\tau}$, where $\tilde{\tau} = \sigma^2\gamma/(k_BT)$  sets the integration time unit. We varied the  number of monomers $N=9, 18, 28, 37$, and number of passive particles $N_t=50,100,250,500,600$.  Unless otherwise specified, the results presented in the manuscript correspond to  $N=37, N_t=600$.

\subsection*{Sweeping Model}

We implemented a minimal model in Python to capture the essential features of the sweeping-induced clustering dynamics. Initially, $N_t$ particles are randomly distributed on a 2D square grid of size $L \times L$, with periodic boundary conditions. 

At each timestep, a \textit{sweep} is performed by selecting a stripe of width $W$ oriented either horizontally or vertically, with equal probability. A random position along the orthogonal direction is chosen to define the stripe location. All particles within the stripe are displaced according to their relative position:

\begin{itemize}
  \item Particles within the \textbf{leading half} of the stripe ($x - W < x_t < x$ or $y - W < y_t < y$) are pushed \textbf{backward} (leftward or downward) to a new position:
  \[
  x_t \rightarrow x - W - 1 \quad \text{or} \quad y_t \rightarrow y - W - 1
  \]
  
  \item Particles within the \textbf{trailing half} ($x \leq x_t < x + W$ or $y \leq y_t < y + W$) are pushed \textbf{forward} (rightward or upward):
  \[
  x_t \rightarrow x + W + 1 \quad \text{or} \quad y_t \rightarrow y + W + 1
  \]
\end{itemize}

If the target site is already occupied, the particle continues moving in the same direction until an unoccupied grid point is found. In rare cases where no free site exists along the sweep direction (e.g., due to high local density), the simulation halts.

After each sweep, clusters are identified as connected groups of adjacent particles using 4-neighbor connectivity. The size distribution of these clusters is recorded. The process then repeats with a new sweep direction and stripe position.
}

%\showmatmethods{} % Display the Materials and Methods section

\section*{acknowledgments}
We thank the Technology Center of the University of Amsterdam for technical support and assistance with the tracking-based machine learning. All simulations were performed using resources provided by the Partnership for an Advanced Computing Environment (PACE) at the Georgia Institute of Technology, USA. We thank Ishant Tiwari for critical reading of the manuscript. S.B. acknowledges funding support from NIH Grant R35GM142588; NSF Grants PHY-2310691; and NSF CAREER iOS-1941933.

% Bibliography
\bibliography{bib.bib}

\begin{thebibliography}{70}%
\makeatletter
\providecommand \@ifxundefined [1]{%
 \@ifx{#1\undefined}
}%
\providecommand \@ifnum [1]{%
 \ifnum #1\expandafter \@firstoftwo
 \else \expandafter \@secondoftwo
 \fi
}%
\providecommand \@ifx [1]{%
 \ifx #1\expandafter \@firstoftwo
 \else \expandafter \@secondoftwo
 \fi
}%
\providecommand \natexlab [1]{#1}%
\providecommand \enquote  [1]{``#1''}%
\providecommand \bibnamefont  [1]{#1}%
\providecommand \bibfnamefont [1]{#1}%
\providecommand \citenamefont [1]{#1}%
\providecommand \href@noop [0]{\@secondoftwo}%
\providecommand \href [0]{\begingroup \@sanitize@url \@href}%
\providecommand \@href[1]{\@@startlink{#1}\@@href}%
\providecommand \@@href[1]{\endgroup#1\@@endlink}%
\providecommand \@sanitize@url [0]{\catcode `\\12\catcode `\$12\catcode `\&12\catcode `\#12\catcode `\^12\catcode `\_12\catcode `\%12\relax}%
\providecommand \@@startlink[1]{}%
\providecommand \@@endlink[0]{}%
\providecommand \url  [0]{\begingroup\@sanitize@url \@url }%
\providecommand \@url [1]{\endgroup\@href {#1}{\urlprefix }}%
\providecommand \urlprefix  [0]{URL }%
\providecommand \Eprint [0]{\href }%
\providecommand \doibase [0]{https://doi.org/}%
\providecommand \selectlanguage [0]{\@gobble}%
\providecommand \bibinfo  [0]{\@secondoftwo}%
\providecommand \bibfield  [0]{\@secondoftwo}%
\providecommand \translation [1]{[#1]}%
\providecommand \BibitemOpen [0]{}%
\providecommand \bibitemStop [0]{}%
\providecommand \bibitemNoStop [0]{.\EOS\space}%
\providecommand \EOS [0]{\spacefactor3000\relax}%
\providecommand \BibitemShut  [1]{\csname bibitem#1\endcsname}%
\let\auto@bib@innerbib\@empty
%</preamble>
\bibitem [{\citenamefont {Deblais}\ \emph {et~al.}(2023)\citenamefont {Deblais}, \citenamefont {Prathyusha}, \citenamefont {Sinaasappel}, \citenamefont {Tuazon}, \citenamefont {Tiwari}, \citenamefont {Patil},\ and\ \citenamefont {Bhamla}}]{ReviewWorm2023}%
  \BibitemOpen
  \bibfield  {author} {\bibinfo {author} {\bibfnamefont {A.}~\bibnamefont {Deblais}}, \bibinfo {author} {\bibfnamefont {K.~R.}\ \bibnamefont {Prathyusha}}, \bibinfo {author} {\bibfnamefont {R.}~\bibnamefont {Sinaasappel}}, \bibinfo {author} {\bibfnamefont {H.}~\bibnamefont {Tuazon}}, \bibinfo {author} {\bibfnamefont {I.}~\bibnamefont {Tiwari}}, \bibinfo {author} {\bibfnamefont {V.~P.}\ \bibnamefont {Patil}},\ and\ \bibinfo {author} {\bibfnamefont {M.~S.}\ \bibnamefont {Bhamla}},\ }\bibfield  {title} {\bibinfo {title} {Worm blobs as entangled living polymers: from topological active matter to flexible soft robot collectives},\ }\href {https://doi.org/10.1039/D3SM00542A} {\bibfield  {journal} {\bibinfo  {journal} {Soft Matter}\ }\textbf {\bibinfo {volume} {19}},\ \bibinfo {pages} {7057} (\bibinfo {year} {2023})}\BibitemShut {NoStop}%
\bibitem [{\citenamefont {Winkler}\ and\ \citenamefont {Gompper}(2020)}]{winkler2020}%
  \BibitemOpen
  \bibfield  {author} {\bibinfo {author} {\bibfnamefont {R.~G.}\ \bibnamefont {Winkler}}\ and\ \bibinfo {author} {\bibfnamefont {G.}~\bibnamefont {Gompper}},\ }\bibfield  {title} {\bibinfo {title} {The physics of active polymers and filaments},\ }\href@noop {} {\bibfield  {journal} {\bibinfo  {journal} {The journal of chemical physics}\ }\textbf {\bibinfo {volume} {153}} (\bibinfo {year} {2020})}\BibitemShut {NoStop}%
\bibitem [{\citenamefont {Ganguly}\ \emph {et~al.}(2012)\citenamefont {Ganguly}, \citenamefont {Williams}, \citenamefont {Palacios},\ and\ \citenamefont {Goldstein}}]{ganguly2012cytoplasmic}%
  \BibitemOpen
  \bibfield  {author} {\bibinfo {author} {\bibfnamefont {S.}~\bibnamefont {Ganguly}}, \bibinfo {author} {\bibfnamefont {L.~S.}\ \bibnamefont {Williams}}, \bibinfo {author} {\bibfnamefont {I.~M.}\ \bibnamefont {Palacios}},\ and\ \bibinfo {author} {\bibfnamefont {R.~E.}\ \bibnamefont {Goldstein}},\ }\bibfield  {title} {\bibinfo {title} {Cytoplasmic streaming in drosophila oocytes varies with kinesin activity and correlates with the microtubule cytoskeleton architecture},\ }\href@noop {} {\bibfield  {journal} {\bibinfo  {journal} {Proc. Natl. Acad. Sci. USA.}\ }\textbf {\bibinfo {volume} {109}},\ \bibinfo {pages} {15109} (\bibinfo {year} {2012})}\BibitemShut {NoStop}%
\bibitem [{\citenamefont {N\'ed\'elec}\ \emph {et~al.}(1997)\citenamefont {N\'ed\'elec}, \citenamefont {Surrey}, \citenamefont {Maggs},\ and\ \citenamefont {Leibler}}]{Nedelec1997}%
  \BibitemOpen
  \bibfield  {author} {\bibinfo {author} {\bibfnamefont {F.}~\bibnamefont {N\'ed\'elec}}, \bibinfo {author} {\bibfnamefont {T.}~\bibnamefont {Surrey}}, \bibinfo {author} {\bibfnamefont {A.~C.}\ \bibnamefont {Maggs}},\ and\ \bibinfo {author} {\bibfnamefont {S.}~\bibnamefont {Leibler}},\ }\bibfield  {title} {\bibinfo {title} {Self-organization of microtubules and motors},\ }\href@noop {} {\bibfield  {journal} {\bibinfo  {journal} {Nature}\ }\textbf {\bibinfo {volume} {389}},\ \bibinfo {pages} {305} (\bibinfo {year} {1997})}\BibitemShut {NoStop}%
\bibitem [{\citenamefont {Sumino}\ \emph {et~al.}(2012)\citenamefont {Sumino}, \citenamefont {Nagai}, \citenamefont {Shitaka}, \citenamefont {Tanaka}, \citenamefont {Yoshikawa}, \citenamefont {Chat{\'e}},\ and\ \citenamefont {Oiwa}}]{sumino-nature-12}%
  \BibitemOpen
  \bibfield  {author} {\bibinfo {author} {\bibfnamefont {Y.}~\bibnamefont {Sumino}}, \bibinfo {author} {\bibfnamefont {K.~H.}\ \bibnamefont {Nagai}}, \bibinfo {author} {\bibfnamefont {Y.}~\bibnamefont {Shitaka}}, \bibinfo {author} {\bibfnamefont {D.}~\bibnamefont {Tanaka}}, \bibinfo {author} {\bibfnamefont {K.}~\bibnamefont {Yoshikawa}}, \bibinfo {author} {\bibfnamefont {H.}~\bibnamefont {Chat{\'e}}},\ and\ \bibinfo {author} {\bibfnamefont {K.}~\bibnamefont {Oiwa}},\ }\bibfield  {title} {\bibinfo {title} {Large-scale vortex lattice emerging from collectively moving microtubules},\ }\href@noop {} {\bibfield  {journal} {\bibinfo  {journal} {Nature}\ }\textbf {\bibinfo {volume} {483}},\ \bibinfo {pages} {448} (\bibinfo {year} {2012})}\BibitemShut {NoStop}%
\bibitem [{\citenamefont {Goff}\ \emph {et~al.}(2002)\citenamefont {Goff}, \citenamefont {Amblard},\ and\ \citenamefont {Furs}}]{LeGoff2002}%
  \BibitemOpen
  \bibfield  {author} {\bibinfo {author} {\bibfnamefont {L.~L.}\ \bibnamefont {Goff}}, \bibinfo {author} {\bibfnamefont {F.}~\bibnamefont {Amblard}},\ and\ \bibinfo {author} {\bibfnamefont {E.~M.}\ \bibnamefont {Furs}},\ }\bibfield  {title} {\bibinfo {title} {Motor-driven dynamics in actin-myosin networks},\ }\href@noop {} {\bibfield  {journal} {\bibinfo  {journal} {Phys. Rev. Lett.}\ }\textbf {\bibinfo {volume} {88}},\ \bibinfo {pages} {018101} (\bibinfo {year} {2002})}\BibitemShut {NoStop}%
\bibitem [{\citenamefont {Sanchez}\ \emph {et~al.}(2012)\citenamefont {Sanchez}, \citenamefont {Chen}, \citenamefont {DeCamp}, \citenamefont {Heymann},\ and\ \citenamefont {Dogic}}]{dogic-activenematic-nature-12}%
  \BibitemOpen
  \bibfield  {author} {\bibinfo {author} {\bibfnamefont {T.}~\bibnamefont {Sanchez}}, \bibinfo {author} {\bibfnamefont {D.~T.}\ \bibnamefont {Chen}}, \bibinfo {author} {\bibfnamefont {S.~J.}\ \bibnamefont {DeCamp}}, \bibinfo {author} {\bibfnamefont {M.}~\bibnamefont {Heymann}},\ and\ \bibinfo {author} {\bibfnamefont {Z.}~\bibnamefont {Dogic}},\ }\bibfield  {title} {\bibinfo {title} {Spontaneous motion in hierarchically assembled active matter},\ }\href@noop {} {\bibfield  {journal} {\bibinfo  {journal} {Nature}\ }\textbf {\bibinfo {volume} {491}},\ \bibinfo {pages} {431} (\bibinfo {year} {2012})}\BibitemShut {NoStop}%
\bibitem [{\citenamefont {Kirchenbuechler}\ \emph {et~al.}(2014)\citenamefont {Kirchenbuechler}, \citenamefont {Guu}, \citenamefont {Kurniawan}, \citenamefont {Koenderink},\ and\ \citenamefont {Lettinga}}]{kirchenbuechler2014direct}%
  \BibitemOpen
  \bibfield  {author} {\bibinfo {author} {\bibfnamefont {I.}~\bibnamefont {Kirchenbuechler}}, \bibinfo {author} {\bibfnamefont {D.}~\bibnamefont {Guu}}, \bibinfo {author} {\bibfnamefont {N.~A.}\ \bibnamefont {Kurniawan}}, \bibinfo {author} {\bibfnamefont {G.~H.}\ \bibnamefont {Koenderink}},\ and\ \bibinfo {author} {\bibfnamefont {M.~P.}\ \bibnamefont {Lettinga}},\ }\bibfield  {title} {\bibinfo {title} {Direct visualization of flow-induced conformational transitions of single actin filaments in entangled solutions},\ }\href@noop {} {\bibfield  {journal} {\bibinfo  {journal} {Nat. Commun.}\ }\textbf {\bibinfo {volume} {5}},\ \bibinfo {pages} {5060} (\bibinfo {year} {2014})}\BibitemShut {NoStop}%
\bibitem [{\citenamefont {Sleigh}(1962)}]{biology_of_cilia_flagella}%
  \BibitemOpen
  \bibfield  {author} {\bibinfo {author} {\bibfnamefont {M.~A.}\ \bibnamefont {Sleigh}},\ }\href@noop {} {\emph {\bibinfo {title} {The Biology of Cilia and Flagella}}}\ (\bibinfo  {publisher} {Macmillan},\ \bibinfo {address} {New York},\ \bibinfo {year} {1962})\BibitemShut {NoStop}%
\bibitem [{\citenamefont {Shah}\ \emph {et~al.}(2009)\citenamefont {Shah}, \citenamefont {Ben-Shahar}, \citenamefont {Moninger}, \citenamefont {Kline},\ and\ \citenamefont {Welsh}}]{shah2009motile}%
  \BibitemOpen
  \bibfield  {author} {\bibinfo {author} {\bibfnamefont {A.~S.}\ \bibnamefont {Shah}}, \bibinfo {author} {\bibfnamefont {Y.}~\bibnamefont {Ben-Shahar}}, \bibinfo {author} {\bibfnamefont {T.~O.}\ \bibnamefont {Moninger}}, \bibinfo {author} {\bibfnamefont {J.~N.}\ \bibnamefont {Kline}},\ and\ \bibinfo {author} {\bibfnamefont {M.~J.}\ \bibnamefont {Welsh}},\ }\bibfield  {title} {\bibinfo {title} {Motile cilia of human airway epithelia are chemosensory},\ }\href@noop {} {\bibfield  {journal} {\bibinfo  {journal} {Science}\ }\textbf {\bibinfo {volume} {325}},\ \bibinfo {pages} {1131} (\bibinfo {year} {2009})}\BibitemShut {NoStop}%
\bibitem [{\citenamefont {Bardy}\ \emph {et~al.}(2003)\citenamefont {Bardy}, \citenamefont {Ng},\ and\ \citenamefont {Jarrell}}]{bardy2003prokaryotic}%
  \BibitemOpen
  \bibfield  {author} {\bibinfo {author} {\bibfnamefont {S.~L.}\ \bibnamefont {Bardy}}, \bibinfo {author} {\bibfnamefont {S.~Y.}\ \bibnamefont {Ng}},\ and\ \bibinfo {author} {\bibfnamefont {K.~F.}\ \bibnamefont {Jarrell}},\ }\bibfield  {title} {\bibinfo {title} {Prokaryotic motility structures},\ }\href@noop {} {\bibfield  {journal} {\bibinfo  {journal} {Microbiol.}\ }\textbf {\bibinfo {volume} {149}},\ \bibinfo {pages} {295} (\bibinfo {year} {2003})}\BibitemShut {NoStop}%
\bibitem [{\citenamefont {Mayfield}\ and\ \citenamefont {Inniss}(1977)}]{mayfield1977rapid}%
  \BibitemOpen
  \bibfield  {author} {\bibinfo {author} {\bibfnamefont {C.~I.}\ \bibnamefont {Mayfield}}\ and\ \bibinfo {author} {\bibfnamefont {W.~E.}\ \bibnamefont {Inniss}},\ }\bibfield  {title} {\bibinfo {title} {A rapid, simple method for staining bacterial flagella},\ }\href@noop {} {\bibfield  {journal} {\bibinfo  {journal} {Can. J. Microbiol.}\ }\textbf {\bibinfo {volume} {23}},\ \bibinfo {pages} {1311} (\bibinfo {year} {1977})}\BibitemShut {NoStop}%
\bibitem [{\citenamefont {Friedrich}\ \emph {et~al.}(2010)\citenamefont {Friedrich}, \citenamefont {Riedel-Kruse}, \citenamefont {Howard},\ and\ \citenamefont {J{\"u}licher}}]{friedrich2010sperm}%
  \BibitemOpen
  \bibfield  {author} {\bibinfo {author} {\bibfnamefont {B.~M.}\ \bibnamefont {Friedrich}}, \bibinfo {author} {\bibfnamefont {I.~H.}\ \bibnamefont {Riedel-Kruse}}, \bibinfo {author} {\bibfnamefont {J.}~\bibnamefont {Howard}},\ and\ \bibinfo {author} {\bibfnamefont {F.}~\bibnamefont {J{\"u}licher}},\ }\bibfield  {title} {\bibinfo {title} {High-precision tracking of sperm swimming fine structure provides strong test of resistive force theory},\ }\href@noop {} {\bibfield  {journal} {\bibinfo  {journal} {J. Exp. Biol.}\ }\textbf {\bibinfo {volume} {213}},\ \bibinfo {pages} {1226} (\bibinfo {year} {2010})}\BibitemShut {NoStop}%
\bibitem [{\citenamefont {Patil}\ \emph {et~al.}(2023)\citenamefont {Patil}, \citenamefont {Tuazon}, \citenamefont {Kaufman}, \citenamefont {Chakrabortty}, \citenamefont {Qin}, \citenamefont {Dunkel},\ and\ \citenamefont {Bhamla}}]{patiltuazon2023}%
  \BibitemOpen
  \bibfield  {author} {\bibinfo {author} {\bibfnamefont {V.~P.}\ \bibnamefont {Patil}}, \bibinfo {author} {\bibfnamefont {H.}~\bibnamefont {Tuazon}}, \bibinfo {author} {\bibfnamefont {E.}~\bibnamefont {Kaufman}}, \bibinfo {author} {\bibfnamefont {T.}~\bibnamefont {Chakrabortty}}, \bibinfo {author} {\bibfnamefont {D.}~\bibnamefont {Qin}}, \bibinfo {author} {\bibfnamefont {J.}~\bibnamefont {Dunkel}},\ and\ \bibinfo {author} {\bibfnamefont {M.~S.}\ \bibnamefont {Bhamla}},\ }\bibfield  {title} {\bibinfo {title} {Ultrafast reversible self-assembly of living tangled matter},\ }\href {https://doi.org/10.1126/science.ade7759} {\bibfield  {journal} {\bibinfo  {journal} {Science}\ }\textbf {\bibinfo {volume} {380}},\ \bibinfo {pages} {392} (\bibinfo {year} {2023})},\ \Eprint {https://arxiv.org/abs/https://www.science.org/doi/pdf/10.1126/science.ade7759} {https://www.science.org/doi/pdf/10.1126/science.ade7759} \BibitemShut {NoStop}%
\bibitem [{\citenamefont {Gray}(1946)}]{gray1946mechanism}%
  \BibitemOpen
  \bibfield  {author} {\bibinfo {author} {\bibfnamefont {J.}~\bibnamefont {Gray}},\ }\bibfield  {title} {\bibinfo {title} {The mechanism of locomotion in snakes},\ }\href@noop {} {\bibfield  {journal} {\bibinfo  {journal} {J. Exp. Biol.}\ }\textbf {\bibinfo {volume} {23}},\ \bibinfo {pages} {101} (\bibinfo {year} {1946})}\BibitemShut {NoStop}%
\bibitem [{\citenamefont {Hern{\'a}ndez-L{\'o}pez}\ \emph {et~al.}(2024)\citenamefont {Hern{\'a}ndez-L{\'o}pez}, \citenamefont {Baconnier}, \citenamefont {Coulais}, \citenamefont {Dauchot},\ and\ \citenamefont {D{\"u}ring}}]{lopezactivesolids2024}%
  \BibitemOpen
  \bibfield  {author} {\bibinfo {author} {\bibfnamefont {C.}~\bibnamefont {Hern{\'a}ndez-L{\'o}pez}}, \bibinfo {author} {\bibfnamefont {P.}~\bibnamefont {Baconnier}}, \bibinfo {author} {\bibfnamefont {C.}~\bibnamefont {Coulais}}, \bibinfo {author} {\bibfnamefont {O.}~\bibnamefont {Dauchot}},\ and\ \bibinfo {author} {\bibfnamefont {G.}~\bibnamefont {D{\"u}ring}},\ }\bibfield  {title} {\bibinfo {title} {Model of active solids: Rigid body motion and shape-changing mechanisms},\ }\href@noop {} {\bibfield  {journal} {\bibinfo  {journal} {Phys. Rev. Lett.}\ }\textbf {\bibinfo {volume} {132}},\ \bibinfo {pages} {238303} (\bibinfo {year} {2024})}\BibitemShut {NoStop}%
\bibitem [{\citenamefont {Zheng}\ \emph {et~al.}(2023)\citenamefont {Zheng}, \citenamefont {Brandenbourger}, \citenamefont {Robinet}, \citenamefont {Schall}, \citenamefont {Lerner},\ and\ \citenamefont {Coulais}}]{zhengrobot-anchor2023}%
  \BibitemOpen
  \bibfield  {author} {\bibinfo {author} {\bibfnamefont {E.}~\bibnamefont {Zheng}}, \bibinfo {author} {\bibfnamefont {M.}~\bibnamefont {Brandenbourger}}, \bibinfo {author} {\bibfnamefont {L.}~\bibnamefont {Robinet}}, \bibinfo {author} {\bibfnamefont {P.}~\bibnamefont {Schall}}, \bibinfo {author} {\bibfnamefont {E.}~\bibnamefont {Lerner}},\ and\ \bibinfo {author} {\bibfnamefont {C.}~\bibnamefont {Coulais}},\ }\bibfield  {title} {\bibinfo {title} {Self-oscillation and synchronization transitions in elastoactive structures},\ }\href@noop {} {\bibfield  {journal} {\bibinfo  {journal} {Phys. Rev. Lett.}\ }\textbf {\bibinfo {volume} {130}},\ \bibinfo {pages} {178202} (\bibinfo {year} {2023})}\BibitemShut {NoStop}%
\bibitem [{\citenamefont {Xi}\ \emph {et~al.}(2024)\citenamefont {Xi}, \citenamefont {Marzin}, \citenamefont {Huang}, \citenamefont {Jones},\ and\ \citenamefont {Brun}}]{brun2024emergent}%
  \BibitemOpen
  \bibfield  {author} {\bibinfo {author} {\bibfnamefont {Y.}~\bibnamefont {Xi}}, \bibinfo {author} {\bibfnamefont {T.}~\bibnamefont {Marzin}}, \bibinfo {author} {\bibfnamefont {R.~B.}\ \bibnamefont {Huang}}, \bibinfo {author} {\bibfnamefont {T.~J.}\ \bibnamefont {Jones}},\ and\ \bibinfo {author} {\bibfnamefont {P.-T.}\ \bibnamefont {Brun}},\ }\bibfield  {title} {\bibinfo {title} {Emergent behaviors of buckling-driven elasto-active structures},\ }\href@noop {} {\bibfield  {journal} {\bibinfo  {journal} {Proc. Natl. Acad. Sci. U.S.A.}\ }\textbf {\bibinfo {volume} {121}},\ \bibinfo {pages} {e2410654121} (\bibinfo {year} {2024})}\BibitemShut {NoStop}%
\bibitem [{\citenamefont {Tuazon}\ \emph {et~al.}(2023)\citenamefont {Tuazon}, \citenamefont {Nguyen}, \citenamefont {Kaufman}, \citenamefont {Tiwari}, \citenamefont {Bermudez}, \citenamefont {Chudasama}, \citenamefont {Peleg},\ and\ \citenamefont {Bhamla}}]{tuazon2023collecting}%
  \BibitemOpen
  \bibfield  {author} {\bibinfo {author} {\bibfnamefont {H.}~\bibnamefont {Tuazon}}, \bibinfo {author} {\bibfnamefont {C.}~\bibnamefont {Nguyen}}, \bibinfo {author} {\bibfnamefont {E.}~\bibnamefont {Kaufman}}, \bibinfo {author} {\bibfnamefont {I.}~\bibnamefont {Tiwari}}, \bibinfo {author} {\bibfnamefont {J.}~\bibnamefont {Bermudez}}, \bibinfo {author} {\bibfnamefont {D.}~\bibnamefont {Chudasama}}, \bibinfo {author} {\bibfnamefont {O.}~\bibnamefont {Peleg}},\ and\ \bibinfo {author} {\bibfnamefont {M.}~\bibnamefont {Bhamla}},\ }\bibfield  {title} {\bibinfo {title} {Collecting--gathering biophysics of the blackworm lumbriculus variegatus},\ }\href@noop {} {\bibfield  {journal} {\bibinfo  {journal} {Integr. Comp. Biol.}\ }\textbf {\bibinfo {volume} {63}},\ \bibinfo {pages} {1474} (\bibinfo {year} {2023})}\BibitemShut {NoStop}%
\bibitem [{\citenamefont {Gokhale}\ \emph {et~al.}(2022)\citenamefont {Gokhale}, \citenamefont {Li}, \citenamefont {Solon}, \citenamefont {Gore},\ and\ \citenamefont {Fakhri}}]{gokhale2022dynamic}%
  \BibitemOpen
  \bibfield  {author} {\bibinfo {author} {\bibfnamefont {S.}~\bibnamefont {Gokhale}}, \bibinfo {author} {\bibfnamefont {J.}~\bibnamefont {Li}}, \bibinfo {author} {\bibfnamefont {A.}~\bibnamefont {Solon}}, \bibinfo {author} {\bibfnamefont {J.}~\bibnamefont {Gore}},\ and\ \bibinfo {author} {\bibfnamefont {N.}~\bibnamefont {Fakhri}},\ }\bibfield  {title} {\bibinfo {title} {Dynamic clustering of passive colloids in dense suspensions of motile bacteria},\ }\href@noop {} {\bibfield  {journal} {\bibinfo  {journal} {Phys. Rev. E}\ }\textbf {\bibinfo {volume} {105}},\ \bibinfo {pages} {054605} (\bibinfo {year} {2022})}\BibitemShut {NoStop}%
\bibitem [{\citenamefont {Dolai}\ \emph {et~al.}(2018)\citenamefont {Dolai}, \citenamefont {Simha},\ and\ \citenamefont {Mishra}}]{pritha_phase_2018}%
  \BibitemOpen
  \bibfield  {author} {\bibinfo {author} {\bibfnamefont {P.}~\bibnamefont {Dolai}}, \bibinfo {author} {\bibfnamefont {A.}~\bibnamefont {Simha}},\ and\ \bibinfo {author} {\bibfnamefont {S.}~\bibnamefont {Mishra}},\ }\bibfield  {title} {\bibinfo {title} {Phase separation in binary mixtures of active and passive particles},\ }\href@noop {} {\bibfield  {journal} {\bibinfo  {journal} {Soft Matter}\ }\textbf {\bibinfo {volume} {14}},\ \bibinfo {pages} {6137} (\bibinfo {year} {2018})}\BibitemShut {NoStop}%
\bibitem [{\citenamefont {Stenhammar}\ \emph {et~al.}(2015)\citenamefont {Stenhammar}, \citenamefont {Wittkowski}, \citenamefont {Marenduzzo},\ and\ \citenamefont {Cates}}]{Cates-mips-AP-mixture-PRL-15}%
  \BibitemOpen
  \bibfield  {author} {\bibinfo {author} {\bibfnamefont {J.}~\bibnamefont {Stenhammar}}, \bibinfo {author} {\bibfnamefont {R.}~\bibnamefont {Wittkowski}}, \bibinfo {author} {\bibfnamefont {D.}~\bibnamefont {Marenduzzo}},\ and\ \bibinfo {author} {\bibfnamefont {M.~E.}\ \bibnamefont {Cates}},\ }\bibfield  {title} {\bibinfo {title} {Activity-induced phase separation and self-assembly in mixtures of active and passive particles},\ }\href@noop {} {\bibfield  {journal} {\bibinfo  {journal} {Phys. Rev. Lett.}\ }\textbf {\bibinfo {volume} {114}},\ \bibinfo {pages} {018301} (\bibinfo {year} {2015})}\BibitemShut {NoStop}%
\bibitem [{\citenamefont {McCandlish}\ \emph {et~al.}(2012)\citenamefont {McCandlish}, \citenamefont {Baskaran},\ and\ \citenamefont {Hagan}}]{aparna2012spontaneous}%
  \BibitemOpen
  \bibfield  {author} {\bibinfo {author} {\bibfnamefont {S.~R.}\ \bibnamefont {McCandlish}}, \bibinfo {author} {\bibfnamefont {A.}~\bibnamefont {Baskaran}},\ and\ \bibinfo {author} {\bibfnamefont {M.~F.}\ \bibnamefont {Hagan}},\ }\bibfield  {title} {\bibinfo {title} {Spontaneous segregation of self-propelled particles with different motilities},\ }\href@noop {} {\bibfield  {journal} {\bibinfo  {journal} {Soft Matter}\ }\textbf {\bibinfo {volume} {8}},\ \bibinfo {pages} {2527} (\bibinfo {year} {2012})}\BibitemShut {NoStop}%
\bibitem [{\citenamefont {Visser}(2007)}]{visser2007biomixing}%
  \BibitemOpen
  \bibfield  {author} {\bibinfo {author} {\bibfnamefont {A.~W.}\ \bibnamefont {Visser}},\ }\bibfield  {title} {\bibinfo {title} {Biomixing of the oceans?},\ }\href@noop {} {\bibfield  {journal} {\bibinfo  {journal} {Science}\ }\textbf {\bibinfo {volume} {316}},\ \bibinfo {pages} {838} (\bibinfo {year} {2007})}\BibitemShut {NoStop}%
\bibitem [{\citenamefont {Prathyusha}(2023)}]{prathyushatransvercargo}%
  \BibitemOpen
  \bibfield  {author} {\bibinfo {author} {\bibfnamefont {K.~R.}\ \bibnamefont {Prathyusha}},\ }\bibfield  {title} {\bibinfo {title} {Passive particle transport using a transversely propelling polymer “sweeper”},\ }\href@noop {} {\bibfield  {journal} {\bibinfo  {journal} {Soft Matter}\ }\textbf {\bibinfo {volume} {19}},\ \bibinfo {pages} {4001} (\bibinfo {year} {2023})}\BibitemShut {NoStop}%
\bibitem [{\citenamefont {Zhang}\ \emph {et~al.}(2021)\citenamefont {Zhang}, \citenamefont {Lei},\ and\ \citenamefont {Zhao}}]{zhangpolymerloop-21}%
  \BibitemOpen
  \bibfield  {author} {\bibinfo {author} {\bibfnamefont {B.}~\bibnamefont {Zhang}}, \bibinfo {author} {\bibfnamefont {T.}~\bibnamefont {Lei}},\ and\ \bibinfo {author} {\bibfnamefont {N.}~\bibnamefont {Zhao}},\ }\bibfield  {title} {\bibinfo {title} {Comparative study of polymer looping kinetics in passive and active environments},\ }\href@noop {} {\bibfield  {journal} {\bibinfo  {journal} {Phys. Chem. Chem. Phys.}\ }\textbf {\bibinfo {volume} {23}},\ \bibinfo {pages} {12171} (\bibinfo {year} {2021})}\BibitemShut {NoStop}%
\bibitem [{\citenamefont {Smrek}\ and\ \citenamefont {Kremer}(2017)}]{smrek_small_2017}%
  \BibitemOpen
  \bibfield  {author} {\bibinfo {author} {\bibfnamefont {J.}~\bibnamefont {Smrek}}\ and\ \bibinfo {author} {\bibfnamefont {K.}~\bibnamefont {Kremer}},\ }\bibfield  {title} {\bibinfo {title} {Small activity differences drive phase separation in active-passive polymer mixtures},\ }\href {https://doi.org/10.1103/PhysRevLett.118.098002} {\bibfield  {journal} {\bibinfo  {journal} {Phys. Rev. Lett.}\ }\textbf {\bibinfo {volume} {118}},\ \bibinfo {pages} {098002} (\bibinfo {year} {2017})}\BibitemShut {NoStop}%
\bibitem [{\citenamefont {Deblais}\ \emph {et~al.}(2018)\citenamefont {Deblais}, \citenamefont {Barois}, \citenamefont {Guerin}, \citenamefont {Delville}, \citenamefont {Vaudaine}, \citenamefont {Lintuvuori}, \citenamefont {Boudet}, \citenamefont {Baret},\ and\ \citenamefont {Kellay}}]{deblais2018boundaries}%
  \BibitemOpen
  \bibfield  {author} {\bibinfo {author} {\bibfnamefont {A.}~\bibnamefont {Deblais}}, \bibinfo {author} {\bibfnamefont {T.}~\bibnamefont {Barois}}, \bibinfo {author} {\bibfnamefont {T.}~\bibnamefont {Guerin}}, \bibinfo {author} {\bibfnamefont {P.-H.}\ \bibnamefont {Delville}}, \bibinfo {author} {\bibfnamefont {R.}~\bibnamefont {Vaudaine}}, \bibinfo {author} {\bibfnamefont {J.~S.}\ \bibnamefont {Lintuvuori}}, \bibinfo {author} {\bibfnamefont {J.-F.}\ \bibnamefont {Boudet}}, \bibinfo {author} {\bibfnamefont {J.-C.}\ \bibnamefont {Baret}},\ and\ \bibinfo {author} {\bibfnamefont {H.}~\bibnamefont {Kellay}},\ }\bibfield  {title} {\bibinfo {title} {Boundaries control collective dynamics of inertial self-propelled robots},\ }\href@noop {} {\bibfield  {journal} {\bibinfo  {journal} {Phys. Rev. Lett.}\ }\textbf {\bibinfo {volume} {120}},\ \bibinfo {pages} {188002} (\bibinfo {year} {2018})}\BibitemShut {NoStop}%
\bibitem [{\citenamefont {Prathyusha}\ \emph {et~al.}(2018)\citenamefont {Prathyusha}, \citenamefont {Henkes},\ and\ \citenamefont {Sknepnek}}]{prathyusha2018PRE}%
  \BibitemOpen
  \bibfield  {author} {\bibinfo {author} {\bibfnamefont {K.~R.}\ \bibnamefont {Prathyusha}}, \bibinfo {author} {\bibfnamefont {S.}~\bibnamefont {Henkes}},\ and\ \bibinfo {author} {\bibfnamefont {R.}~\bibnamefont {Sknepnek}},\ }\bibfield  {title} {\bibinfo {title} {Dynamically generated patterns in dense suspensions of active filaments},\ }\href@noop {} {\bibfield  {journal} {\bibinfo  {journal} {Phys. Rev. E}\ }\textbf {\bibinfo {volume} {97}},\ \bibinfo {pages} {022606} (\bibinfo {year} {2018})}\BibitemShut {NoStop}%
\bibitem [{\citenamefont {Isele-Holder}\ \emph {et~al.}(2015)\citenamefont {Isele-Holder}, \citenamefont {Elgeti},\ and\ \citenamefont {Gompper}}]{riseleholder-15}%
  \BibitemOpen
  \bibfield  {author} {\bibinfo {author} {\bibfnamefont {R.~E.}\ \bibnamefont {Isele-Holder}}, \bibinfo {author} {\bibfnamefont {J.}~\bibnamefont {Elgeti}},\ and\ \bibinfo {author} {\bibfnamefont {G.}~\bibnamefont {Gompper}},\ }\bibfield  {title} {\bibinfo {title} {Self-propelled worm-like filaments: spontaneous spiral formation, structure, and dynamics},\ }\href@noop {} {\bibfield  {journal} {\bibinfo  {journal} {Soft Matter}\ }\textbf {\bibinfo {volume} {11}},\ \bibinfo {pages} {7181} (\bibinfo {year} {2015})}\BibitemShut {NoStop}%
\bibitem [{\citenamefont {Fazelzadeh}\ \emph {et~al.}(2023)\citenamefont {Fazelzadeh}, \citenamefont {Di}, \citenamefont {Irani}, \citenamefont {Mokhtari},\ and\ \citenamefont {Jabbari-Farouji}}]{fazelzadeh2023}%
  \BibitemOpen
  \bibfield  {author} {\bibinfo {author} {\bibfnamefont {M.}~\bibnamefont {Fazelzadeh}}, \bibinfo {author} {\bibfnamefont {Q.}~\bibnamefont {Di}}, \bibinfo {author} {\bibfnamefont {E.}~\bibnamefont {Irani}}, \bibinfo {author} {\bibfnamefont {Z.}~\bibnamefont {Mokhtari}},\ and\ \bibinfo {author} {\bibfnamefont {S.}~\bibnamefont {Jabbari-Farouji}},\ }\bibfield  {title} {\bibinfo {title} {Active motion of tangentially driven polymers in periodic array of obstacles},\ }\href@noop {} {\bibfield  {journal} {\bibinfo  {journal} {J. Chem. Phys.}\ }\textbf {\bibinfo {volume} {159}} (\bibinfo {year} {2023})}\BibitemShut {NoStop}%
\bibitem [{\citenamefont {Ozkan-Aydin}\ \emph {et~al.}(2021)\citenamefont {Ozkan-Aydin}, \citenamefont {Goldman},\ and\ \citenamefont {Bhamla}}]{Ozkan2021}%
  \BibitemOpen
  \bibfield  {author} {\bibinfo {author} {\bibfnamefont {Y.}~\bibnamefont {Ozkan-Aydin}}, \bibinfo {author} {\bibfnamefont {D.~I.}\ \bibnamefont {Goldman}},\ and\ \bibinfo {author} {\bibfnamefont {M.~S.}\ \bibnamefont {Bhamla}},\ }\bibfield  {title} {\bibinfo {title} {Collective dynamics in entangled worm and robot blobs},\ }\href {https://doi.org/10.1073/pnas.2010542118} {\bibfield  {journal} {\bibinfo  {journal} {Proc. Natl. Acad. Sci. U. S. A..}\ }\textbf {\bibinfo {volume} {118}},\ \bibinfo {pages} {e2010542118} (\bibinfo {year} {2021})}\BibitemShut {NoStop}%
\bibitem [{Supplementary movies()}]{Movies}%
  \BibitemOpen
  Supplementary movies,\ \href {https://github.com/deb-lab/Supplementary-Movies-collecting} {}\bibinfo {howpublished} {\url{https://github.com/deb-lab/Supplementary-Movies-collecting}}\BibitemShut {NoStop}%
\bibitem [{\citenamefont {Ginot}\ \emph {et~al.}(2018)\citenamefont {Ginot}, \citenamefont {Theurkauff}, \citenamefont {Detcheverry}, \citenamefont {Ybert},\ and\ \citenamefont {Cottin-Bizonne}}]{ginot2018}%
  \BibitemOpen
  \bibfield  {author} {\bibinfo {author} {\bibfnamefont {F.}~\bibnamefont {Ginot}}, \bibinfo {author} {\bibfnamefont {I.}~\bibnamefont {Theurkauff}}, \bibinfo {author} {\bibfnamefont {F.}~\bibnamefont {Detcheverry}}, \bibinfo {author} {\bibfnamefont {C.}~\bibnamefont {Ybert}},\ and\ \bibinfo {author} {\bibfnamefont {C.}~\bibnamefont {Cottin-Bizonne}},\ }\bibfield  {title} {\bibinfo {title} {Aggregation-fragmentation and individual dynamics of active clusters},\ }\href@noop {} {\bibfield  {journal} {\bibinfo  {journal} {Nat. Commun.}\ }\textbf {\bibinfo {volume} {9}},\ \bibinfo {pages} {696} (\bibinfo {year} {2018})}\BibitemShut {NoStop}%
\bibitem [{\citenamefont {Redner}\ \emph {et~al.}(2016)\citenamefont {Redner}, \citenamefont {Wagner}, \citenamefont {Baskaran},\ and\ \citenamefont {Hagan}}]{Redner2016}%
  \BibitemOpen
  \bibfield  {author} {\bibinfo {author} {\bibfnamefont {G.~S.}\ \bibnamefont {Redner}}, \bibinfo {author} {\bibfnamefont {C.~G.}\ \bibnamefont {Wagner}}, \bibinfo {author} {\bibfnamefont {A.}~\bibnamefont {Baskaran}},\ and\ \bibinfo {author} {\bibfnamefont {M.~F.}\ \bibnamefont {Hagan}},\ }\bibfield  {title} {\bibinfo {title} {Classical nucleation theory description of active colloid assembly},\ }\href {https://doi.org/10.1103/PhysRevLett.117.148002} {\bibfield  {journal} {\bibinfo  {journal} {Phys. Rev. Lett.}\ }\textbf {\bibinfo {volume} {117}},\ \bibinfo {pages} {148002} (\bibinfo {year} {2016})}\BibitemShut {NoStop}%
\bibitem [{\citenamefont {Martín-Roca}\ \emph {et~al.}(2024)\citenamefont {Martín-Roca}, \citenamefont {Locatelli}, \citenamefont {Bianco}, \citenamefont {Malgaretti},\ and\ \citenamefont {Valeriani}}]{locatelli-24}%
  \BibitemOpen
  \bibfield  {author} {\bibinfo {author} {\bibfnamefont {J.}~\bibnamefont {Martín-Roca}}, \bibinfo {author} {\bibfnamefont {E.}~\bibnamefont {Locatelli}}, \bibinfo {author} {\bibfnamefont {V.}~\bibnamefont {Bianco}}, \bibinfo {author} {\bibfnamefont {P.}~\bibnamefont {Malgaretti}},\ and\ \bibinfo {author} {\bibfnamefont {C.}~\bibnamefont {Valeriani}},\ }\bibfield  {title} {\bibinfo {title} {Tangentially active polymers in cylindrical channels},\ }\href {https://doi.org/10.21468/SciPostPhys.17.4.107} {\bibfield  {journal} {\bibinfo  {journal} {SciPost Phys.}\ }\textbf {\bibinfo {volume} {17}},\ \bibinfo {pages} {107} (\bibinfo {year} {2024})}\BibitemShut {NoStop}%
\bibitem [{\citenamefont {Sinaasappel}\ \emph {et~al.}(2025)\citenamefont {Sinaasappel}, \citenamefont {Fazelzadeh}, \citenamefont {Hooijschuur}, \citenamefont {Di}, \citenamefont {Jabbari-Farouji},\ and\ \citenamefont {Deblais}}]{sinaasappel2025}%
  \BibitemOpen
  \bibfield  {author} {\bibinfo {author} {\bibfnamefont {R.}~\bibnamefont {Sinaasappel}}, \bibinfo {author} {\bibfnamefont {M.}~\bibnamefont {Fazelzadeh}}, \bibinfo {author} {\bibfnamefont {T.}~\bibnamefont {Hooijschuur}}, \bibinfo {author} {\bibfnamefont {Q.}~\bibnamefont {Di}}, \bibinfo {author} {\bibfnamefont {S.}~\bibnamefont {Jabbari-Farouji}},\ and\ \bibinfo {author} {\bibfnamefont {A.}~\bibnamefont {Deblais}},\ }\bibfield  {title} {\bibinfo {title} {Locomotion of active polymerlike worms in porous media},\ }\href {https://doi.org/10.1103/PhysRevLett.134.128303} {\bibfield  {journal} {\bibinfo  {journal} {Phys. Rev. Lett.}\ }\textbf {\bibinfo {volume} {134}},\ \bibinfo {pages} {128303} (\bibinfo {year} {2025})}\BibitemShut {NoStop}%
\bibitem [{\citenamefont {Deseigne}\ \emph {et~al.}(2010)\citenamefont {Deseigne}, \citenamefont {Dauchot},\ and\ \citenamefont {Chat{\'e}}}]{Deseigne2010a}%
  \BibitemOpen
  \bibfield  {author} {\bibinfo {author} {\bibfnamefont {J.}~\bibnamefont {Deseigne}}, \bibinfo {author} {\bibfnamefont {O.}~\bibnamefont {Dauchot}},\ and\ \bibinfo {author} {\bibfnamefont {H.}~\bibnamefont {Chat{\'e}}},\ }\bibfield  {title} {\bibinfo {title} {Collective {{Motion}} of {{Vibrated Polar Disks}}},\ }\href {https://doi.org/10.1103/PhysRevLett.105.098001} {\bibfield  {journal} {\bibinfo  {journal} {Phys. Rev. Lett.}\ }\textbf {\bibinfo {volume} {105}},\ \bibinfo {pages} {098001} (\bibinfo {year} {2010})}\BibitemShut {NoStop}%
\bibitem [{\citenamefont {Deseigne}\ \emph {et~al.}(2012)\citenamefont {Deseigne}, \citenamefont {L{\'e}onard}, \citenamefont {Dauchot},\ and\ \citenamefont {Chat{\'e}}}]{deseigne2012}%
  \BibitemOpen
  \bibfield  {author} {\bibinfo {author} {\bibfnamefont {J.}~\bibnamefont {Deseigne}}, \bibinfo {author} {\bibfnamefont {S.}~\bibnamefont {L{\'e}onard}}, \bibinfo {author} {\bibfnamefont {O.}~\bibnamefont {Dauchot}},\ and\ \bibinfo {author} {\bibfnamefont {H.}~\bibnamefont {Chat{\'e}}},\ }\bibfield  {title} {\bibinfo {title} {Vibrated polar disks: spontaneous motion, binary collisions, and collective dynamics},\ }\href@noop {} {\bibfield  {journal} {\bibinfo  {journal} {Soft Matter}\ }\textbf {\bibinfo {volume} {8}},\ \bibinfo {pages} {5629} (\bibinfo {year} {2012})}\BibitemShut {NoStop}%
\bibitem [{\citenamefont {Patterson}\ \emph {et~al.}(2017)\citenamefont {Patterson}, \citenamefont {Fierens}, \citenamefont {Sangiuliano~Jimka}, \citenamefont {K{\"o}nig}, \citenamefont {Garcimart{\'\i}n}, \citenamefont {Zuriguel}, \citenamefont {Pugnaloni},\ and\ \citenamefont {Parisi}}]{patterson2017}%
  \BibitemOpen
  \bibfield  {author} {\bibinfo {author} {\bibfnamefont {G.~A.}\ \bibnamefont {Patterson}}, \bibinfo {author} {\bibfnamefont {P.~I.}\ \bibnamefont {Fierens}}, \bibinfo {author} {\bibfnamefont {F.}~\bibnamefont {Sangiuliano~Jimka}}, \bibinfo {author} {\bibfnamefont {P.}~\bibnamefont {K{\"o}nig}}, \bibinfo {author} {\bibfnamefont {{\'A}.}~\bibnamefont {Garcimart{\'\i}n}}, \bibinfo {author} {\bibfnamefont {I.}~\bibnamefont {Zuriguel}}, \bibinfo {author} {\bibfnamefont {L.~A.}\ \bibnamefont {Pugnaloni}},\ and\ \bibinfo {author} {\bibfnamefont {D.~R.}\ \bibnamefont {Parisi}},\ }\bibfield  {title} {\bibinfo {title} {Clogging transition of vibration-driven vehicles passing through constrictions},\ }\href@noop {} {\bibfield  {journal} {\bibinfo  {journal} {Phys. Rev. Lett.}\ }\textbf {\bibinfo {volume} {119}},\ \bibinfo {pages} {248301} (\bibinfo {year} {2017})}\BibitemShut {NoStop}%
\bibitem [{\citenamefont {Dauchot}\ and\ \citenamefont {D{\'e}mery}(2019)}]{dauchot2019}%
  \BibitemOpen
  \bibfield  {author} {\bibinfo {author} {\bibfnamefont {O.}~\bibnamefont {Dauchot}}\ and\ \bibinfo {author} {\bibfnamefont {V.}~\bibnamefont {D{\'e}mery}},\ }\bibfield  {title} {\bibinfo {title} {Dynamics of a self-propelled particle in a harmonic trap},\ }\href@noop {} {\bibfield  {journal} {\bibinfo  {journal} {Phys. Rev. Lett.}\ }\textbf {\bibinfo {volume} {122}},\ \bibinfo {pages} {068002} (\bibinfo {year} {2019})}\BibitemShut {NoStop}%
\bibitem [{\citenamefont {Krapivsky}\ \emph {et~al.}(2010)\citenamefont {Krapivsky}, \citenamefont {Redner},\ and\ \citenamefont {{Ben-Naim}}}]{Krapivsky2010}%
  \BibitemOpen
  \bibfield  {author} {\bibinfo {author} {\bibfnamefont {P.~L.}\ \bibnamefont {Krapivsky}}, \bibinfo {author} {\bibfnamefont {S.}~\bibnamefont {Redner}},\ and\ \bibinfo {author} {\bibfnamefont {E.}~\bibnamefont {{Ben-Naim}}},\ }\href@noop {} {\emph {\bibinfo {title} {A Kinetic View of Statistical Physics}}}\ (\bibinfo  {publisher} {Cambridge University Press},\ \bibinfo {year} {2010})\BibitemShut {NoStop}%
\bibitem [{\citenamefont {Ziff}(1980)}]{ziff1980}%
  \BibitemOpen
  \bibfield  {author} {\bibinfo {author} {\bibfnamefont {R.~M.}\ \bibnamefont {Ziff}},\ }\bibfield  {title} {\bibinfo {title} {Kinetics of polymerization},\ }\href@noop {} {\bibfield  {journal} {\bibinfo  {journal} {J. Stat. Phys.}\ }\textbf {\bibinfo {volume} {23}},\ \bibinfo {pages} {241} (\bibinfo {year} {1980})}\BibitemShut {NoStop}%
\bibitem [{\citenamefont {Bouvard}\ \emph {et~al.}(2023)\citenamefont {Bouvard}, \citenamefont {Moisy},\ and\ \citenamefont {Auradou}}]{bouvard2023}%
  \BibitemOpen
  \bibfield  {author} {\bibinfo {author} {\bibfnamefont {J.}~\bibnamefont {Bouvard}}, \bibinfo {author} {\bibfnamefont {F.}~\bibnamefont {Moisy}},\ and\ \bibinfo {author} {\bibfnamefont {H.}~\bibnamefont {Auradou}},\ }\bibfield  {title} {\bibinfo {title} {Ostwald-like ripening in the two-dimensional clustering of passive particles induced by swimming bacteria},\ }\href@noop {} {\bibfield  {journal} {\bibinfo  {journal} {Phys. Rev. E}\ }\textbf {\bibinfo {volume} {107}},\ \bibinfo {pages} {044607} (\bibinfo {year} {2023})}\BibitemShut {NoStop}%
\bibitem [{\citenamefont {Smoluchowski}\ and\ \citenamefont {im~unbegrenzten Raum}(1906)}]{Smoluchowski1906}%
  \BibitemOpen
  \bibfield  {author} {\bibinfo {author} {\bibfnamefont {V.}~\bibnamefont {Smoluchowski}}\ and\ \bibinfo {author} {\bibfnamefont {I.~D.}\ \bibnamefont {im~unbegrenzten Raum}},\ }\bibfield  {title} {\bibinfo {title} {Zusammenfassende bearbeitungen},\ }\href@noop {} {\bibfield  {journal} {\bibinfo  {journal} {Ann. Phys.}\ }\textbf {\bibinfo {volume} {21}},\ \bibinfo {pages} {756} (\bibinfo {year} {1906})}\BibitemShut {NoStop}%
\bibitem [{\citenamefont {Darwin}(1898)}]{darwin1898formation}%
  \BibitemOpen
  \bibfield  {author} {\bibinfo {author} {\bibfnamefont {C.}~\bibnamefont {Darwin}},\ }\href@noop {} {\emph {\bibinfo {title} {The formation of vegetable mould through the action of worms}}},\ Vol.~\bibinfo {volume} {16}\ (\bibinfo  {publisher} {D. Appleton},\ \bibinfo {year} {1898})\BibitemShut {NoStop}%
\bibitem [{\citenamefont {Seleb}\ \emph {et~al.}(2025)\citenamefont {Seleb}, \citenamefont {Chatterjee},\ and\ \citenamefont {Bhamla}}]{seleb2025moving}%
  \BibitemOpen
  \bibfield  {author} {\bibinfo {author} {\bibfnamefont {B.}~\bibnamefont {Seleb}}, \bibinfo {author} {\bibfnamefont {A.}~\bibnamefont {Chatterjee}},\ and\ \bibinfo {author} {\bibfnamefont {S.}~\bibnamefont {Bhamla}},\ }\bibfield  {title} {\bibinfo {title} {Moving mountains: grazing agents drive terracette formation on steep hillslopes},\ }\href@noop {} {\bibfield  {journal} {\bibinfo  {journal} {arXiv preprint arXiv:2504.17496}\ } (\bibinfo {year} {2025})}\BibitemShut {NoStop}%
\bibitem [{\citenamefont {Menaut}\ and\ \citenamefont {Walker}(2001)}]{menaut2001banded}%
  \BibitemOpen
  \bibfield  {author} {\bibinfo {author} {\bibfnamefont {J.-C.}\ \bibnamefont {Menaut}}\ and\ \bibinfo {author} {\bibfnamefont {B.}~\bibnamefont {Walker}},\ }\href@noop {} {\emph {\bibinfo {title} {Banded vegetation patterning in arid and semiarid environments: ecological processes and consequences for management}}},\ Vol.\ \bibinfo {volume} {149}\ (\bibinfo  {publisher} {Springer Science \& Business Media},\ \bibinfo {year} {2001})\BibitemShut {NoStop}%
\bibitem [{\citenamefont {Louw}\ \emph {et~al.}(2017)\citenamefont {Louw}, \citenamefont {Le~Roux}, \citenamefont {Meyer-Milne},\ and\ \citenamefont {Haussmann}}]{louw2017mammal}%
  \BibitemOpen
  \bibfield  {author} {\bibinfo {author} {\bibfnamefont {M.}~\bibnamefont {Louw}}, \bibinfo {author} {\bibfnamefont {P.}~\bibnamefont {Le~Roux}}, \bibinfo {author} {\bibfnamefont {E.}~\bibnamefont {Meyer-Milne}},\ and\ \bibinfo {author} {\bibfnamefont {N.}~\bibnamefont {Haussmann}},\ }\bibfield  {title} {\bibinfo {title} {Mammal burrowing in discrete landscape patches further increases soil and vegetation heterogeneity in an arid environment},\ }\href@noop {} {\bibfield  {journal} {\bibinfo  {journal} {J. Arid Environ.}\ }\textbf {\bibinfo {volume} {141}},\ \bibinfo {pages} {68} (\bibinfo {year} {2017})}\BibitemShut {NoStop}%
\bibitem [{\citenamefont {Pringle}\ and\ \citenamefont {Tarnita}(2017)}]{pringle2017spatial}%
  \BibitemOpen
  \bibfield  {author} {\bibinfo {author} {\bibfnamefont {R.~M.}\ \bibnamefont {Pringle}}\ and\ \bibinfo {author} {\bibfnamefont {C.~E.}\ \bibnamefont {Tarnita}},\ }\bibfield  {title} {\bibinfo {title} {Spatial self-organization of ecosystems: integrating multiple mechanisms of regular-pattern formation},\ }\href@noop {} {\bibfield  {journal} {\bibinfo  {journal} {Annu. Rev. Entomol.}\ }\textbf {\bibinfo {volume} {62}},\ \bibinfo {pages} {359} (\bibinfo {year} {2017})}\BibitemShut {NoStop}%
\bibitem [{\citenamefont {Cummins}\ and\ \citenamefont {Klug}(1979)}]{cummins1979feeding}%
  \BibitemOpen
  \bibfield  {author} {\bibinfo {author} {\bibfnamefont {K.~W.}\ \bibnamefont {Cummins}}\ and\ \bibinfo {author} {\bibfnamefont {M.~J.}\ \bibnamefont {Klug}},\ }\bibfield  {title} {\bibinfo {title} {Feeding ecology of stream invertebrates},\ }\href@noop {} {\bibfield  {journal} {\bibinfo  {journal} {Annu. Rev. Ecol. Evol. Syst.}\ }\textbf {\bibinfo {volume} {10}},\ \bibinfo {pages} {147} (\bibinfo {year} {1979})}\BibitemShut {NoStop}%
\bibitem [{\citenamefont {Kudrolli}\ and\ \citenamefont {Ramirez}(2019)}]{kudrolli2019burrowing}%
  \BibitemOpen
  \bibfield  {author} {\bibinfo {author} {\bibfnamefont {A.}~\bibnamefont {Kudrolli}}\ and\ \bibinfo {author} {\bibfnamefont {B.}~\bibnamefont {Ramirez}},\ }\bibfield  {title} {\bibinfo {title} {Burrowing dynamics of aquatic worms in soft sediments},\ }\href@noop {} {\bibfield  {journal} {\bibinfo  {journal} {Proc. Natl. Acad. Sci. U. S. A.}\ }\textbf {\bibinfo {volume} {116}},\ \bibinfo {pages} {25569} (\bibinfo {year} {2019})}\BibitemShut {NoStop}%
\bibitem [{\citenamefont {Grill}\ and\ \citenamefont {Dorgan}(2015)}]{grill2015burrowing}%
  \BibitemOpen
  \bibfield  {author} {\bibinfo {author} {\bibfnamefont {S.}~\bibnamefont {Grill}}\ and\ \bibinfo {author} {\bibfnamefont {K.~M.}\ \bibnamefont {Dorgan}},\ }\bibfield  {title} {\bibinfo {title} {Burrowing by small polychaetes--mechanics, behavior and muscle structure of capitella sp.},\ }\href@noop {} {\bibfield  {journal} {\bibinfo  {journal} {J. Exp. Biol.}\ }\textbf {\bibinfo {volume} {218}},\ \bibinfo {pages} {1527} (\bibinfo {year} {2015})}\BibitemShut {NoStop}%
\bibitem [{\citenamefont {Marvi}\ \emph {et~al.}(2014)\citenamefont {Marvi}, \citenamefont {Gong}, \citenamefont {Gravish}, \citenamefont {Astley}, \citenamefont {Travers}, \citenamefont {Hatton}, \citenamefont {Mendelson~III}, \citenamefont {Choset}, \citenamefont {Hu},\ and\ \citenamefont {Goldman}}]{marvi2014sidewinding}%
  \BibitemOpen
  \bibfield  {author} {\bibinfo {author} {\bibfnamefont {H.}~\bibnamefont {Marvi}}, \bibinfo {author} {\bibfnamefont {C.}~\bibnamefont {Gong}}, \bibinfo {author} {\bibfnamefont {N.}~\bibnamefont {Gravish}}, \bibinfo {author} {\bibfnamefont {H.}~\bibnamefont {Astley}}, \bibinfo {author} {\bibfnamefont {M.}~\bibnamefont {Travers}}, \bibinfo {author} {\bibfnamefont {R.~L.}\ \bibnamefont {Hatton}}, \bibinfo {author} {\bibfnamefont {J.~R.}\ \bibnamefont {Mendelson~III}}, \bibinfo {author} {\bibfnamefont {H.}~\bibnamefont {Choset}}, \bibinfo {author} {\bibfnamefont {D.~L.}\ \bibnamefont {Hu}},\ and\ \bibinfo {author} {\bibfnamefont {D.~I.}\ \bibnamefont {Goldman}},\ }\bibfield  {title} {\bibinfo {title} {Sidewinding with minimal slip: Snake and robot ascent of sandy slopes},\ }\href@noop {} {\bibfield  {journal} {\bibinfo  {journal} {Science}\ }\textbf {\bibinfo {volume} {346}},\ \bibinfo {pages} {224} (\bibinfo {year} {2014})}\BibitemShut {NoStop}%
\bibitem [{\citenamefont {Juarez}\ \emph {et~al.}(2010)\citenamefont {Juarez}, \citenamefont {Lu}, \citenamefont {Sznitman},\ and\ \citenamefont {Arratia}}]{juarez-celegance-10}%
  \BibitemOpen
  \bibfield  {author} {\bibinfo {author} {\bibfnamefont {G.}~\bibnamefont {Juarez}}, \bibinfo {author} {\bibfnamefont {K.}~\bibnamefont {Lu}}, \bibinfo {author} {\bibfnamefont {J.}~\bibnamefont {Sznitman}},\ and\ \bibinfo {author} {\bibfnamefont {P.~E.}\ \bibnamefont {Arratia}},\ }\bibfield  {title} {\bibinfo {title} {Motility of small nematodes in wet granular media},\ }\href@noop {} {\bibfield  {journal} {\bibinfo  {journal} {Euro. Phys. Lett.}\ }\textbf {\bibinfo {volume} {92}},\ \bibinfo {pages} {44002} (\bibinfo {year} {2010})}\BibitemShut {NoStop}%
\bibitem [{\citenamefont {Pedersen}\ \emph {et~al.}(2024)\citenamefont {Pedersen}, \citenamefont {Mukherjee}, \citenamefont {Doostmohammadi}, \citenamefont {Mondal},\ and\ \citenamefont {Thijssen}}]{amin-knead-prl-24}%
  \BibitemOpen
  \bibfield  {author} {\bibinfo {author} {\bibfnamefont {M.~C.}\ \bibnamefont {Pedersen}}, \bibinfo {author} {\bibfnamefont {S.}~\bibnamefont {Mukherjee}}, \bibinfo {author} {\bibfnamefont {A.}~\bibnamefont {Doostmohammadi}}, \bibinfo {author} {\bibfnamefont {C.}~\bibnamefont {Mondal}},\ and\ \bibinfo {author} {\bibfnamefont {K.}~\bibnamefont {Thijssen}},\ }\bibfield  {title} {\bibinfo {title} {Active particles knead three-dimensional gels into porous structures},\ }\href@noop {} {\bibfield  {journal} {\bibinfo  {journal} {Phys. Rev. Lett.}\ }\textbf {\bibinfo {volume} {133}},\ \bibinfo {pages} {228301} (\bibinfo {year} {2024})}\BibitemShut {NoStop}%
\bibitem [{\citenamefont {Grober}\ \emph {et~al.}(2023)\citenamefont {Grober}, \citenamefont {Palaia}, \citenamefont {U{\c{c}}ar}, \citenamefont {Hannezo}, \citenamefont {{\v{S}}ari{\'c}},\ and\ \citenamefont {Palacci}}]{aggregation-bacteria-23}%
  \BibitemOpen
  \bibfield  {author} {\bibinfo {author} {\bibfnamefont {D.}~\bibnamefont {Grober}}, \bibinfo {author} {\bibfnamefont {I.}~\bibnamefont {Palaia}}, \bibinfo {author} {\bibfnamefont {M.~C.}\ \bibnamefont {U{\c{c}}ar}}, \bibinfo {author} {\bibfnamefont {E.}~\bibnamefont {Hannezo}}, \bibinfo {author} {\bibfnamefont {A.}~\bibnamefont {{\v{S}}ari{\'c}}},\ and\ \bibinfo {author} {\bibfnamefont {J.}~\bibnamefont {Palacci}},\ }\bibfield  {title} {\bibinfo {title} {Unconventional colloidal aggregation in chiral bacterial baths},\ }\href@noop {} {\bibfield  {journal} {\bibinfo  {journal} {Nat. Phys.}\ }\textbf {\bibinfo {volume} {19}},\ \bibinfo {pages} {1680} (\bibinfo {year} {2023})}\BibitemShut {NoStop}%
\bibitem [{\citenamefont {Peterson}\ \emph {et~al.}(2021)\citenamefont {Peterson}, \citenamefont {Baskaran},\ and\ \citenamefont {Hagan}}]{peterson2021vesicle}%
  \BibitemOpen
  \bibfield  {author} {\bibinfo {author} {\bibfnamefont {M.~S.}\ \bibnamefont {Peterson}}, \bibinfo {author} {\bibfnamefont {A.}~\bibnamefont {Baskaran}},\ and\ \bibinfo {author} {\bibfnamefont {M.~F.}\ \bibnamefont {Hagan}},\ }\bibfield  {title} {\bibinfo {title} {Vesicle shape transformations driven by confined active filaments},\ }\href@noop {} {\bibfield  {journal} {\bibinfo  {journal} {Nat. commun.}\ }\textbf {\bibinfo {volume} {12}},\ \bibinfo {pages} {7247} (\bibinfo {year} {2021})}\BibitemShut {NoStop}%
\bibitem [{\citenamefont {Prathyusha}\ \emph {et~al.}(2022)\citenamefont {Prathyusha}, \citenamefont {Ziebert},\ and\ \citenamefont {Golestanian}}]{prathyusha2022transverse}%
  \BibitemOpen
  \bibfield  {author} {\bibinfo {author} {\bibfnamefont {K.~R.}\ \bibnamefont {Prathyusha}}, \bibinfo {author} {\bibfnamefont {F.}~\bibnamefont {Ziebert}},\ and\ \bibinfo {author} {\bibfnamefont {R.}~\bibnamefont {Golestanian}},\ }\bibfield  {title} {\bibinfo {title} {Emergent conformational properties of end-tailored transversely propelling polymers},\ }\href@noop {} {\bibfield  {journal} {\bibinfo  {journal} {Soft Matter}\ }\textbf {\bibinfo {volume} {18}},\ \bibinfo {pages} {2928} (\bibinfo {year} {2022})}\BibitemShut {NoStop}%
\bibitem [{\citenamefont {Theeyancheri}\ \emph {et~al.}(2022)\citenamefont {Theeyancheri}, \citenamefont {Chaki}, \citenamefont {Bhattacharjee},\ and\ \citenamefont {Chakrabarti}}]{ligesh-rajarshi-2022}%
  \BibitemOpen
  \bibfield  {author} {\bibinfo {author} {\bibfnamefont {L.}~\bibnamefont {Theeyancheri}}, \bibinfo {author} {\bibfnamefont {S.}~\bibnamefont {Chaki}}, \bibinfo {author} {\bibfnamefont {T.}~\bibnamefont {Bhattacharjee}},\ and\ \bibinfo {author} {\bibfnamefont {R.}~\bibnamefont {Chakrabarti}},\ }\bibfield  {title} {\bibinfo {title} {Migration of active rings in porous media},\ }\href@noop {} {\bibfield  {journal} {\bibinfo  {journal} {Phys. Rev. E}\ }\textbf {\bibinfo {volume} {106}},\ \bibinfo {pages} {014504} (\bibinfo {year} {2022})}\BibitemShut {NoStop}%
\bibitem [{\citenamefont {Wen}\ \emph {et~al.}(2022)\citenamefont {Wen}, \citenamefont {Zhu}, \citenamefont {Peng}, \citenamefont {Kumar},\ and\ \citenamefont {Laradji}}]{wen2022collective}%
  \BibitemOpen
  \bibfield  {author} {\bibinfo {author} {\bibfnamefont {H.}~\bibnamefont {Wen}}, \bibinfo {author} {\bibfnamefont {Y.}~\bibnamefont {Zhu}}, \bibinfo {author} {\bibfnamefont {C.}~\bibnamefont {Peng}}, \bibinfo {author} {\bibfnamefont {P.~S.}\ \bibnamefont {Kumar}},\ and\ \bibinfo {author} {\bibfnamefont {M.}~\bibnamefont {Laradji}},\ }\bibfield  {title} {\bibinfo {title} {Collective motion of cells modeled as ring polymers},\ }\href@noop {} {\bibfield  {journal} {\bibinfo  {journal} {Soft Matter}\ }\textbf {\bibinfo {volume} {18}},\ \bibinfo {pages} {1228} (\bibinfo {year} {2022})}\BibitemShut {NoStop}%
\bibitem [{\citenamefont {Mokhtari}\ and\ \citenamefont {Zippelius}(2019)}]{mokhtari2019dynamics}%
  \BibitemOpen
  \bibfield  {author} {\bibinfo {author} {\bibfnamefont {Z.}~\bibnamefont {Mokhtari}}\ and\ \bibinfo {author} {\bibfnamefont {A.}~\bibnamefont {Zippelius}},\ }\bibfield  {title} {\bibinfo {title} {Dynamics of active filaments in porous media},\ }\href@noop {} {\bibfield  {journal} {\bibinfo  {journal} {Phys. Rev. Lett.}\ }\textbf {\bibinfo {volume} {123}},\ \bibinfo {pages} {028001} (\bibinfo {year} {2019})}\BibitemShut {NoStop}%
\bibitem [{\citenamefont {Kurzthaler}\ \emph {et~al.}(2021)\citenamefont {Kurzthaler}, \citenamefont {Mandal}, \citenamefont {Bhattacharjee}, \citenamefont {L{\"o}wen}, \citenamefont {Datta},\ and\ \citenamefont {Stone}}]{kurzthaler2021geometric}%
  \BibitemOpen
  \bibfield  {author} {\bibinfo {author} {\bibfnamefont {C.}~\bibnamefont {Kurzthaler}}, \bibinfo {author} {\bibfnamefont {S.}~\bibnamefont {Mandal}}, \bibinfo {author} {\bibfnamefont {T.}~\bibnamefont {Bhattacharjee}}, \bibinfo {author} {\bibfnamefont {H.}~\bibnamefont {L{\"o}wen}}, \bibinfo {author} {\bibfnamefont {S.~S.}\ \bibnamefont {Datta}},\ and\ \bibinfo {author} {\bibfnamefont {H.~A.}\ \bibnamefont {Stone}},\ }\bibfield  {title} {\bibinfo {title} {A geometric criterion for the optimal spreading of active polymers in porous media},\ }\href@noop {} {\bibfield  {journal} {\bibinfo  {journal} {Nat. Commun.}\ }\textbf {\bibinfo {volume} {12}},\ \bibinfo {pages} {7088} (\bibinfo {year} {2021})}\BibitemShut {NoStop}%
\bibitem [{\citenamefont {Theeyancheri}\ \emph {et~al.}(2023)\citenamefont {Theeyancheri}, \citenamefont {Chaki}, \citenamefont {Bhattacharjee},\ and\ \citenamefont {Chakrabarti}}]{ligesh-rajarshi2023active}%
  \BibitemOpen
  \bibfield  {author} {\bibinfo {author} {\bibfnamefont {L.}~\bibnamefont {Theeyancheri}}, \bibinfo {author} {\bibfnamefont {S.}~\bibnamefont {Chaki}}, \bibinfo {author} {\bibfnamefont {T.}~\bibnamefont {Bhattacharjee}},\ and\ \bibinfo {author} {\bibfnamefont {R.}~\bibnamefont {Chakrabarti}},\ }\bibfield  {title} {\bibinfo {title} {Active dynamics of linear chains and rings in porous media},\ }\href@noop {} {\bibfield  {journal} {\bibinfo  {journal} {J. Chem. Phys.}\ }\textbf {\bibinfo {volume} {159}} (\bibinfo {year} {2023})}\BibitemShut {NoStop}%
\bibitem [{\citenamefont {Boudet}\ \emph {et~al.}(2021)\citenamefont {Boudet}, \citenamefont {Lintuvuori}, \citenamefont {Lacouture}, \citenamefont {Barois}, \citenamefont {Deblais}, \citenamefont {Xie}, \citenamefont {Cassagnere}, \citenamefont {Tregon}, \citenamefont {Br{\"u}ckner}, \citenamefont {Baret} \emph {et~al.}}]{boudet2021}%
  \BibitemOpen
  \bibfield  {author} {\bibinfo {author} {\bibfnamefont {J.-F.}\ \bibnamefont {Boudet}}, \bibinfo {author} {\bibfnamefont {J.}~\bibnamefont {Lintuvuori}}, \bibinfo {author} {\bibfnamefont {C.}~\bibnamefont {Lacouture}}, \bibinfo {author} {\bibfnamefont {T.}~\bibnamefont {Barois}}, \bibinfo {author} {\bibfnamefont {A.}~\bibnamefont {Deblais}}, \bibinfo {author} {\bibfnamefont {K.}~\bibnamefont {Xie}}, \bibinfo {author} {\bibfnamefont {S.}~\bibnamefont {Cassagnere}}, \bibinfo {author} {\bibfnamefont {B.}~\bibnamefont {Tregon}}, \bibinfo {author} {\bibfnamefont {D.~B.}\ \bibnamefont {Br{\"u}ckner}}, \bibinfo {author} {\bibfnamefont {J.-C.}\ \bibnamefont {Baret}}, \emph {et~al.},\ }\bibfield  {title} {\bibinfo {title} {From collections of independent, mindless robots to flexible, mobile, and directional superstructures},\ }\href@noop {} {\bibfield  {journal} {\bibinfo  {journal} {Science Robotics}\ }\textbf {\bibinfo {volume} {6}},\ \bibinfo {pages} {eabd0272} (\bibinfo {year} {2021})}\BibitemShut {NoStop}%
\bibitem [{\citenamefont {Veenstra}\ \emph {et~al.}(2025)\citenamefont {Veenstra}, \citenamefont {Scheibner}, \citenamefont {Brandenbourger}, \citenamefont {Binysh}, \citenamefont {Souslov}, \citenamefont {Vitelli},\ and\ \citenamefont {Coulais}}]{veenstra2025}%
  \BibitemOpen
  \bibfield  {author} {\bibinfo {author} {\bibfnamefont {J.}~\bibnamefont {Veenstra}}, \bibinfo {author} {\bibfnamefont {C.}~\bibnamefont {Scheibner}}, \bibinfo {author} {\bibfnamefont {M.}~\bibnamefont {Brandenbourger}}, \bibinfo {author} {\bibfnamefont {J.}~\bibnamefont {Binysh}}, \bibinfo {author} {\bibfnamefont {A.}~\bibnamefont {Souslov}}, \bibinfo {author} {\bibfnamefont {V.}~\bibnamefont {Vitelli}},\ and\ \bibinfo {author} {\bibfnamefont {C.}~\bibnamefont {Coulais}},\ }\bibfield  {title} {\bibinfo {title} {Adaptive locomotion of active solids},\ }\href@noop {} {\bibfield  {journal} {\bibinfo  {journal} {Nature}\ ,\ \bibinfo {pages} {1}} (\bibinfo {year} {2025})}\BibitemShut {NoStop}%
\bibitem [{\citenamefont {Tuazon}\ \emph {et~al.}(2022)\citenamefont {Tuazon}, \citenamefont {Kaufman}, \citenamefont {Goldman},\ and\ \citenamefont {Bhamla}}]{tuazon2022oxygenation}%
  \BibitemOpen
  \bibfield  {author} {\bibinfo {author} {\bibfnamefont {H.}~\bibnamefont {Tuazon}}, \bibinfo {author} {\bibfnamefont {E.}~\bibnamefont {Kaufman}}, \bibinfo {author} {\bibfnamefont {D.~I.}\ \bibnamefont {Goldman}},\ and\ \bibinfo {author} {\bibfnamefont {M.}~\bibnamefont {Bhamla}},\ }\bibfield  {title} {\bibinfo {title} {Oxygenation-controlled collective dynamics in aquatic worm blobs},\ }\href@noop {} {\bibfield  {journal} {\bibinfo  {journal} {Integr. Comp. Biol.}\ }\textbf {\bibinfo {volume} {62}},\ \bibinfo {pages} {890} (\bibinfo {year} {2022})}\BibitemShut {NoStop}%
\bibitem [{\citenamefont {Tuazon}\ \emph {et~al.}(2024)\citenamefont {Tuazon}, \citenamefont {David}, \citenamefont {Ma},\ and\ \citenamefont {Bhamla}}]{2024Tuazon}%
  \BibitemOpen
  \bibfield  {author} {\bibinfo {author} {\bibfnamefont {H.}~\bibnamefont {Tuazon}}, \bibinfo {author} {\bibfnamefont {S.}~\bibnamefont {David}}, \bibinfo {author} {\bibfnamefont {K.}~\bibnamefont {Ma}},\ and\ \bibinfo {author} {\bibfnamefont {S.}~\bibnamefont {Bhamla}},\ }\bibfield  {title} {\bibinfo {title} {Leeches {Predate} on {Fast}-{Escaping} and {Entangling} {Blackworms} by {Spiral} {Entombment}},\ }\href {https://doi.org/10.1093/icb/icae118} {\bibfield  {journal} {\bibinfo  {journal} {Integr. Comp. Biol.}\ }\textbf {\bibinfo {volume} {64}},\ \bibinfo {pages} {1408} (\bibinfo {year} {2024})}\BibitemShut {NoStop}%
\bibitem [{\citenamefont {Weeks}\ \emph {et~al.}(1971)\citenamefont {Weeks}, \citenamefont {Chandler},\ and\ \citenamefont {Andersen}}]{weeks-jcp-1971}%
  \BibitemOpen
  \bibfield  {author} {\bibinfo {author} {\bibfnamefont {J.~D.}\ \bibnamefont {Weeks}}, \bibinfo {author} {\bibfnamefont {D.}~\bibnamefont {Chandler}},\ and\ \bibinfo {author} {\bibfnamefont {H.~C.}\ \bibnamefont {Andersen}},\ }\bibfield  {title} {\bibinfo {title} {Role of repulsive forces in determining the equilibrium structure of simple liquids},\ }\href@noop {} {\bibfield  {journal} {\bibinfo  {journal} {J. Chem. Phys.}\ }\textbf {\bibinfo {volume} {54}},\ \bibinfo {pages} {5237} (\bibinfo {year} {1971})}\BibitemShut {NoStop}%
\bibitem [{\citenamefont {Kremer}\ and\ \citenamefont {Grest}(1990)}]{kkremer-90}%
  \BibitemOpen
  \bibfield  {author} {\bibinfo {author} {\bibfnamefont {K.}~\bibnamefont {Kremer}}\ and\ \bibinfo {author} {\bibfnamefont {G.~S.}\ \bibnamefont {Grest}},\ }\bibfield  {title} {\bibinfo {title} {Dynamics of entangled linear polymer melts: A molecular‐dynamics simulation},\ }\href {https://doi.org/10.1063/1.458541} {\bibfield  {journal} {\bibinfo  {journal} {J. Chem. Phys.}\ }\textbf {\bibinfo {volume} {92}},\ \bibinfo {pages} {5057} (\bibinfo {year} {1990})}\BibitemShut {NoStop}%
\end{thebibliography}%


\begin{thebibliography}{11}%
\makeatletter
\providecommand \@ifxundefined [1]{%
 \@ifx{#1\undefined}
}%
\providecommand \@ifnum [1]{%
 \ifnum #1\expandafter \@firstoftwo
 \else \expandafter \@secondoftwo
 \fi
}%
\providecommand \@ifx [1]{%
 \ifx #1\expandafter \@firstoftwo
 \else \expandafter \@secondoftwo
 \fi
}%
\providecommand \natexlab [1]{#1}%
\providecommand \enquote  [1]{``#1''}%
\providecommand \bibnamefont  [1]{#1}%
\providecommand \bibfnamefont [1]{#1}%
\providecommand \citenamefont [1]{#1}%
\providecommand \href@noop [0]{\@secondoftwo}%
\providecommand \href [0]{\begingroup \@sanitize@url \@href}%
\providecommand \@href[1]{\@@startlink{#1}\@@href}%
\providecommand \@@href[1]{\endgroup#1\@@endlink}%
\providecommand \@sanitize@url [0]{\catcode `\\12\catcode `\$12\catcode `\&12\catcode `\#12\catcode `\^12\catcode `\_12\catcode `\%12\relax}%
\providecommand \@@startlink[1]{}%
\providecommand \@@endlink[0]{}%
\providecommand \url  [0]{\begingroup\@sanitize@url \@url }%
\providecommand \@url [1]{\endgroup\@href {#1}{\urlprefix }}%
\providecommand \urlprefix  [0]{URL }%
\providecommand \Eprint [0]{\href }%
\providecommand \doibase [0]{https://doi.org/}%
\providecommand \selectlanguage [0]{\@gobble}%
\providecommand \bibinfo  [0]{\@secondoftwo}%
\providecommand \bibfield  [0]{\@secondoftwo}%
\providecommand \translation [1]{[#1]}%
\providecommand \BibitemOpen [0]{}%
\providecommand \bibitemStop [0]{}%
\providecommand \bibitemNoStop [0]{.\EOS\space}%
\providecommand \EOS [0]{\spacefactor3000\relax}%
\providecommand \BibitemShut  [1]{\csname bibitem#1\endcsname}%
\let\auto@bib@innerbib\@empty
%</preamble>
\bibitem [{\citenamefont {Deblais}\ \emph {et~al.}(2020{\natexlab{a}})\citenamefont {Deblais}, \citenamefont {Woutersen},\ and\ \citenamefont {Bonn}}]{Deblais2020a}%
  \BibitemOpen
  \bibfield  {author} {\bibinfo {author} {\bibfnamefont {A.}~\bibnamefont {Deblais}}, \bibinfo {author} {\bibfnamefont {S.}~\bibnamefont {Woutersen}},\ and\ \bibinfo {author} {\bibfnamefont {D.}~\bibnamefont {Bonn}},\ }\bibfield  {title} {\bibinfo {title} {Rheology of entangled active polymer-like t. tubifex worms},\ }\href@noop {} {\bibfield  {journal} {\bibinfo  {journal} {Phys. Rev. Lett.}\ }\textbf {\bibinfo {volume} {124}},\ \bibinfo {pages} {188002} (\bibinfo {year} {2020}{\natexlab{a}})}\BibitemShut {NoStop}%
\bibitem [{\citenamefont {Deblais}\ \emph {et~al.}(2020{\natexlab{b}})\citenamefont {Deblais}, \citenamefont {Maggs}, \citenamefont {Bonn},\ and\ \citenamefont {Woutersen}}]{Deblais2020b}%
  \BibitemOpen
  \bibfield  {author} {\bibinfo {author} {\bibfnamefont {A.}~\bibnamefont {Deblais}}, \bibinfo {author} {\bibfnamefont {A.}~\bibnamefont {Maggs}}, \bibinfo {author} {\bibfnamefont {D.}~\bibnamefont {Bonn}},\ and\ \bibinfo {author} {\bibfnamefont {S.}~\bibnamefont {Woutersen}},\ }\bibfield  {title} {\bibinfo {title} {Phase separation by entanglement of active polymerlike worms},\ }\href@noop {} {\bibfield  {journal} {\bibinfo  {journal} {Phys. Rev. Lett.}\ }\textbf {\bibinfo {volume} {124}},\ \bibinfo {pages} {208006} (\bibinfo {year} {2020}{\natexlab{b}})}\BibitemShut {NoStop}%
\bibitem [{\citenamefont {Heeremans}\ \emph {et~al.}(2022)\citenamefont {Heeremans}, \citenamefont {Deblais}, \citenamefont {Bonn},\ and\ \citenamefont {Woutersen}}]{Heeremans2022}%
  \BibitemOpen
  \bibfield  {author} {\bibinfo {author} {\bibfnamefont {T.}~\bibnamefont {Heeremans}}, \bibinfo {author} {\bibfnamefont {A.}~\bibnamefont {Deblais}}, \bibinfo {author} {\bibfnamefont {D.}~\bibnamefont {Bonn}},\ and\ \bibinfo {author} {\bibfnamefont {S.}~\bibnamefont {Woutersen}},\ }\bibfield  {title} {\bibinfo {title} {Chromatographic separation of active polymer like worm mixtures by contour length and activity},\ }\href {https://doi.org/10.1126/sciadv.abj7918} {\bibfield  {journal} {\bibinfo  {journal} {Sci. Adv.}\ }\textbf {\bibinfo {volume} {8}},\ \bibinfo {pages} {eabj7918} (\bibinfo {year} {2022})}\BibitemShut {NoStop}%
\bibitem [{\citenamefont {Deblais}\ \emph {et~al.}(2023)\citenamefont {Deblais}, \citenamefont {Prathyusha}, \citenamefont {Sinaasappel}, \citenamefont {Tuazon}, \citenamefont {Tiwari}, \citenamefont {Patil},\ and\ \citenamefont {Bhamla}}]{ReviewWorm2023}%
  \BibitemOpen
  \bibfield  {author} {\bibinfo {author} {\bibfnamefont {A.}~\bibnamefont {Deblais}}, \bibinfo {author} {\bibfnamefont {K.~R.}\ \bibnamefont {Prathyusha}}, \bibinfo {author} {\bibfnamefont {R.}~\bibnamefont {Sinaasappel}}, \bibinfo {author} {\bibfnamefont {H.}~\bibnamefont {Tuazon}}, \bibinfo {author} {\bibfnamefont {I.}~\bibnamefont {Tiwari}}, \bibinfo {author} {\bibfnamefont {V.~P.}\ \bibnamefont {Patil}},\ and\ \bibinfo {author} {\bibfnamefont {M.~S.}\ \bibnamefont {Bhamla}},\ }\bibfield  {title} {\bibinfo {title} {Worm blobs as entangled living polymers: from topological active matter to flexible soft robot collectives},\ }\href {https://doi.org/10.1039/D3SM00542A} {\bibfield  {journal} {\bibinfo  {journal} {Soft Matter}\ }\textbf {\bibinfo {volume} {19}},\ \bibinfo {pages} {7057} (\bibinfo {year} {2023})}\BibitemShut {NoStop}%
\bibitem [{\citenamefont {Tuazon}\ \emph {et~al.}(2022)\citenamefont {Tuazon}, \citenamefont {Kaufman}, \citenamefont {Goldman},\ and\ \citenamefont {Bhamla}}]{tuazon2022oxygenation}%
  \BibitemOpen
  \bibfield  {author} {\bibinfo {author} {\bibfnamefont {H.}~\bibnamefont {Tuazon}}, \bibinfo {author} {\bibfnamefont {E.}~\bibnamefont {Kaufman}}, \bibinfo {author} {\bibfnamefont {D.~I.}\ \bibnamefont {Goldman}},\ and\ \bibinfo {author} {\bibfnamefont {M.}~\bibnamefont {Bhamla}},\ }\bibfield  {title} {\bibinfo {title} {Oxygenation-controlled collective dynamics in aquatic worm blobs},\ }\href@noop {} {\bibfield  {journal} {\bibinfo  {journal} {Integr. Comp. Biol.}\ }\textbf {\bibinfo {volume} {62}},\ \bibinfo {pages} {890} (\bibinfo {year} {2022})}\BibitemShut {NoStop}%
\bibitem [{\citenamefont {Tuazon}\ \emph {et~al.}(2023)\citenamefont {Tuazon}, \citenamefont {Nguyen}, \citenamefont {Kaufman}, \citenamefont {Tiwari}, \citenamefont {Bermudez}, \citenamefont {Chudasama}, \citenamefont {Peleg},\ and\ \citenamefont {Bhamla}}]{tuazon2023collecting}%
  \BibitemOpen
  \bibfield  {author} {\bibinfo {author} {\bibfnamefont {H.}~\bibnamefont {Tuazon}}, \bibinfo {author} {\bibfnamefont {C.}~\bibnamefont {Nguyen}}, \bibinfo {author} {\bibfnamefont {E.}~\bibnamefont {Kaufman}}, \bibinfo {author} {\bibfnamefont {I.}~\bibnamefont {Tiwari}}, \bibinfo {author} {\bibfnamefont {J.}~\bibnamefont {Bermudez}}, \bibinfo {author} {\bibfnamefont {D.}~\bibnamefont {Chudasama}}, \bibinfo {author} {\bibfnamefont {O.}~\bibnamefont {Peleg}},\ and\ \bibinfo {author} {\bibfnamefont {M.}~\bibnamefont {Bhamla}},\ }\bibfield  {title} {\bibinfo {title} {Collecting--gathering biophysics of the blackworm lumbriculus variegatus},\ }\href@noop {} {\bibfield  {journal} {\bibinfo  {journal} {Integr. Comp. Biol.}\ }\textbf {\bibinfo {volume} {63}},\ \bibinfo {pages} {1474} (\bibinfo {year} {2023})}\BibitemShut {NoStop}%
\bibitem [{\citenamefont {Tuazon}\ \emph {et~al.}(2024)\citenamefont {Tuazon}, \citenamefont {David}, \citenamefont {Ma},\ and\ \citenamefont {Bhamla}}]{2024Tuazon}%
  \BibitemOpen
  \bibfield  {author} {\bibinfo {author} {\bibfnamefont {H.}~\bibnamefont {Tuazon}}, \bibinfo {author} {\bibfnamefont {S.}~\bibnamefont {David}}, \bibinfo {author} {\bibfnamefont {K.}~\bibnamefont {Ma}},\ and\ \bibinfo {author} {\bibfnamefont {S.}~\bibnamefont {Bhamla}},\ }\bibfield  {title} {\bibinfo {title} {Leeches {Predate} on {Fast}-{Escaping} and {Entangling} {Blackworms} by {Spiral} {Entombment}},\ }\href {https://doi.org/10.1093/icb/icae118} {\bibfield  {journal} {\bibinfo  {journal} {Integr. Comp. Biol.}\ }\textbf {\bibinfo {volume} {64}},\ \bibinfo {pages} {1408} (\bibinfo {year} {2024})}\BibitemShut {NoStop}%
\bibitem [{\citenamefont {Weeks}\ \emph {et~al.}(1971)\citenamefont {Weeks}, \citenamefont {Chandler},\ and\ \citenamefont {Andersen}}]{weeks-jcp-1971}%
  \BibitemOpen
  \bibfield  {author} {\bibinfo {author} {\bibfnamefont {J.~D.}\ \bibnamefont {Weeks}}, \bibinfo {author} {\bibfnamefont {D.}~\bibnamefont {Chandler}},\ and\ \bibinfo {author} {\bibfnamefont {H.~C.}\ \bibnamefont {Andersen}},\ }\bibfield  {title} {\bibinfo {title} {Role of repulsive forces in determining the equilibrium structure of simple liquids},\ }\href@noop {} {\bibfield  {journal} {\bibinfo  {journal} {J. Chem. Phys.}\ }\textbf {\bibinfo {volume} {54}},\ \bibinfo {pages} {5237} (\bibinfo {year} {1971})}\BibitemShut {NoStop}%
\bibitem [{\citenamefont {Kremer}\ and\ \citenamefont {Grest}(1990)}]{kkremer-90}%
  \BibitemOpen
  \bibfield  {author} {\bibinfo {author} {\bibfnamefont {K.}~\bibnamefont {Kremer}}\ and\ \bibinfo {author} {\bibfnamefont {G.~S.}\ \bibnamefont {Grest}},\ }\bibfield  {title} {\bibinfo {title} {Dynamics of entangled linear polymer melts: A molecular‐dynamics simulation},\ }\href {https://doi.org/10.1063/1.458541} {\bibfield  {journal} {\bibinfo  {journal} {J. Chem. Phys.}\ }\textbf {\bibinfo {volume} {92}},\ \bibinfo {pages} {5057} (\bibinfo {year} {1990})}\BibitemShut {NoStop}%
\bibitem [{\citenamefont {Zheng}\ \emph {et~al.}(2023)\citenamefont {Zheng}, \citenamefont {Brandenbourger}, \citenamefont {Robinet}, \citenamefont {Schall}, \citenamefont {Lerner},\ and\ \citenamefont {Coulais}}]{zhengrobot-anchor2023}%
  \BibitemOpen
  \bibfield  {author} {\bibinfo {author} {\bibfnamefont {E.}~\bibnamefont {Zheng}}, \bibinfo {author} {\bibfnamefont {M.}~\bibnamefont {Brandenbourger}}, \bibinfo {author} {\bibfnamefont {L.}~\bibnamefont {Robinet}}, \bibinfo {author} {\bibfnamefont {P.}~\bibnamefont {Schall}}, \bibinfo {author} {\bibfnamefont {E.}~\bibnamefont {Lerner}},\ and\ \bibinfo {author} {\bibfnamefont {C.}~\bibnamefont {Coulais}},\ }\bibfield  {title} {\bibinfo {title} {Self-oscillation and synchronization transitions in elastoactive structures},\ }\href@noop {} {\bibfield  {journal} {\bibinfo  {journal} {Phys. Rev. Lett.}\ }\textbf {\bibinfo {volume} {130}},\ \bibinfo {pages} {178202} (\bibinfo {year} {2023})}\BibitemShut {NoStop}%
\bibitem [{\citenamefont {Smoluchowski}\ and\ \citenamefont {im~unbegrenzten Raum}(1906)}]{Smoluchowski1906}%
  \BibitemOpen
  \bibfield  {author} {\bibinfo {author} {\bibfnamefont {V.}~\bibnamefont {Smoluchowski}}\ and\ \bibinfo {author} {\bibfnamefont {I.~D.}\ \bibnamefont {im~unbegrenzten Raum}},\ }\bibfield  {title} {\bibinfo {title} {Zusammenfassende bearbeitungen},\ }\href@noop {} {\bibfield  {journal} {\bibinfo  {journal} {Ann. Phys.}\ }\textbf {\bibinfo {volume} {21}},\ \bibinfo {pages} {756} (\bibinfo {year} {1906})}\BibitemShut {NoStop}%
\end{thebibliography}%
\end{document}

\title{Supplementary Materials for\\ ``Collecting Particles in Confined Spaces by Active Filamentous Matter''}
%\date{\today}

\author{R.~Sinaasappel}
\thanks{These two authors contributed equally}
\affiliation{Van der Waals-Zeeman Institute, Institute of Physics, University of Amsterdam, 1098XH Amsterdam, The Netherlands.}
\author{K.~R.~Prathyusha}
\thanks{These two authors contributed equally}
\affiliation{School of Chemical and Biomolecular Engineering, Georgia Institute of Technology, Atlanta, GA 30332, USA.}
\author{Harry Tuazon}
\affiliation{School of Chemical and Biomolecular Engineering, Georgia Institute of Technology, Atlanta, GA 30332, USA.}
\author{E.~Mirzahossein}
\affiliation{Van der Waals-Zeeman Institute, Institute of Physics, University of Amsterdam, 1098XH Amsterdam, The Netherlands.}
\author{P.~Illien}
\affiliation{Sorbonne Universit´e, CNRS, Physicochimie des Electrolytes et Nanosyst`emes Interfaciaux (PHENIX), Paris, France.}
\author{M.~Saad Bhamla}
\email{saadb@chbe.gatech.edu}
\affiliation{School of Chemical and Biomolecular Engineering, Georgia Institute of Technology, Atlanta, GA 30332, USA.}
\author{A.~Deblais}
\email{a.deblais@uva.nl}
\affiliation{Van der Waals-Zeeman Institute, Institute of Physics, University of Amsterdam, 1098XH Amsterdam, The Netherlands.}

\date{\today}

\begin{abstract}
This Supplementary Material provides additional information on the different active filamentous systems investigated, the experimental setup, and methodologies employed in this study, and on the polymer model and simulation details used for our analysis.
\end{abstract}
\maketitle
\tableofcontents
\newpage
%\newline
%\section{}

\section{Living Worm experiments}
\begin{figure}
    \centering
\includegraphics[width=0.65\linewidth]{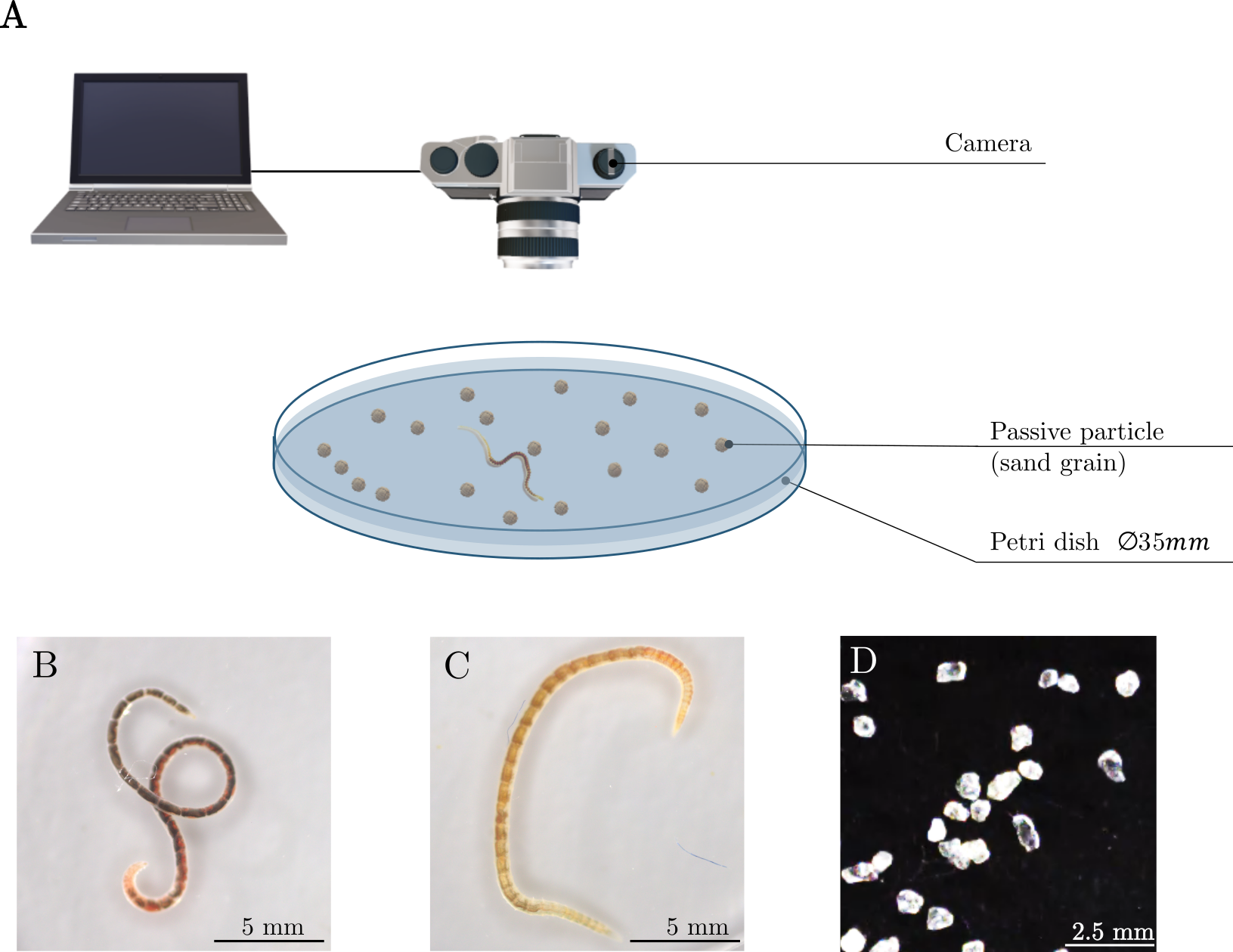}
\caption{(\textbf{A}) Experimental setup for investigating particle collection by worms.
 A 35 mm diameter Petri dish is filled with fresh water and \( 50 \) mg of sand particles with a typical size of 1 mm.   A webcam records all experiments at a frame rate of 20 FPS for four hours with fixed lighting
(\textbf{B}) A \textit{T. tubifex} worm or  
(\textbf{C}) A \textit{California blackworm} is introduced into the dish to study particle collection.  
(\textbf{D}) Snapshot of the sand particles used in the experiments.} 
    \label{fig:worm_diagram}
\end{figure}

We investigated particle collection using two different species of annelid worms, \textit{Tubifex tubifex} (\textit{T.~tubifex}) and California Blackworms  (\textit{Lumbriculus variegatus}). These species are widely available from commercial supplies and  differ in their aspect ratio, persistence length, and motility dynamics, providing a valuable basis for comparison. The schematic of the experimental setup and images of worms are presented in Fig.~\ref{fig:worm_diagram}.~(\textbf{A})-(\textbf{C}).
\subsection{\textit{Tubifex tubifex} worms}
One of the species of living worms that we studied was \textit{Tubifex tubifex}, a living biological system that has been the focus of several recent studies~\cite{Deblais2020a,Deblais2020b,Heeremans2022,ReviewWorm2023}(See Fig.~\ref{fig:worm_diagram}.~(\textbf{B})). All batches of \textit{T.~tubifex} worms analyzed in this work were purchased from the provider Aquadip (\url{https://www.aquadip.nl/}) and ordered in a prepacked configuration, where the worms were at adult size. 
Their contour length $\ell_c$ varies between 20 and 40 mm and their width is typically 0.8 mm.
The worms were maintained in an aquarium at room temperature, constantly under filtered flow, with water consisting of demineralized water mixed with salt solutions optimized for their needs. 
The salt solution consists of a mixture of: 50~g/L of \ce{NaHCO3};  10~g/L of \ce{KHCO3}; 100~g/L of \ce{CaCl2}; 90~g/L of \ce{MgSO4}.  
Worms were fed weekly with standard goldfish food, and the water was refreshed once per week or more frequently if needed.

\subsection{California blackworms}
We purchased California blackworms (\textit{Lumbriculus variegatus}) from Ward's Science and were reared similarly as described in past publications (Tuazon \textit{et al.}\cite{tuazon2022oxygenation, tuazon2023collecting, 2024Tuazon}) (See Fig.~\ref{fig:worm_diagram}.~(\textbf{C})).
\newpage
\subsection{Experimental setup}
\begin{figure}[h]
    \centering
    \includegraphics[width=\linewidth]{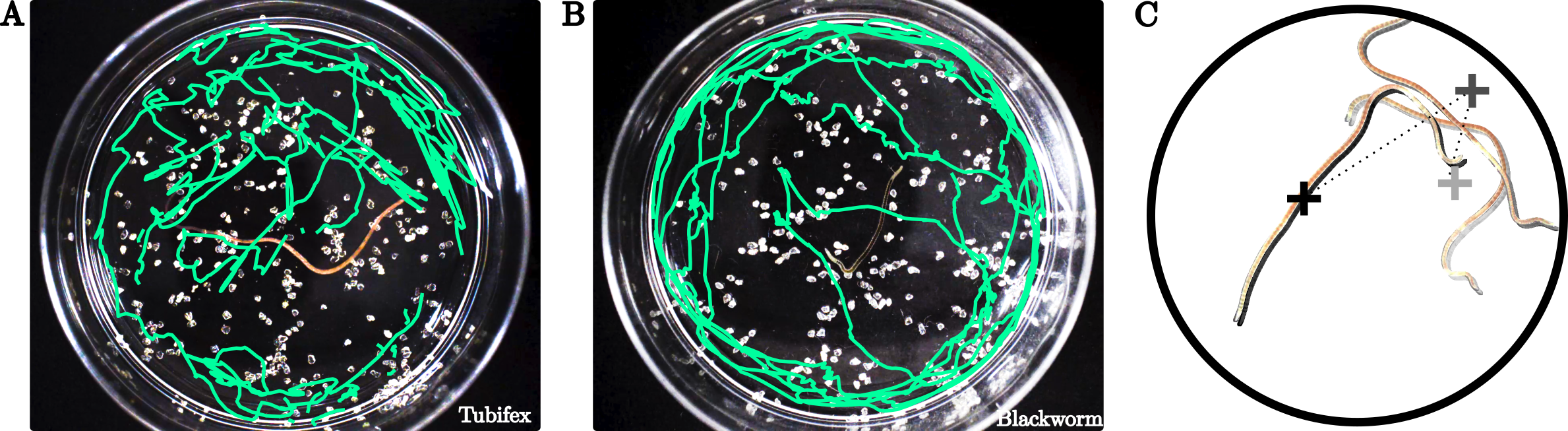}
    \caption{(\textbf{A}) Center of mass trajectory of a \textit{T. tubifex} worm over a period of 30 minutes. The worm spends most of its time near the boundary but occasionally moves inward, crossing through the central region of the Petri dish.
    (\textbf{B}) Center of mass Trajectory of a blackworm over a period of 40 minutes, showing movement along the periphery as well as crossings through the center of the Petri dish.  
    (\textbf{C}) Superimposed images of a blackworm, showing how the worms deflects from the boundary of the Petri dish. The gray crosses indicate the center of mass and get darker at consecutive timesteps.}
    \label{fig:worm_trajectory}
\end{figure}

To investigate particle collection by worms, we mounted an ImageSource DFK 33UX264 camera (Charlotte, NC) on an optical table using 80/20 parts (See the schematic in Fig.~\ref{fig:worm_diagram}.~(\textbf{A})). The experimental protocol is described similarly in Tuazon, et al.~\cite{tuazon2023collecting}. We recorded each experiment at a frame rate of 20 FPS for four hours with fixed lighting, where the first three hours were used for data analysis. 50$\pm$0.01 mg of 20\# palmetto pool filtered sand (Woodruff, SC) was used for the test materials where $\sim$~0.7 mm grains were isolated using nylon mesh (See~Fig.~\ref{fig:worm_diagram}.~(\textbf{D})). The sand grains were transferred to a 35 mm Petri Dish and submerged in 2 mm of filtered water. Before worm transfer, the sand grains were manually dispersed using a pipet (to ensure that the grains were evenly distributed). After undergoing  one-hour habituation period, worms were transferred to the Petri Dish with the sand grains. We repeated this experiment 11 times for blackworms and 3 times for \textit{T.~tubifex}. The center of mass trajectory of both worms is shown in Fig.~\ref{fig:worm_trajectory}.~(\textbf{C}).

%\newpage

\subsection{Image analyses}
In our image analysis, we pursued two main objectives: (i) measuring the size and location of the passive particles that eventually form clusters and (ii) extracting the contour of the filament. In both worm and robotic experiments, the filaments and tracers have distinct colors, allowing for standard thresholding and contour detection algorithms from the Python OpenCV library (\url{https://github.com/opencv/opencv-python}) to be largely effective. However, in certain worm experiments, achieving sufficient color contrast proved challenging. In these cases, we employed machine-learning-based segmentation using the YOLO package developed by Ultralytics (\url{https://github.com/ultralytics/ultralytics}), which enabled high-precision filament contour detection (model available upon request).

Once the filament contour was identified, the midline—required for measuring the effective persistence length—was extracted using the skeletonization algorithm implemented in the Python scikit-image library (\url{https://github.com/scikit-image/scikit-image}). Clusters of tracer particles were detected by identifying the contours of segmented tracers; adjacent tracers sharing a common contour were classified as a single cluster.
\begin{figure}[h]
    \centering
    \includegraphics[width=\linewidth]{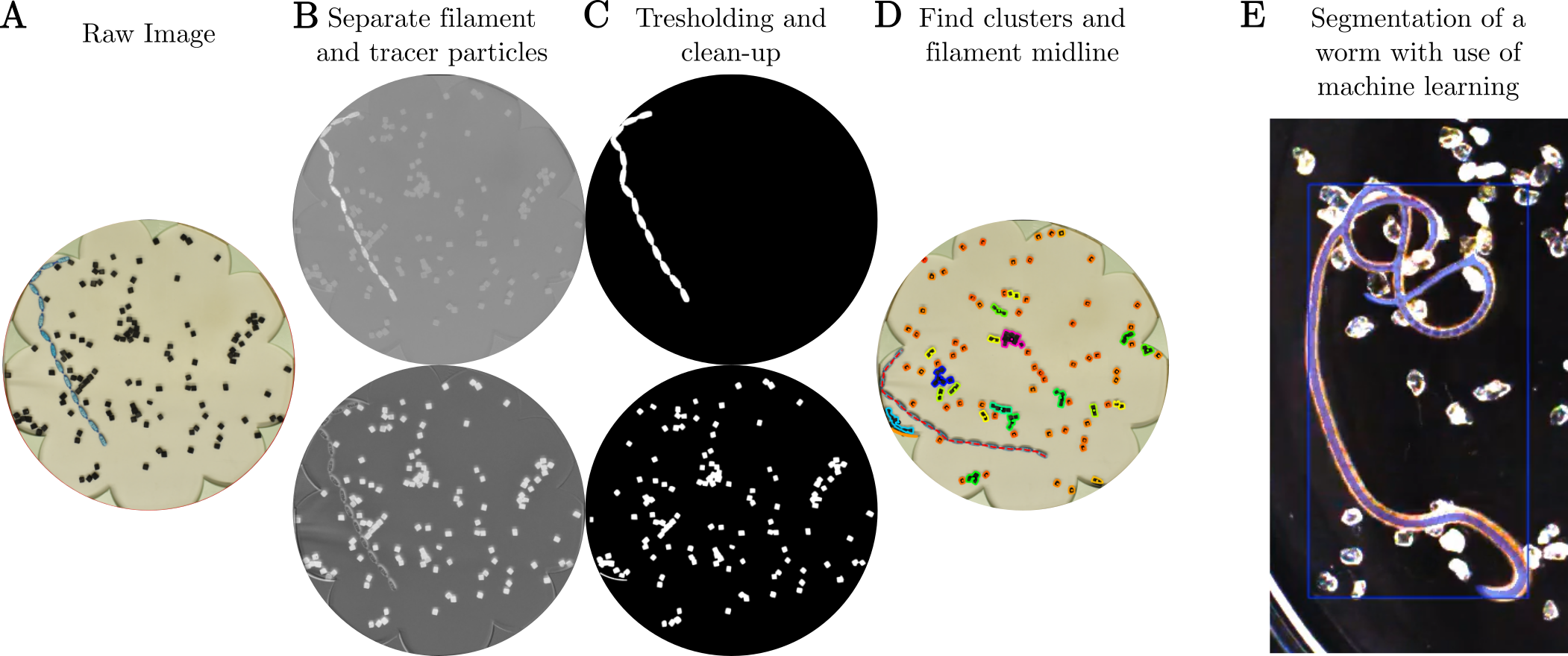}
\caption{\textbf{Image analysis protocol for particle tracking and active filament conformation and dynamics.}  
(\textbf{A}) Original experimental image with a white background, black tracer particles, and a blue robotic chain.  
(\textbf{B}) By selecting the appropriate color channels, the robotic chain is separated from the tracer particles and background.  
(\textbf{C}) After thresholding and noise filtering, a binary mask of both the filament and tracer particles is obtained.  
(\textbf{D}) Clusters are identified by detecting the contours of connected tracer particles, while the filament’s midline is extracted using a skeletonization algorithm.
(\textbf{E}) Example of the segmentation of a worm, with the use of a machine learning model.}  
    \label{fig:segmentation}
\end{figure}
\clearpage
\section{Simulation details:}

In this section, we discuss the details of the simulation used to study a system of passive particles and active polymer. We consider a $2D$ polymer made of $N$ active monomers and $N_t$ passive particles confined to a circular wall made of stationary $N_b$ particles.

\subsection{Active polymer}
The equation of motion for the particles is, 
\\
\begin{equation}
\gamma{{\dot {\bf r}}_i}=-\nabla_i{{ U}^{WCA}} +{\mathcal A}( -\nabla_i{U}^{stretch}{-\nabla_i U}^{bend}+{\bf F}_i^{noise}+ {\bf F}_i^{active})  
 \label{eq:eqofmotion}
\end{equation}

The first term corresponds to the  steric interactions between  all the particles in the systems and is modelled via Weeks-Chandler-Anderson potential  \cite{weeks-jcp-1971}, 
\begin{equation}
U^{WCA}(r)=4\varepsilon\left[\left(\frac{\sigma}{r}\right)^{12}-\left(\frac{\sigma}{r}\right)^{6}+\frac{1}{4}\right],
\label{eqn:wca}
\end{equation}
which vanishes  beyond any distance greater than  $2^{1/6}\sigma$.
Here $\varepsilon$ measures the strength of the repulsive interaction and is set to $1.0$.  $\sigma$ is the size of the particles,  we set to monomer size to $1$ and the size of passive particle to $0.5 \sigma$. $r\equiv\left|\mathbf{r}_{ij}\right|=\left|\mathbf{r}_i-\mathbf{r}_j\right|$ is the distance between two beads. The interaction prevents the  polymer from crossing different segments of its chain as well as makes passive beads repulsive. Even though wall particles are stationary, when any other particles come within  the distance ($2^{1/6}\sigma$), they experience the $U^{WCA}$ interaction and avoid any leakage of the particles through the wall. 

Now, we discuss the last term in Eqn.~\ref{eq:eqofmotion}  containing contributions from four interactions and is experienced by the polymer beads solely as $\mathcal A$ is set to 1 for the polymers and $\mathcal A = 0$ for tracer particles. Since these interactions are absent for passive particles, we set ${\mathcal A}$  to zero for such particles. Each connected pair of the polymer interacts via a  harmonic potential \cite{kkremer-90}, 
\begin{equation}
U_{stretch}(r)=k_b \big(r- R_0\big)^2 \label{eq:bond}.
\end{equation}
  Here $R_0=1.0\sigma$ is the maximum bond length, and $k_b=4000~k_BT/\sigma^2$
is the bond stiffness. 
Bending interaction   is experienced by any connected triplet in the polymer, and it is given by  the potential,
\begin{equation}
U_{bend}= \frac{\kappa }{2}\big(\theta -\theta_0 \big)^2
\end{equation}
 Here $\kappa$ is the bending stiffness and is related to the continuum bending stiffness, $\tilde \kappa$, as $\kappa \approx \tilde \kappa /2\sigma$,  $\theta$ is the angle between any consecutive bond vectors and $\theta_0$ is set to $\pi$. These parameters make the chain effectively inextensible with bond length $b\sim 1 \sigma$ and polymer length $ L =(N-1)b \sigma$.
 
The stochastic term ${\bf F}^{noise}_i$ represents the thermal fluctuations and is modelled as white noise with zero mean and variance proportional to $ \sqrt{k_BT\gamma/\Delta t }$.  

We measure length in units of particle size $\sigma$, energies in units of $k_BT$.

Finally, the last term represents self-propulsion force for the polymer, 
\begin{equation}
{\bf F}_i^{active}=\frac{f_p}{2}({\bf \hat  t}_{i-1,i}+{\bf \hat  t}_{i,i+1})
\end{equation}
Here, $f_p$ is the strength of the force and ${\bf \hat  t}_{i,i +1} = {\bf r}_{i,i +1} / {\bf r}_{i,i +1}$ is the unit tangent vector along the bond connecting beads i and i + 1. This is experienced by the beads ($i=2,3,4... N-1$) of the polymer. 
The active forces for the end beads ($i=1$ and $i=N$), as they  have only  one nearest neighbour,   are 
\begin{equation}
{\bf F}_{1}^{active}=   \ \hat{\bf t}_{1}   f_{p} \; \& \; \;
 \\
 {\bf F}_{N}^{active}=   \hat{\bf t}_{N-1} f_{p}
\label{activeforce-end}
\end{equation}

We set the friction coefficient $\gamma$ for polymers to be 1.0 and $\gamma=10.0$ for the particles.

We also create a confinement using a set of repulsive particles that are stationary. In our simulations, the circular confinement is modeled using stationary boundary particles that interact with active filament and passive particles via $U^{WCA}$ potential. These boundary particles are fixed in space and do not respond to collisions, effectively acting as a rigid wall. It captures the essential physics of the experimental setup involving worms in a Petri dish, where the walls exert forces without recoiling and the reaction forces are absorbed by the surrounding medium.

In experiments, sand particles are stationary until moved by the worm. Thus, the equation of motion of such particles, modeled as an athermal passive particle of size $0.5 \sigma$, is given by,
\begin{equation}
\gamma \frac{d\mathbf{r}_i}{dt} = 
\sum_j 
\begin{cases}
-\nabla_{\mathbf{r}_i} U^{\mathrm{WCA}}(r_{ij}),& r_{ij} < 2^{1/6} \sigma_{ij} \\
0, & r_{ij} \ge 2^{1/6} \sigma_{ij}
\end{cases}
\end{equation}

In our simulation, we used $\Delta t=0.0001 \tilde{\tau}$, where $\tilde{\tau} = \sigma^2\gamma/(k_BT)$  sets the integration time unit. We varied the  number of monomers $N=9, 18, 28, 37$, and the number of passive particles $N_t=50,100,250,500,600$.  Unless otherwise specified, the results presented in the manuscript correspond to  $N=37, N_t=600$. Magnitude of $f_p=1$

\subsection{Influence of boundary-located "Re-injectors" on Particle collection}

\begin{figure}[h]
	    \centering
\includegraphics[width=0.8\linewidth]{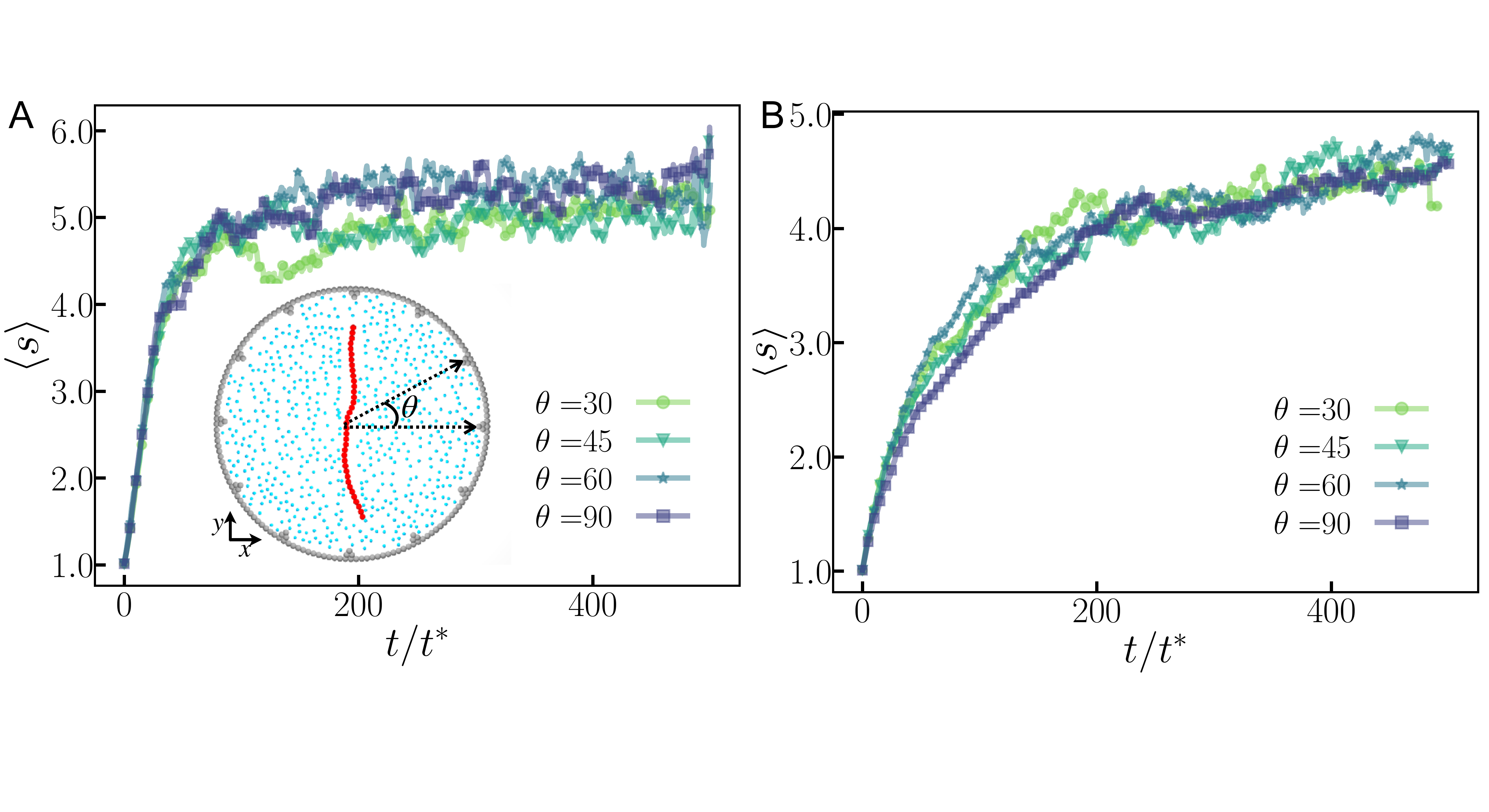}
\caption{\textbf{Effect of the number of boundary reflectors on particle collection.}  
(\textbf{A}) Particle collection dynamics for a flexible filament $\ell_p/\ell_c=0.08$ for different number re-injectors, The inset illustrates the simulation setup and indicates the angular spacing $\theta$ between re-injectors.
(\textbf{B}) Particle collection dynamics when $\ell_p/\ell_c=0.24 $   }  
	\label{fig:bump-number}
\end{figure}

In order to prevent the active filament from spending prolonged time near the boundary, we introduced  triangular re-injectors at the boundary  to deflect it back toward the center. We investigated the impact of varying the number of such re-injectors on particle collection in our simulations. Each re-injector consists of three stationary particles arranged in a triangular configuration. They were positioned at regular angular intervals, $\theta$, measured from the x-axis.
As shown in ~Fig.~\ref{fig:bump-number}, the number of re-injectors does not have a significant effect on the passive particle collections. It  suggests that while the re-injectors locally redirect the active filament, they do not substantially influence the overall collecting behavior.

\subsection{Effect of particle density}

\begin{figure}[h]
    \centering
\includegraphics[width=0.8\linewidth]{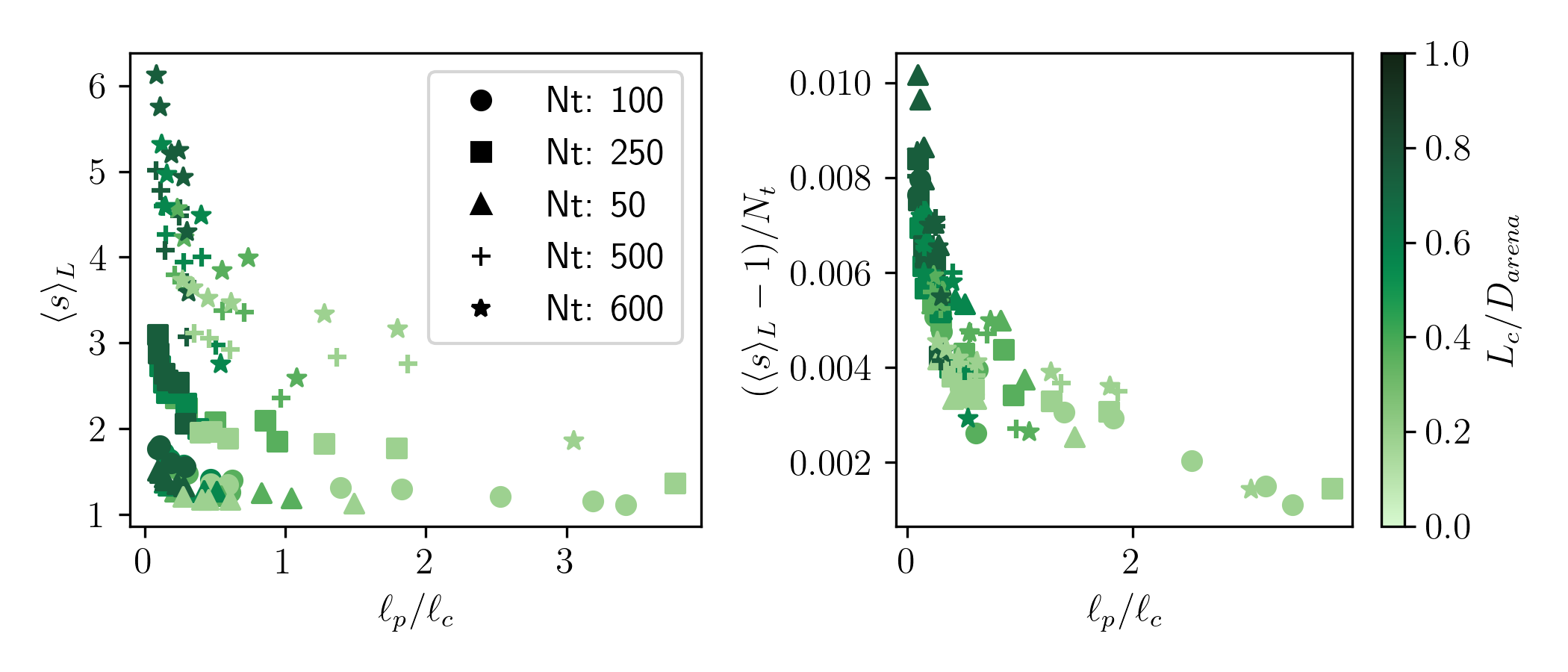}
\caption{\textbf{Effect of initial particle density on the long-time average cluster size.} (Left) The long-time average cluster size from simulations is plotted as a function of the normalized persistence length $\ell_p /\ell_c$. We examined densities ranging from 50 to 600 passive particles. (Right) Same data as in (Left), but rescaled as $(\langle s \rangle_L -1)/N_t$, allowing collapse of all the data on the same curve.} 
\label{fig:Effect_of_ParticleDensity}
\end{figure}

We examined the effect of initial passive particle density on the long-time average cluster size \( \langle s \rangle_L \) using simulations. The results are presented in~Fig.~\ref{fig:Effect_of_ParticleDensity}. By normalizing the long-time average cluster size as  
\((\langle s \rangle_L -1)/N_t\), we achieve a collapse of all data onto a single curve, effectively capturing the influence of the initial total number of passive particles within the confinement.

\clearpage
\begin{figure}
    \centering
    \includegraphics[width=0.8\linewidth]{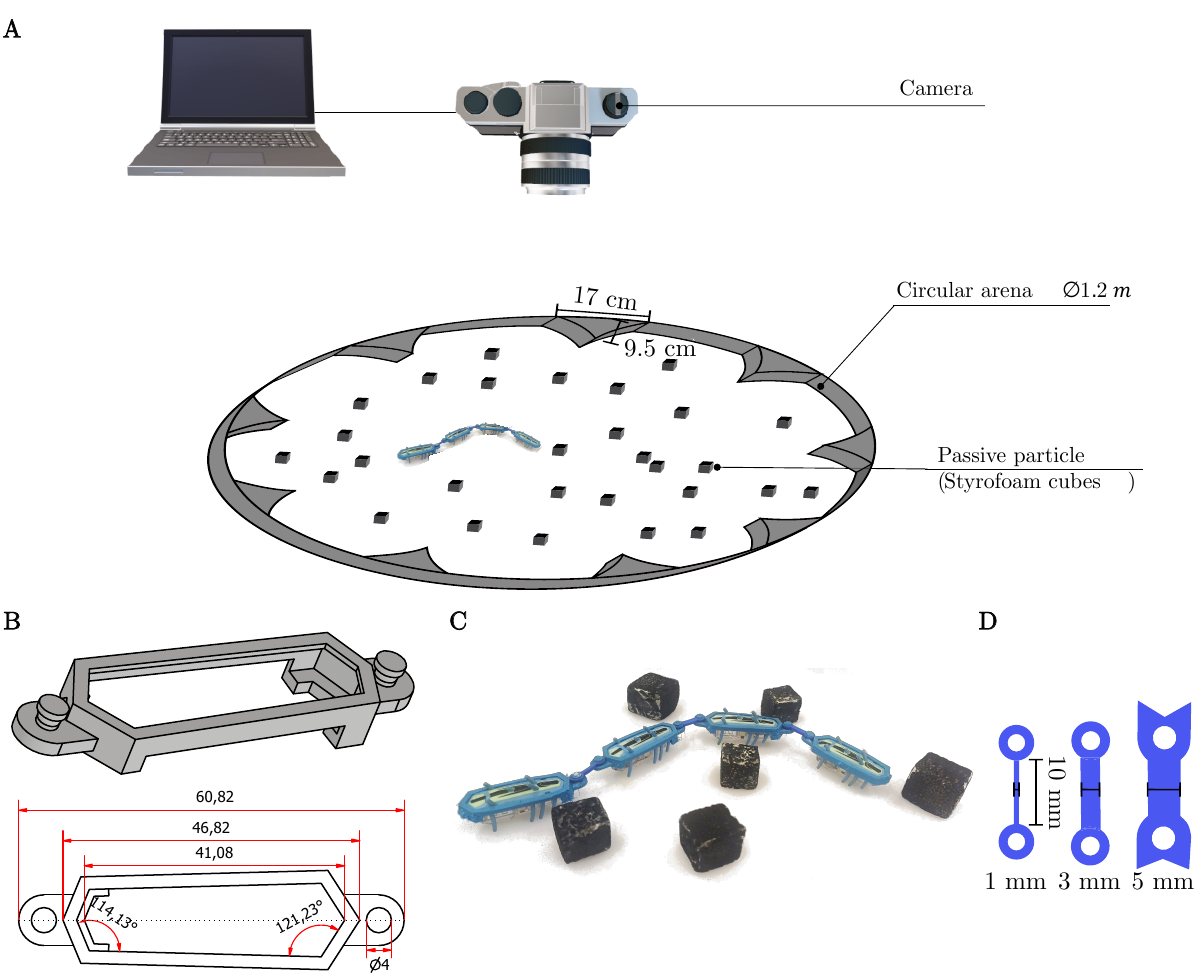}
\caption{\textbf{Experimental setup for particle collection with a robotic chain.}  
(\textbf{A}) The circular arena has a diameter of 1.20 m, with nine ``re-injectors'' evenly spaced around the perimeter. A high-resolution camera is mounted on the ceiling to capture the chain's conformation, dynamics, and the position of passive particles over time.  
(\textbf{B}) Schematic with dimensions of the plastic casing that houses the robots.  
(\textbf{C}) Close-up view of a small chain of robots interacting with the styrofoam cubic particles.  
(\textbf{D}) The three different types of connectors used in the experiments to modify the design and stiffness of the robotic chain.}  
    \label{fig:robot_diagram}
\end{figure}

\section{Robotic filament}
As mentioned in the main text, the robots that are at the basis of our robotic filaments are commercially available bristle bots ``Hexbug'', the Hexbug Nano Nitro. (\url{https://www.hexbug.com/}) Bristle bots are a class of self-propelled robots that exhibit active Brownian motion. They consist out of a body housing some electronics supported by bristles. The bristle bots are able to locomote due to a battery powered off-balance flywheel that causes the robot to vibrate. This vibration causes the weight pressing on the bristles to fluctuate, propelling the bot in a random direction at each oscillation. The bristles underneath the Hexbug nano are slightly curved backward, which causes a bias in the direction the bot is propelled and allows the movement of the bot to be more persistently forward. When constrained at zero velocity, each bot exerts an active force $f^a \sim 15$ mN in the direction of its polarization, tangential to the filament’s axis \cite{zhengrobot-anchor2023}.

To construct our robotic filaments, individual Hexbug units were enclosed in a custom-designed 3D-printed housing. Each housing was equipped with small protrusions at the front and back, allowing for the attachment of flexible sillicone rubber connectors. These connectors, laser-cut to precise dimensions, served as elastic joints between adjacent robots, enabling controlled variations in the filament's flexibility. A schematic representation of the housing and the experimental arena, along with relevant dimensions, is provided in Sup.~Fig.~\ref{fig:robot_diagram}.

\begin{figure}
    \centering
    \includegraphics[width=0.8\linewidth]{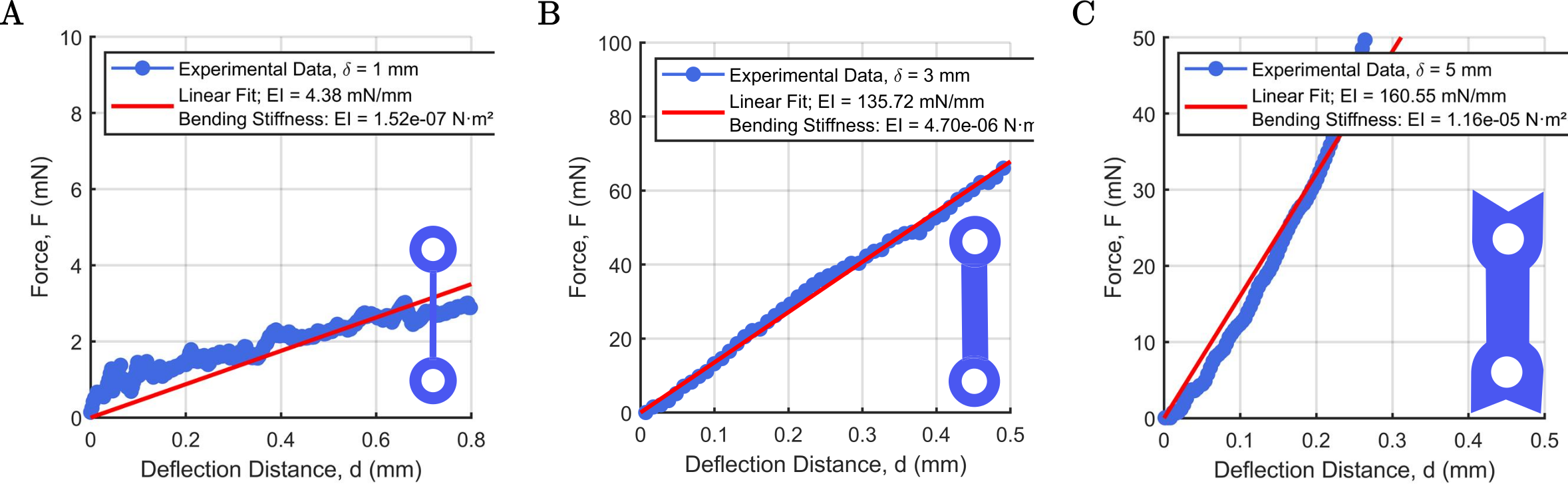}
    \caption{\textbf{Bending stiffness of all elastic bonds used to tune the robotic filaments' flexibility.} The bending stiffness is measured with a rheometer by imposing the deflection of the elastic bond and measuring the force for three different widths: (\textbf{A}) $\delta$ = 1 mm; (\textbf{B}) $\delta$ = 3 mm; (\textbf{C}) $\delta$ = 5 mm.}
    \label{fig:kappa_robot}
\end{figure}

The bending stiffness of the rubber connectors was systematically tuned by adjusting their width, allowing us to investigate robotic filaments with a broad range of persistence lengths. By varying both the number of bots in the chain (contour length) and the stiffness of the connectors, we were able to explore the effects of filament flexibility on clustering dynamics. A summary of the effective persistence lengths and corresponding bending stiffness values for all tested configurations is provided in the Sup.~Table \ref{tab:robot_perisitence} and Sup.~Fig.~\ref{fig:kappa_robot}.

\begin{figure}
    \centering
    \includegraphics[width=0.8\linewidth]{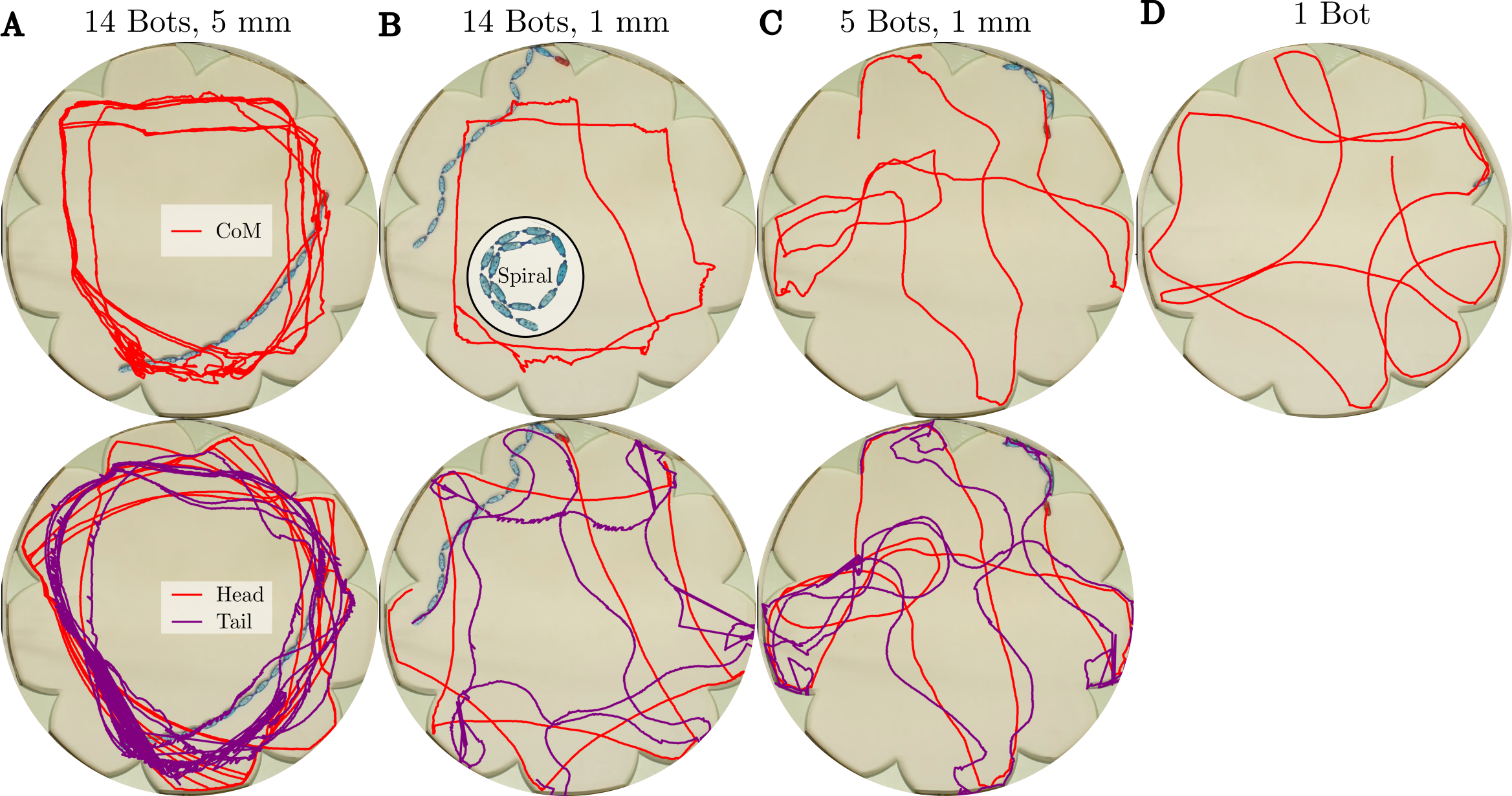}
\caption{\textbf{Trajectories of the head, tail, and center of mass for three representative robotic chains.}  
(\textbf{A}) A long, stiff robotic chain predominantly follows the arena boundary, with the tail bot closely tracing the path of the leading bot.  
(\textbf{B}) A long, flexible robotic chain also remains near the boundary, but the tail bot exhibits pronounced undulations around the path of the leading bot.  
(\textbf{C}) A short robotic chain navigates more frequently through the center of the arena.
(\textbf{D}) Center of mass trajectory of a single Hexbug}  
    \label{fig:robot_trajectories}
\end{figure}

Across different persistence and contour lengths, we observed a variety of distinct motion patterns. Stiffer filaments exhibited more persistent trajectories, often skimming along the arena boundary, whereas more flexible filaments displayed increased curvature in their paths. Extremely long and highly flexible filaments occasionally became trapped in self-induced spirals. This effect was mitigated by reinforcing the frontal connector, reducing excessive bending at the leading end. Representative trajectories illustrating these behaviors are shown in Sup.~Fig.~\ref{fig:robot_trajectories}.

We note that individual bristle bots can exhibit a slight turning bias in their trajectories, despite careful preselection to mitigate this effect. This bias arises from minor asymmetries in their internal construction. While visible in recorded trajectories, its overall impact on clustering efficiency remains negligible.

\subsection{Experimental setup and protocol}

The experiments were conducted in an arena enclosed by a metal barrier, with re-injection defects positioned around the inside perimeter. One hundred styrofoam cubes (20 mm per side) were painted black to serve as passive particles. A camera was positioned above the arena to capture the experiment from a top view perspective (see Sup.~Fig.~\ref{fig:robot_diagram}).

Before each experiment, the tracer particles were evenly distributed within the arena. The robotic filament was then turned on and carefully placed into the arena to minimize disturbance to the particles before the experiment commenced. The recording began immediately after the filament was introduced.

\begin{table}\centering
\begin{tabular}{l|lllll|l}
     & 2 bots  & 6 bots  & 10 bots & 14 bots & 18 bots & bending stiffness \\ \hline
1 mm & 0.64 & 0.57  & 0.33 & 0.24 & 0.31 &   1.52 $\times 10^{-7}$ $Nm^2$   \\
3 mm &         & 0.76 & 0.72 & 0.47 & 0.33 &  4.70 $\times 10^{-6}$ $Nm^2$  \\
5 mm & 0.67 & 0.79 & 0.78 & 0.50 & 0.47 &   1.16 $\times 10^{-5}$ $Nm^2$  
\end{tabular}
\caption{\textbf{Characteristics of the robotic chains.} The average persistence length and bending stiffness of the connectors are reported for the different chain lengths investigated in this work.}
\label{tab:robot_perisitence}
\end{table}
\begin{figure}
    \centering
    \includegraphics[width=\linewidth]{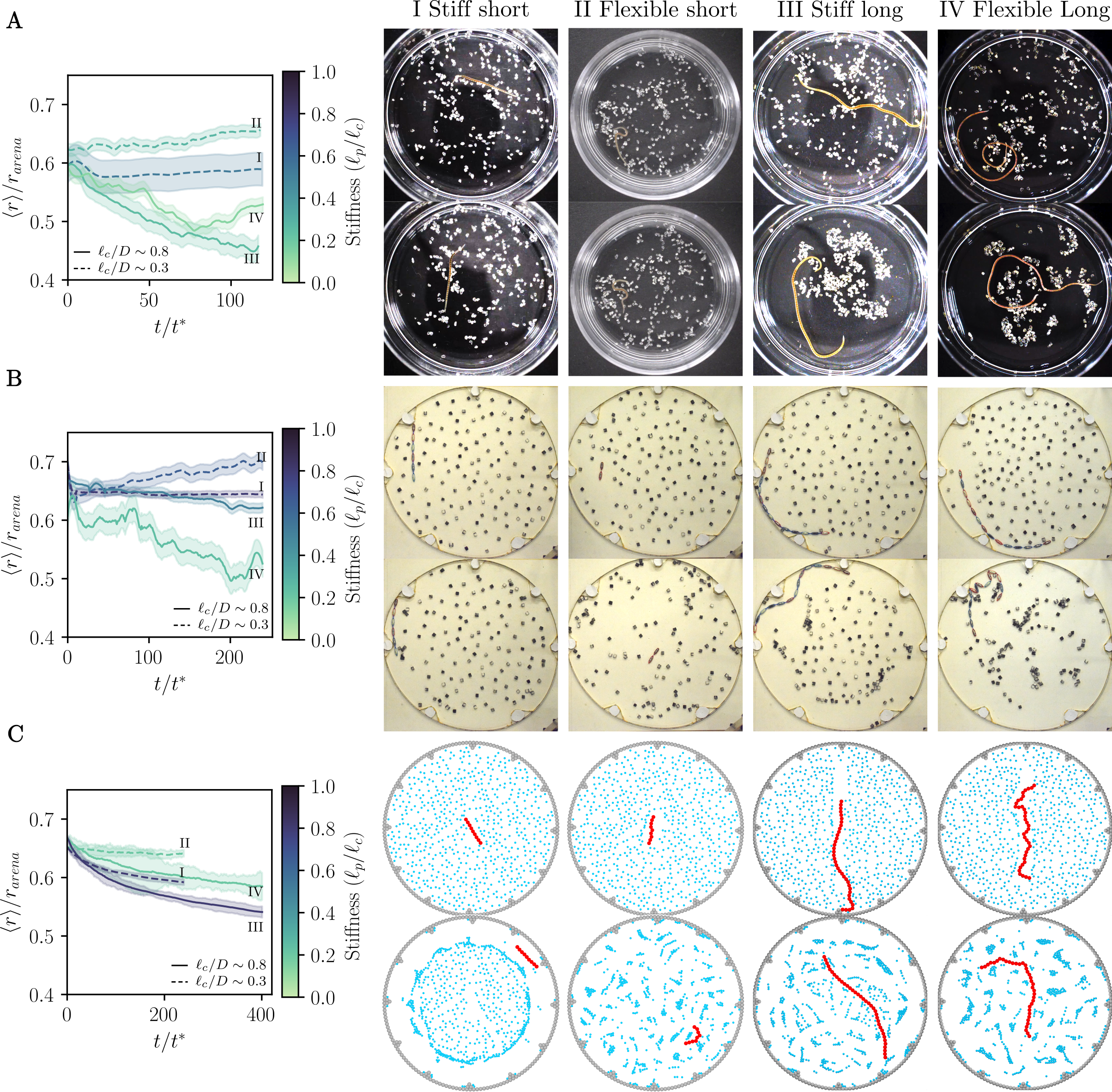}
    \caption{\textbf{Snapshots of the start and end of the experiments and simulations corresponding to figure 3 of the main text.} (\textbf{A}) Start and end of 4 worm experiments. \textbf{A.I} Blackworm, $\ell_c=15$ mm, $\ell_p/\ell_c=0.53$. \textbf{A.II} Blackworm, $\ell_c=16$ mm, $\ell_p/\ell_c=0.28$. \textbf{A.III} Blackworm, $\ell_c=34$ mm, $\ell_p/\ell_c=0.25$. \textbf{A.IV} Tubifex, $\ell_c=46$ mm, $\ell_p/\ell_c=0.13$. Diameter of the Petri dish is 35 mm.
    (\textbf{B}) Start and end of 4 robot experiments. \textbf{B.I} 6 bots, 5 mm connector. \textbf{B.II} 2 bots, 1 mm connector. \textbf{B.III} 14 bots, 5 mm connector. \textbf{B.IV} 14 bots, 1 mm connector. Diameter of the circular arena is 120 cm.
    (\textbf{C}) Start and end of 4 robot experiments. \textbf{C.I} 9 monomers, $\kappa=5.0$. \textbf{C.II} 9 monomers, $\kappa=0.1$. \textbf{C.III} 37 monomers, $\kappa=5.0$. \textbf{C.IV} 37 monomers, $\kappa=0.1$.
    }
    \label{fig:enter-label}
\end{figure}

\newpage

\section{Measuring the sweeping width}
\begin{figure}[h]
    \centering
    \includegraphics[width=0.9\linewidth]{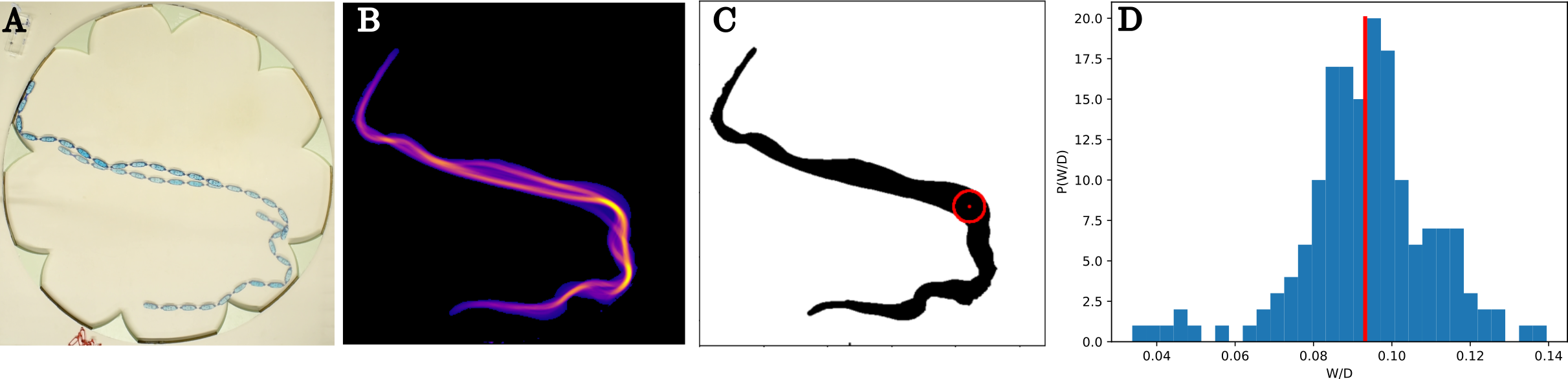}
\caption{\textbf{Method for measuring the sweeping width, \( W \).}  
(\textbf{A}) Superimposed images over a time interval equal to the typical crossing time \( t^* \) of a filament (showing only the start, middle, and end frames for clarity).  
(\textbf{B}) Heatmap of the filament's positions over the time window \( t^* \).  
(\textbf{C}) Binary mask of the heatmap shown in (\textbf{B}), with the largest possible inscribed circle defining the sweeping width \( w \) highlighted in red.  
(\textbf{D}) Distribution of measured \( W \) values over an entire experiment for the filament shown in (\textbf{A}) (14 bots, 1 mm connector). The red line represents the median of this distribution.}  
    \label{fig:w_definition}
\end{figure}

To determine the sweep width $W$ as defined in the main text, we superimposed consecutive images until the filament had traversed its full contour length. In the resulting composite image, we identified the largest inscribed circle fully contained within the contour, taking its diameter as the relevant measure of the sweep width. Repeating this procedure over the full duration of an experiment yielded a distribution of $W$ values, from which we report the median as the characteristic sweep width in the main text.(See Fig.~\ref{fig:w_definition}).

\subsection{Effect of persistence length on the sweep width $W$}

\begin{figure}[h]
    \centering
    \includegraphics[width=0.5\linewidth]{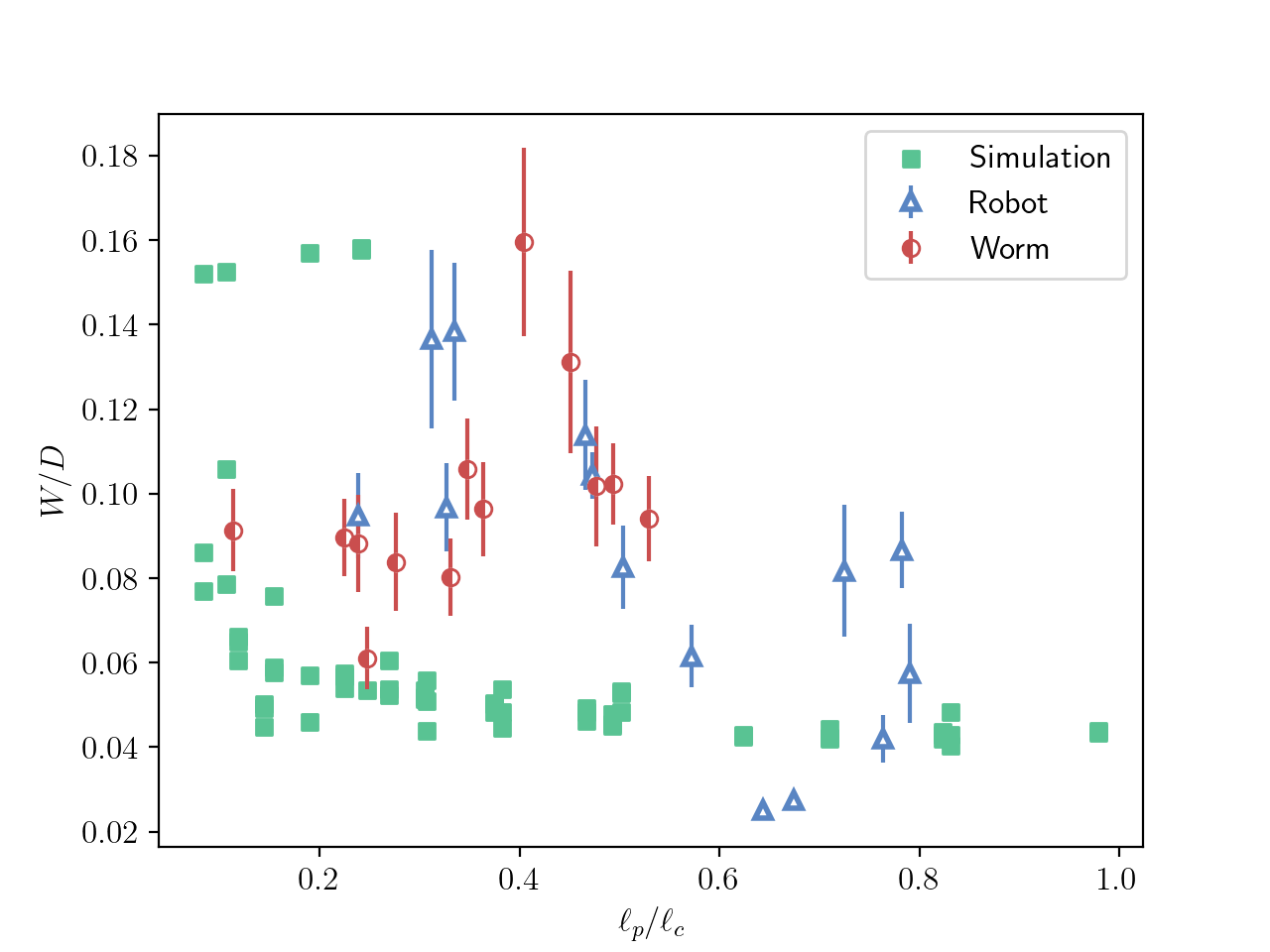}
\caption{\textbf{Sweep width $W$ as a function of the normalized persistence length for all active filament systems.} The same trend is observed between $W$ and filament flexibility across all systems studied.}
    \label{fig:lp_vs_w}
\end{figure}

Longer and more flexible active filaments exhibit larger transverse fluctuations, leading to an increased sweep width $W$. $W$ is measured as explained in section C above. This trend is confirmed in Fig.~\ref{fig:lp_vs_w}, which shows a systematic increase of the sweep width $W$ with filament length and flexibility. However, one can note that it varies across the different systems.

\newpage

\section{Toy model for particle collecting}  

To gain insight into the underlying mechanisms of particle collection observed in our experiments, we develop a minimal computational model, as discussed in Figure~4 of the main text. This simplified model simulates the collection process in an enclosed arena of size $L \times L$, where passive particles are randomly distributed and displaced by repeated sweeping motions.  

We initialize the system as follows:
\begin{itemize}
    \item The arena is represented as a discrete $L \times L$ grid. 
    \item A total of $N_t$ pixels are randomly selected to represent passive particles, ensuring a dilute regime. 
    \item The sweep width is set to $w = 2w^* + 1$, where $w^*$ defines a characteristic half-width of the sweeping motion.  
\end{itemize}

After initialization, the simulation proceeds through the following iterative steps:  
\begin{enumerate}  
    \item A sweep direction (horizontal or vertical) is chosen randomly with equal probability.  
    \item A random position $x$ (for vertical sweeps) or $y$ (for horizontal sweeps) is selected within the arena to place the sweep.  
    \item All tracer particles within the sweep region are displaced:  
        \begin{itemize}  
            \item If a particle's position $x_t$ (or $y_t$) satisfies $x-w < x_t < x$, it is moved leftward (or downward) to $x_t \rightarrow x - w - 1$.  
            \item If a particle's position satisfies $x \leq x_t < x+w$, it is moved rightward (or upward) to $x_t \rightarrow x + w + 1$.  
        \end{itemize}  
    \item If the new location is already occupied, the particle continues moving in the same direction until it reaches an empty spot. In rare cases where no empty spot is found, the simulation terminates.  
    \item The sizes of particle clusters, defined as aggregates of directly neighboring tracer particles, are measured.  
    \item A new direction and position are selected, and the process repeats.  
\end{enumerate}  

For all simulations, the arena size is fixed at $L = 100$, while the number of tracer particles $N_t$ is varied from 100 to 2000 to explore different densities.  

This toy model provides a simple yet effective framework for capturing the essential features of particle aggregation and redistribution driven by sweeping motions.  

\subsection{Comparison to the other systems}

\begin{figure}[h]
    \centering
    \includegraphics[width=0.9\linewidth]{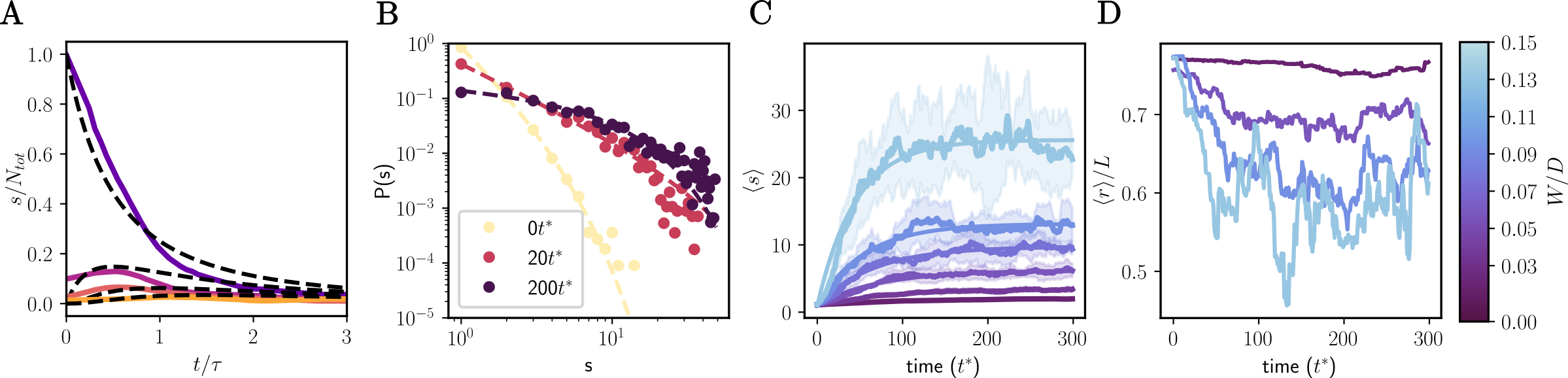}
\caption{\textbf{Results from the toy-model simulations}
\textbf{A} Abundance of particle clusters of sizes 1 to 4 in the minimal model, fitted with Smoluchowski coagulation for a constant kernel. \textbf{B} Cluster size distribution at early, intermediate, and long times, revealing a clear power-law behavior with an exponential cutoff.\textbf{C} Average cluster size in time for different sweeping widths. \textbf{D} Average particle distance from the center of the box for different sweeping widths.}
    \label{fig:sup_model}
\end{figure}

To validate our minimal simulation model in replicating the sweeping process that leads to particle clustering (Fig.~4 of the main text), we compare the same quantities as in the experimental systems. Figure~\ref{fig:sup_model} highlights the relevance of our model in capturing the key clustering dynamics.

\newpage

\section{Smoluchowski-coagulation theory}
\begin{figure}[h]
    \centering
    \includegraphics[width=0.8\linewidth]{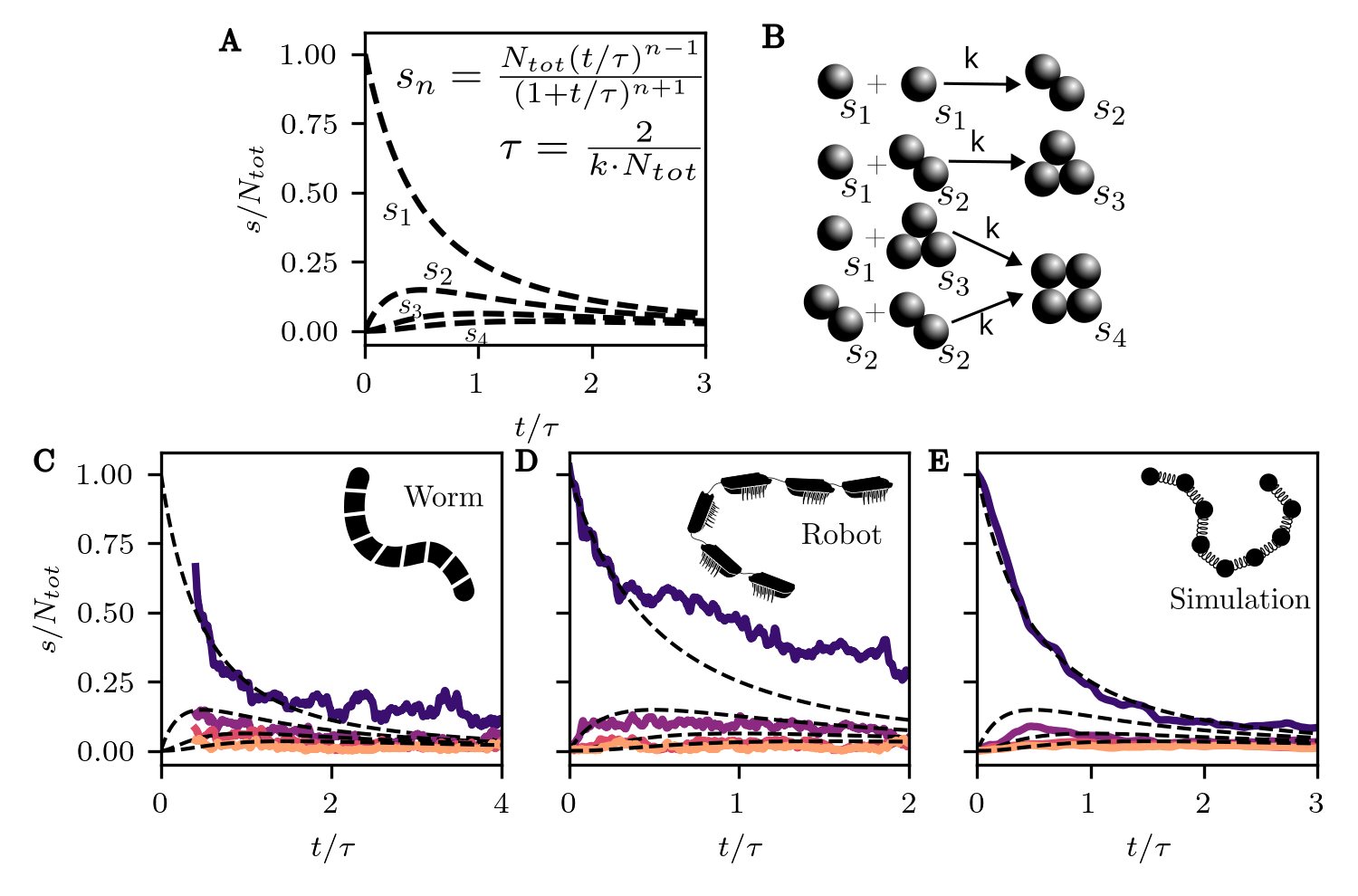}
    \caption{\textbf{Smoluchowski coagluation for a constant kernel.} \textbf{A} Theoretical prediction. \textbf{B} Sketch explaining the terms used. $s_n$ denotes a cluster of size $n$, k is the reaction rate. 
    \textbf{C} Fit to the clustering measured for one of the worms \textbf{D} Fit to the clustering measured for one of the robots
    \textbf{E} Fit to the clustering measured for one of the worms}
    \label{fig:smo}
\end{figure}

A natural framework to describe the observed aggregation is Smoluchowski aggregation theory \cite{Smoluchowski1906}. Assuming no fragmentation and that clusters of all sizes interact equally with any other cluster in the system, the dynamics follow:

\begin{align}
    s_n &= \frac{N_{tot}(t/\tau)^{n-1}}{(1+t/\tau)^{n+1}}, \\
    \tau &= \frac{2}{k\cdot N_{tot}},
\end{align}

where \( s_n \) represents a cluster of size \( n \) and \( \tau \) is the characteristic aggregation timescale. This relation accurately describes the initial stages of aggregation across all systems (see Fig.~\ref{fig:smo}). However, as time progresses, fragmentation becomes significant, rendering this model insufficient to fully capture the observed dynamics.

\begin{table}[h]\centering
\caption{The exponents obtained from the fit of \( P(s) = s^{-\gamma} \exp{(-s/s^{*})} \), as reported in Figure 1 of the main text}
\begin{tabular}{l|llll}
                 & Worm & Simulation & Robot & Model \\ \hline
$0t^*$           & -2.1 & -7.2       & -6.7 & -2.4 \\
$\approx 15 t^*$ & -1.6 & -2.3       & -2.6 & -1.3 \\
$<100 t^*$       & -1.6 & -1.3       & -1.9 & -0.2
\end{tabular}

\end{table}

%%% Add this line AFTER all your figures and tables

\newpage
\noindent \textbf{Movie S1} A \textit{Tubifex~tubifex} worm and a \textit{California~Blackworm} collecting dispersed sand particles in a Petri dish. As the worms move through the enclosed space, they aggregate individual sand grains into clusters and occasionally fragment them. Initially, aggregation dominates, but over time a dynamic balance is reached between aggregation and fragmentation. This evolution is reflected in the average particle size $\langle s\rangle$ shown in the lower panel.
\\
\\
\noindent \textbf{Movie S2} Brownian dynamics simulations of an active polymer aggregating randomly distributed passive athermal particles. The simulation includes $N_t=600$ tracer particles and an active polymer composed of $N=37$ monomers. The bottom panel shows the evolution of the average cluster size $\langle s\rangle$ over time. Two cases are shown: a flexible polymer with persistence length $\ell_p/\ell_c=0.08$ (left), and a stiffer polymer with $\ell_p/\ell_c=0.3$ (right).
\\
\\
\noindent \textbf{Movie S3} Robotic filaments collecting particles. Left: robot with 5 mm wide connectors ($\ell_p/\ell_c=0.5$). Right: robot with 1 mm wide connectors ($\ell_p/\ell_c=0.24$). Both robots consist of 14 monomers. The bottom panel shows the corresponding growth of the average cluster size $\langle s\rangle$ over time.
\\
\\
\noindent \textbf{Movie S4} Example of the sweeping model. The system has box dimensions $L \times L = 100$, with $N_t = 800$ passive particles of size 1 pixel ($0.01 \times 0.01~L$), and a sweep width $W = 11 = 0.11~L$.
\\
\\
\noindent \textbf{Movie S2} Examples of alternative robotic filament designs. By tuning the active force or introducing features such as branching, we can modify their behavior and particle collection efficiency.

\bibliography{bib}